\documentclass[12pt]{article}

\usepackage{amssymb}

\newcommand{\newsection}{\setcounter{equation}{0}\section}

\def\appendix#1{\addtocounter{section}{1}\setcounter{equation}{0}
\renewcommand{\thesection}{\Alph{section}}
\section*{Appendix\thesection\protect\indent \parbox[t]{11.715cm} {#1}}
\addcontentsline{toc}{section}{Appendix \thesection\ \ \ #1} }
\newcommand{\complex}{{\bb C}} 
\newcommand{\zed}{{\bb Z}} 
\newcommand{\real}{{\bb R}} 
\newcommand{\realbm}{{\bbbm R}} 
\newcommand{\zeds}{{\bbs Z}} 
\newcommand{\zedbm}{{\bbbm Z}} 

\def\hil{{\cal H}}

\font\mybbbm=msbm10 at 14pt
\def\bbbm#1{\hbox{\mybbbm#1}}
\font\mybb=msbm10 at 12pt
\def\bb#1{\hbox{\mybb#1}}

\font\mybbs=msbm10 at 9pt
\def\bbs#1{\hbox{\mybbs#1}}

\def\nn{\nonumber}

\newcommand{\Tr}[1]{\:{\rm Tr}\,#1}
\def\e{{\,\rm e}\,}

\newcommand{\non}{\nonumber \\ \nopagebreak}
\newcommand{\nopg}{\\ \nopagebreak}

\def\be{\begin{equation}}
\def\ee{\end{equation}}
\def\bea{\begin{eqnarray}}
\def\eea{\end{eqnarray}}
\def\bd{\begin{displaymath}}
\def\ed{\end{displaymath}}

\def\d{\partial}

\def\dd{{\rm d}}

\def\ii{{\,{\rm i}\,}}

\def\scT{{\sf T}}

\def\scJ{{\sf J}}
\def\scQ{{\sf Q}}

\def\sc\Phi{{\sf \Phi}}
\def\scV{{\sf V}}
\def\scz{{\sf z}}
\def\scw{{\sf w}}
\def\scC{{\sf C}}
\def\scD{{\sf D}}

\def\scOm{{\sf\Omega}}

\def\scx{{\sf x}}
\def\scX{{\sf X}}
\def\scb{{\sf b}}
\def\scc{{\sf c}}
\def\deriv{{\cal D}}

\makeatletter
\newdimen\normalarrayskip              
\newdimen\minarrayskip                 
\normalarrayskip\baselineskip
\minarrayskip\jot
\newif\ifold             \oldtrue            
\def\arraymode{\ifold\relax\else\displaystyle\fi} 
\def\@arrayskip{\ifold\baselineskip\z@\lineskip\z@
     \else
     \baselineskip\minarrayskip\lineskip2\minarrayskip\fi}
\def\@arrayclassz{\ifcase \@lastchclass \@acolampacol \or
\@ampacol \or \or \or \@addamp \or
   \@acolampacol \or \@firstampfalse \@acol \fi
\edef\@preamble{\@preamble
  \ifcase \@chnum
     \hfil$\relax\arraymode\@sharp$\hfil
     \or $\relax\arraymode\@sharp$\hfil
     \or \hfil$\relax\arraymode\@sharp$\fi}}
\def\@array[#1]#2{\setbox\@arstrutbox=\hbox{\vrule
     height\arraystretch \ht\strutbox
     depth\arraystretch \dp\strutbox
     width\z@}\@mkpream{#2}\edef\@preamble{\halign \noexpand\@halignto
\bgroup \tabskip\z@ \@arstrut \@preamble \tabskip\z@ \cr}%
\let\@startpbox\@@startpbox \let\@endpbox\@@endpbox
  \if #1t\vtop \else \if#1b\vbox \else \vcenter \fi\fi
  \bgroup \let\par\relax
  \let\@sharp##\let\protect\relax
  \@arrayskip\@preamble}
\makeatother

\newcommand{\beq}{\begin{eqnarray}}
\newcommand{\eeq}{\end{eqnarray}}

\begin{document}
\begin{titlepage}
\begin{flushright}

\baselineskip=12pt
HWM--02--23\\ EMPG--02--15\\ hep-th/0207273\\
\hfill{ }\\ July 2002
\end{flushright}

\vspace{0.5cm}

\begin{center}

\baselineskip=24pt

{\Large\bf The Neveu-Schwarz and Ramond Algebras of\\ Logarithmic
Superconformal Field Theory}

\baselineskip=14pt

\vspace{2cm}

{\bf N.E. Mavromatos}
\\[3mm]
{\it Department of Physics -- Theoretical Physics\\ King's College
  London\\ Strand, London WC2R 2LS, U.K.}\\ {\tt Nikolaos.Mavromatos@cern.ch}
\\[10mm]
{\bf R.J. Szabo}
\\[3mm]
{\it Department of Mathematics\\ Heriot-Watt University\\ Riccarton,
Edinburgh EH14 4AS, U.K.}\\ {\tt R.J.Szabo@ma.hw.ac.uk}
\\[40mm]

\end{center}

\begin{abstract}

\baselineskip=12pt

We describe the general features of the Neveu-Schwarz and Ramond sectors of
logarithmic conformal field theories with $N=1$ supersymmetry. Three
particular systems are examined in some detail -- D-brane recoil, a
superconformal extension of the $c=-2$ model, and the supersymmetric
$SU(2)_2$ WZW model.

\end{abstract}

\end{titlepage}

{\baselineskip=12pt
\tableofcontents}

\newpage
\setcounter{page}{1}

\newsection{Introduction and Summary}

Logarithmic conformal field theories~\cite{gurarie,lcftfurther} have
recently been attracting a lot of attention because of their diverse range of
applications, from condensed matter models of disorder~\cite{ckt,disorder} to
applications involving gravitational dressing of two-dimensional field
theories~\cite{liouv}, a general analysis of target space symmetries
in string theory~\cite{km}, D-brane recoil~\cite{kmw,ms}, and $AdS$
backgrounds in string theory and also M-theory~\cite{logads}
(see~\cite{LCFTRev} for reviews and more exhaustive lists of
references). They lie on the border between conformally invariant and
general renormalizable field theories in two dimensions. A logarithmic
conformal field theory is characterized by the property that its correlation
functions differ from the standard conformal field theoretic
ones by terms which contain logarithmic branch cuts. Nevertheless, it
is a limiting case of an ordinary conformal field theory which is
still compatible with conformal invariance and which can
still be classified to a certain extent by means of conformal data.

The current understanding of logarithmic conformal field theories
lacks the depth and generality that characterizes the conventional
conformally invariant field theories. Most of the analyses so far
pertain to specific models, and usually to those involving free field
realizations. Nevertheless, some general properties of logarithmic
conformal field theories are now very well understood. For example,
an important deviation from standard conformal field theory is the
non-diagonalizable spectrum of the Virasoro Hamiltonian operator
$L_0$, which connects vectors in a Jordan cell of a certain size. This
implies that the logarithmic operators of the theory, whose
correlation functions exhibit logarithmic scaling violations, come in
pairs, and they appear in the spectrum of a conformal field theory
when two primary operators become degenerate. It would be most
desirable to develop methods that would
classify and analyse the origin of logarithmic singularities in these
models in as general a way as possible, and in particular beyond
the free field prescriptions. Some modest steps in this direction have
been undertaken recently using different approaches. For instance, an
algebraic approach is advocated in~\cite{GabKausch1,Gab1} and used to
classify the logarithmic triplet theory as well as certain
non-unitary, fractional level Wess-Zumino-Witten (WZW) models. The
characteristic features of logarithmic conformal field theories are
described within this setting in terms of the representation theory of
the Virasoro algebra. An alternative approach to the construction of
logarithmic conformal field theories starting from conformally invariant
ones is proposed in~\cite{fuchs}. In this setting, logarithmic behaviour arises
in extended models obtained by appropriately deforming the fields, including
the energy-momentum tensor, in the chiral algebra of an ordinary
conformal field theory.

{}From whatever point of view one wishes to look at logarithmic
conformal field theory, an important issue concerns the nature of the
extensions of these models to include worldsheet supersymmetry. In many
applications, most notably in string theory, supersymmetry plays a
crucial role in ensuring the overall stability of the target space theory.
The purpose of this paper is to analyse in some detail the general
characteristics of the $N=1$ supersymmetric extension of logarithmic
conformal field theory. These models were introduced
in~\cite{CKLT}--\cite{MavSz}, where some features of the
Neveu-Schwarz~(NS) sector of the superconformal algebra were
described. In the following we will extend and elaborate on these
studies, and further incorporate the Ramond~(R) sector of the
theory. In addition to unveiling some general features of logarithmic
superconformal field theories, we shall study in detail how these
novel structures emerge in three explicit realizations.

\subsection{Outline and Summary of Results}

In the remainder of this section we shall outline the structure of the
rest of this paper, and summarize in detail the main
accomplishments of the present work section by section.

\bigskip

\noindent
$\circ~\underline{\it Section~2:~Generalities}$

\noindent
We introduce the relations of the $N=1$ logarithmic superconformal
algebra and define its highest weight representations in both the
Neveu-Schwarz and Ramond sectors of the worldsheet field theory. We
pay particular attention to the structure of the spin operators which
connect the Ramond highest weight states with the vacuum
representation of the Neveu-Schwarz algebra. We show that, in addition
to the standard R sector spin fields of superconformal field theory~\cite{FQS},
there are excited spin fields produced by the logarithmic pairs which
generate, from their action on the NS vacuum module, representations
of an ordinary logarithmic conformal algebra inside the Ramond
algebra. Generally, this yields two orthogonal Jordan cells for the
action of $L_0$ on the Ramond highest weight representation. The
degeneracy is lifted by projecting onto the sector of unbroken Ramond
supersymmetry, leaving a single copy of the logarithmic conformal
field theory inside the R~algebra, analogously to the NS~sector.

\bigskip

\noindent
$\circ~\underline{\it Section~3:~Correlators}$

\noindent
We examine the problem of determining the correlation functions of a
generic logarithmic superconformal field theory. We derive the
appropriate superconformal Ward identities and present explicit
expressions for the two-point and three-point Green's functions of the
logarithmic superfields in the NS sector~\cite{KAG,MavSz}. Using the
local monodromy constraints provided by the spin fields, we derive the
two-point functions of the logarithmic operators in the R sector. We
also derive the two-point functions of all spin fields in the theory,
as well as the mixed three-point correlators among the two types of spin
operators and the fermionic components of the logarithmic
superfields.

\bigskip

\noindent
$\circ~\underline{\it Section~4:~Singular~vectors}$

\noindent
We briefly discuss some issues related to null
states of the logarithmic superconformal field theory. We show that
logarithmic superfields of vanishing superconformal dimension generate
a non-trivial, dynamical fermionic symmetry of the theory. We then
describe the three possible ways to generate local superconformal
algebras involving the spin operators. One way is to project onto the
NS sector, another is to project onto the supersymmetric Ramond ground
state. A third way is to project the chiral symmetry algebra onto
operators of positive fermion parity~\cite{FQS}, which leaves a local quantum
field theory in which the spin operators together completely determine
the entire field content of the logarithmic superconformal field theory.

\bigskip

\noindent
$\circ~\underline{\it Section~5:~D-brane~recoil}$

\noindent
We study the first of three explicit examples and revisit, within the
context of sections~2--4, the well-studied supersymmetric impulse
operators~\cite{MavSz} which describe the recoil of D-particles in string
theory. We complete the program of constructing quantum states which
describe the back-reaction of a D-brane background to the scattering of closed
string states. We exploit the relative simplicity of this model to
give a detailed account of the construction of spin operators,
introducing techniques that are further used in subsequent
sections. We then explicitly construct the target space fermionic
vertex operator in this problem, and thereby complete the construction
of supersymmetric recoil states for D-particles in Type~IIA superstring theory.

\bigskip

\noindent
$\circ~\underline{\it Section~6:~Symplectic~fermions}$

\noindent
We develop the superconformal extension of the best understood
logarithmically extended $c_{p,q}$ minimal model, the $c=-2$ model,
through its local formulation in terms of symplectic
fermions~\cite{Kausch1}. This example is responsible for the onset of
logarithmic behaviour in many problems. The superconformal symplectic
fermion is formulated using the standard expressions for chiral scalar
superfields (with opposite statistics). We construct spin operators
and show that, in contrast to the other examples of this paper which
each possess a unique state of unbroken Ramond supersymmetry, there is
a whole continuum of Ramond ground states labelled by their $U(1)$
charge in the underlying non-compact, fermionic current algebra of the
model. They generate infinitely many inequivalent highest-weight
representations of the Ramond algebra in this case. This infinity
appears to be related to some integrability property of the
superconformal $c=-2$ model. We then examine the superconformal
versions of some of the logarithmically extended algebras that are
normally used to classify the bosonic sector of this theory. We derive
the supersymmetric extension of the ``logarithmic'' $sl(2,\real)$
Kac-Moody symmetry~\cite{Kausch1,KogNich1}, and hence show that the
supersymmetric triplet model is properly classified in terms of a
superconformal enhancement of its underlying $W$-algebra in the
bosonic sector. We also study the fusion algebra in the
$\zed_2$-twisted version of the model~\cite{GabKausch1}, and derive
the fusion rules for the supersymmetric extension which give rise to
the appropriate indecomposable
representations~\cite{GabKausch1,rohsiepe} of the $N=1$ super Virasoro
algebra. This formulates the model in purely algebraic terms, beyond
its explicit realization in terms of symplectic fermions.

\newpage

\noindent
$\circ~\underline{\it Section~7:~WZW~models}$

\noindent
We consider a third and final example which takes us beyond the
minimal models, the $N=1$ supersymmetric WZW model based on the group
$SU(2)$ at level $k\in\zed_+$. These models constitute the simplest
interacting quantum field theories within the present context. By
employing current algebra methods and using an appropriate Coulomb gas
representation, we begin by recalling how the fermionic fields of the
theory decouple such that the bosonic sector of the model is
equivalent to the bosonic $su(2)$ Kac-Moody algebra at level
$k-2$. This representation enables the construction of the spin
operators that define the Ramond sector of the worldsheet field
theory. It further enables a very explicit construction of logarithmic
operators in the theory via a special deformation of
certain primary states of the quantum field theory. Superficially, we
find that these operators exist in the theory for any values of the
level number of the form $k=n(n+1)$, where $n$ is any positive
integer. For these values of $k$ the deformation is exactly marginal,
and hence produces an isomorphic superconformal algebra. Furthermore,
for these levels we can use the fermionic screening operator of the
particular Coulomb gas representation to define a deformation of the
underlying chiral symmetry algebra of the theory, along the lines
of~\cite{fuchs}, and hence recover the logarithmic states through a
precise extension of the Virasoro algebra. Through the supersymmetric
formalism with the given special free field realization of the model,
we are thereby able to exhibit, for the first time, explicitly the
logarithmic operators for this class of models. Furthermore, our
construction, when compared with other models, hints at the fact that
logarithmic operators may be generically present in ordinary
conformal field theories through ``hidden'' deformations and
extensions, as also suggested in~\cite{fuchs}.

Although at present we have found no obstruction to the construction
of these operators at the stated level numbers, we are only able to
rigorously prove that they yield a bonafide logarithmic pair for the
very special value $k=2$. This is the level at which logarithmic
behaviour has been previously observed in these models~\cite{km,CKLT}. In fact,
within the present formalism we can shed some new light on the special
properties of the supersymmetric $SU(2)_2$ model and its logarithmic
behaviour. Let us summarize our findings for the special
characteristics of $k=2$ and contrast them to the cases of affine
$su(2)_k$ algebras with $k>2$:
\begin{itemize}
\item The supersymmetric $su(2)_2$ Kac-Moody algebra is determined by
  a single scalar superfield, and as a consequence it has a unique
  ground state of unbroken Ramond supersymmetry, as anticipated from
  the general arguments of section~2. For $k>2$ the affine algebra is
  described by means of three scalar superfields and the Ramond
  supersymmetry is broken.
\item For $k=2$ the logarithmic operators so constructed can be shown
  to naturally arise through the fusion products of spin $\frac12$
  primary fields, as they are expected to~\cite{km,CKLT}. For $k>2$ it
  is not clear which fusion relations they are associated to, if any.
\item At $k=2$ we can naturally embed the bosonic sector of the WZW
  model into a twisted $N=2$ superconformal algebra~\cite{Gerasimov}
  such that the fermionic screening charge of the WZW model becomes
  the BRST operator of the Felder complexes of Fock space
  representations of the $c=0$ Virasoro algebra. Using the deformation
  provided by the screening, we are then able to explicitly construct
  indecomposable representations in which $L_0$ acquires Jordan
  blocks~\cite{fuchs}. For $k>2$ this construction does not work.
\item For $k=2$ additional generators of a $c=-2$ Virasoro algebra
  appear in the spectrum of the quantum field theory. These operators
  do not generate the logarithmic behaviour associated with the
  deformation, but rather they dress the logarithmic operators to
  ensure that the deformations are marginal and hence that an
  isomorphic superconformal algebra is generated. Instead, through a
  similar deformation process~\cite{fuchs}, the $c=-2$ operators
  generate an independent logarithmic sector of the supersymmetric
  $SU(2)_2$ WZW model. For $k>2$, one may use a Wakimoto
  representation~\cite{wakimoto} of the current algebra to construct
  $c=-2$ sectors of the WZW model for generic values of the level
  $k$~\cite{nichols}. Logarithmic structures then emerge again through
  appropriate deformations of the underlying chiral algebras. However, these
  deformations do not seem to preserve the superconformal algebra.
\end{itemize}
We conclude with a comparison of the $su(2)_k$ affine
algebras and the $sl(2,\real)_k$ WZW models, in which logarithmic
operators are known to emerge for any
$k$~\cite{fuchs,lcftgeneral,giribet}, and also briefly discuss some
extensions and applications to coset current algebras.

\newsection{Definition of the $N=1$ Logarithmic Superconformal
  Algebra\label{SLCFTgen}}

We will start by looking at an abstract logarithmic superconformal field theory
to see what some of the general features are. Throughout we
will deal for simplicity with situations in which the two-dimensional
field theory contains only a single Jordan cell of rank~2, but our
considerations easily extend to more general situations. In this
section we shall begin by discussing how to properly incorporate the
Ramond sector of the theory.

\subsection{Operator Product Expansions}

Consider a logarithmic superconformal field theory defined on the complex plane
$\complex$ (or the Riemann sphere $\complex\cup\{\infty\}$) with
coordinate $z$. For the most part we will only write formulas
explicitly for the holomorphic sector of the two-dimensional field
theory. We will also use a superspace notation, with complex
supercoordinates $\scz=(z,\theta)$, where $\theta$ is a complex
Grassmann variable, $\theta^2=0$. The superconformal algebra is
generated by the holomorphic super energy-momentum tensor
\beq
\scT(\scz)=G(z)+\theta\,T(z)
\label{superEMtensor}\eeq
which is a chiral superfield of dimension $\frac32$. Here $T(z)$ is
the bosonic energy-momentum tensor of conformal dimension 2,
while $G(z)$ is the fermionic supercurrent of dimension $\frac32$ with the
boundary conditions
\beq
G(\e^{2\pi\ii}\,z)=\e^{\pi\ii\lambda}\,G(z) \ ,
\label{Gbcs}\eeq
where $\lambda=0$ in the NS sector of the theory (corresponding to periodic
boundary conditions on the fermion fields) and $\lambda=1$ in the R sector
(corresponding to anti-periodic boundary conditions).

The $N=1$ superconformal algebra may then be characterized by the
anomalous operator product expansion
\beq
\scT(\scz_1)\,\scT(\scz_2)=\frac{\hat c}4\,\frac1{(\scz_{12})^3}+
\frac{2\theta_{12}}{(\scz_{12})^2}\,\scT(\scz_2)+\frac12\,\frac1{\scz_{12}}
\,\deriv_{\scz_2}\scT(\scz_2)+\frac{\theta_{12}}{\scz_{12}}\,
\partial_{z_2}\scT(\scz_2)+\dots \ ,
\label{TOPE}\eeq
where in general we introduce the variables
\beq
\scz_{ij}=z_i-z_j-\theta_i\theta_j \ , ~~
\theta_{ij}=\theta_i-\theta_j
\label{zijthetaijdef}\eeq
corresponding to any set of holomorphic superspace coordinates
$\scz_i=(z_i,\theta_i)$. Here
\beq
\deriv_{\scz}=\partial_\theta+\theta\,\partial_z \ , ~~
\deriv_\scz^2=\partial_z
\label{supercovderivdef}\eeq
is the superspace covariant derivative, and $\hat
c=2c/3$ is the superconformal central charge
with $c$ the ordinary Virasoro central charge. An ellipsis will always
denote terms which are regular in the operator product expansion
as $\scz_1\to\scz_2$. By introducing the usual mode expansions
\bea
T(z)&=&\sum_{n=-\infty}^\infty L_n~z^{-n-2} \ , \non
G(z)&=&\sum_{n=-\infty}^\infty\,\frac12\,G_{n+(1-\lambda)/2}
{}~z^{-n-2+\lambda/2}
\label{TGmodes}\eea
with $L_n^\dag=L_{-n}$ and $G_r^\dag=G_{-r}$, the operator product expansion
(\ref{TOPE}) is equivalent to the usual relations of the $N=1$ supersymmetric
extension of the Virasoro algebra,
\bea
[L_m,L_n]&=&(m-n)L_{m+n}+\frac{\hat c}8\,\Bigl(m^3-m\Bigr)\,\delta_{m+n,0}
\ , \non  {~}[L_m,G_r]&=&\left(\frac m2-r\right)G_{m+r} \ ,
\non \{G_r,G_s\}&=&2L_{r+s}+\frac{\hat c}2\,\left(r^2-\frac14\right)\,
\delta_{r+s,0} \ ,
\label{SUSYViralg}\eea
where $m,n\in\zed$, and $r,s\in\zed+\frac12$ for the NS algebra while
$r,s\in\zed$ for the R algebra. In particular, the five operators
$L_0$, $L_{\pm1}$ and $G_{\pm1/2}$ generate the orthosymplectic Lie
algebra of the global superconformal group $OSp(2,1)$.

In the simplest instance, logarithmic superconformal operators of
weight $\Delta_C$ correspond to a pair of superfields
\bea
\scC(\scz)&=&C(z)+\theta\,\chi^{~}_{C}(z) \ , \non
\scD(\scz)&=&D(z)+\theta\,\chi^{~}_{D}(z)
\label{CDsuperfields}\eea
which have operator product expansions with the super
energy-momentum tensor given by~\cite{KAG,MavSz}
\bea
\scT(\scz_1)\,\scC(\scz_2)&=&\frac{\Delta_C\,\theta_{12}}{(\scz_{12})^2}\,
\scC(\scz_2)+\frac12\,\frac1{\scz_{12}}\,\deriv_{\scz_2}\scC(\scz_2)+
\frac{\theta_{12}}{\scz_{12}}\,\partial_{z_2}\scC(\scz_2)+\dots \ , \non
\scT(\scz_1)\,\scD(\scz_2)&=&\frac{\Delta_C\,\theta_{12}}{(\scz_{12})^2}\,
\scD(\scz_2)+\frac{\theta_{12}}{(\scz_{12})^2}\,\scC(\scz_2)+
\frac12\,\frac1{\scz_{12}}\,\deriv_{\scz_2}\scD(\scz_2)+
\frac{\theta_{12}}{\scz_{12}}\,\partial_{z_2}\scD(\scz_2)+\dots \ . \non &&
\label{CDsuperOPE}\eea
Note that $\scC(\scz)$ is a primary superfield of the superconformal
algebra of dimension $\Delta_C$, which is necessarily an
integer~\cite{ckt}. The appropriately normalized superfield
$\scD(\scz)$ is its quasi-primary logarithmic partner. This latter
assumption, i.e. that $[L_n,\scD(z)]=[G_r,\scD(z)]=0$ for $n,r>0$, is
not necessary, but it will simplify some of the arguments which
follow. The operators $C(z)$ and $D(z)$ correspond to an ordinary
logarithmic pair and their superpartners $\chi_C^{~}(z)$ and
$\chi_D^{~}(z)$ are generated through the operator products with the
fermionic supercurrent as
\bea
G(z)\,C(z)&=&\frac{1/2}{z-w}\,\chi_C^{~}(w)+\dots \ , \non
G(z)\,D(z)&=&\frac{1/2}{z-w}\,\chi_D^{~}(w)+\dots \ .
\label{GCDchiexpl}\eea
In particular, in the NS algebra we may write the superpartners as
$\chi_C^{~}(z)=[G_{-1/2},C(z)]$ and $\chi_D^{~}(z)=[G_{-1/2},D(z)]$.

\subsection{Highest-Weight Representations\label{HighestWeight}}

The quantum Hilbert space $\cal H$ of the superconformal field theory
decomposes into two subspaces,
\beq
{\cal H}={\cal H}_{\rm NS}\oplus{\cal H}_{\rm R} \ ,
\label{Hilbertsplit}\eeq
corresponding to the two types of boundary conditions obeyed by the fermionic
fields. They carry the representations of the NS and R algebras, respectively.
In this space, we assume that some of the highest-weight representations of the
$N=1$ superconformal algebra are
indecomposable~\cite{GabKausch1,rohsiepe}. Then a (rank 2)
highest-weight Jordan cell of energy $\Delta_C$ is generated by a pair of
appropriately normalized states $|C\rangle$, $|D\rangle$ obeying the conditions
\bea
L_0|C\rangle&=&\Delta_C|C\rangle \ , \nn \\L_0|D\rangle&=&\Delta_C|D\rangle+
|C\rangle \ , \non L_n|C\rangle&=&L_n|D\rangle~=~0 \ , ~~ n>0 \ , \non
G_r|C\rangle&=&G_r|D\rangle~=~0 \ , ~~ r>0 \ .
\label{highestwtdef}\eea
A highest-weight representation of the logarithmic superconformal algebra is
then generated by applying the raising operators $L_n$, $G_r$, $n,r<0$ to these
vectors giving rise to the descendant states of the theory. Note that
$|C\rangle$ is a highest-weight state of the irreducible sub-representation of
the superconformal algebra contained in the Jordan cell.

\subsubsection*{Neveu-Schwarz Sector}

The NS sector ${\cal H}_{\rm NS}$ of the Hilbert space contains the
normalized, $OSp(2,1)$-invariant vacuum state $|0\rangle$ which is the unique
state of lowest energy $\Delta=0$ in a unitary theory,
\beq
L_0|0\rangle=0 \ .
\label{L0vacuum}\eeq
In this sector, the states defined by (\ref{highestwtdef}) are in a one-to-one
correspondence with the logarithmic operators satisfying the operator product
expansions (\ref{CDsuperOPE}). Namely, under the usual operator-state
correspondence of local quantum field theory, the superfields $\scC(\scz)$
and $\scD(\scz)$ are associated with highest weight states of energy
$\Delta_C$ through
\bea
C(0)|0\rangle&=&|C\rangle^{~}_{\rm NS} \ , \non
\chi^{~}_C(0)|0\rangle&=&G_{-1/2}|C\rangle^{~}_{\rm NS} \ , \non
D(0)|0\rangle&=&|D\rangle^{~}_{\rm NS} \ , \non
\chi^{~}_D(0)|0\rangle&=&G_{-1/2}|D\rangle^{~}_{\rm NS} \ .
\label{NSopstate}\eea
In this way, the NS sector is formally analogous to an ordinary, bosonic
logarithmic conformal field theory. Note that the vacuum state
$|0\rangle$ itself corresponds to the identity operator $I$.

\subsubsection*{Ramond Sector}

Things are quite different in the R sector $\hil_{\rm R}$. Consider a
highest weight state $|\Delta\rangle^{~}_{\rm R}$ of energy $\Delta$,
\beq
L_0|\Delta\rangle^{~}_{\rm R}=\Delta|\Delta\rangle^{~}_{\rm R} \ .
\label{L0h}\eeq
{}From the superconformal algebra (\ref{SUSYViralg}), we see that the operators
$L_0$ and $G_0$ commute in the R sector, so that the supercurrent zero
mode $G_0$ acts on the highest weight states. As a consequence, the state
$G_0|\Delta\rangle^{~}_{\rm R}$ also has energy $\Delta$. Therefore,
the highest weight states of the R sector $\hil_{\rm R}$ come in
orthogonal pairs $|\Delta\rangle^{~}_{\rm R}$,
$G_0|\Delta\rangle^{~}_{\rm R}$ of the same energy. Under the
operator-state correspondence, the Ramond highest weight states are
created from the vacuum $|0\rangle$ by the application of spin fields
$\Sigma^\pm_\Delta(z)$~\cite{FQS} which are ordinary conformal fields
of dimension $\Delta$,
\bea
\Sigma_\Delta^+(0)|0\rangle&=&|\Delta\rangle^{~}_{\rm R} \ , \non
\Sigma_\Delta^-(0)|0\rangle&=&G_0|\Delta\rangle^{~}_{\rm R} \ .
\label{hpmspinfield}\eeq

The operator product expansions of the spin fields with the super
energy-momentum tensor may be computed from (\ref{L0h}) and
(\ref{hpmspinfield}) and are given by
\bea
T(z)\,\Sigma_\Delta^\pm(w)&=&\frac\Delta{(z-w)^2}\,\Sigma_\Delta^\pm(w)+
\frac1{z-w}\,\partial_w\Sigma_\Delta^\pm(w)+\dots \ ,
\label{TSigmaOPE}\nopg G(z)\,\Sigma_\Delta^+(w)&=&\frac12\,
\frac1{(z-w)^{3/2}}\,\Sigma_\Delta^-(w)+\dots \ ,
\label{GSigma+OPE}\nopg G(z)\,\Sigma_\Delta^-(w)&=&\frac12\,
\left(\Delta-\frac{\hat c}{16}\right)\,
\frac1{(z-w)^{3/2}}\,\Sigma_\Delta^+(w)+\dots \ ,
\label{GSigma-OPE}\eea
where we have used the super-Virasoro algebra (\ref{SUSYViralg}) to
write
\beq
G_0^2=L_0-\frac{\hat c}{16} \ .
\label{G02L0}\eeq
The operator product (\ref{TSigmaOPE}) merely states that
$\Sigma_\Delta^\pm(z)$ is a dimension $\Delta$ primary field
of the ordinary, bosonic Virasoro algebra, while (\ref{GSigma+OPE})
and (\ref{GSigma-OPE}) show that the fermionic supercurrent
$G(z)$ is double-valued with respect to the spin fields, since they are
equivalent to the monodromy conditions
\beq
G(\e^{2\pi\ii}\,z)\,\Sigma_\Delta^\pm(w)=-G(z)\,\Sigma_\Delta^\pm(w) \ .
\label{GSigmamonodromy}\eeq
It follows that Ramond
boundary conditions can be regarded as due to a branch cut in the complex plane
connecting the spin fields $\Sigma_\Delta^\pm(z)$ at $z=0$ and
$z=\infty$. The spin fields make the entire superconformal field
theory non-local, and correspond to the irreducible representations of
the Ramond algebra. Note that the ordinary superfields are block
diagonal with respect to the decomposition (\ref{Hilbertsplit}),
i.e. they are operators on $\hil_{\rm NS}\to\hil_{\rm NS}$ and
$\hil_{\rm R}\to\hil_{\rm R}$, while the spin fields
$\Sigma_\Delta^\pm:\hil_{\rm NS}\to\hil_{\rm R}$ are block
off-diagonal.

The spin fields $\Sigma_\Delta^\pm(z)$ do not affect the integer weight fields
$C(z)$ and $D(z)$, while their operator product expansions with the
fermionic partners to the logarithmic operators in the R sector are given by
\bea
\chi^{~}_C(z)\,\Sigma_\Delta^\pm(w)&=&\frac1{\sqrt{z-w}}\,
\widetilde{\Sigma}^\pm_{C,\Delta}(w)+\dots
\ , \non\chi^{~}_D(z)\,\Sigma_\Delta^\pm(w)&=&\frac1{\sqrt{z-w}}\,
\widetilde{\Sigma}^\pm_{D,\Delta}(w)+\dots \ .
\label{chiCDSigma}\eea
The relations (\ref{chiCDSigma}) define two different excited twist fields
$\widetilde{\Sigma}^\pm_{C,\Delta}(z)$ and
$\widetilde{\Sigma}^\pm_{D,\Delta}(z)$ which are conjugate to the spin fields
$\Sigma_\Delta^\pm(z)$. They are also double-valued with respect to
$\chi^{~}_C$ and $\chi^{~}_D$, respectively, and they each act within
the Ramond sector as operators on ${\cal H}_{\rm NS}\to{\cal H}_{\rm
  R}$. The relative non-locality of the operator product
expansions (\ref{chiCDSigma}) yields the global $\zed_2$-twists in the
boundary conditions required of the R sector fermionic fields.

While $\widetilde{\Sigma}_{C,\Delta}^\pm(z)$ are primary fields of
conformal dimension $\Delta_C+\Delta$, the conjugate spin fields
$\widetilde{\Sigma}^\pm_{D,\Delta}(z)$ exhibit
logarithmic mixing behaviour. This can be seen explicitly by applying the
operator product expansions to both sides of (\ref{chiCDSigma}) using
(\ref{CDsuperOPE}) and (\ref{TSigmaOPE})--(\ref{GSigma-OPE}) to get
\bea
T(z)\,\widetilde{\Sigma}_{C,\Delta}^\pm(w)&=&\frac{\Delta_C+\Delta}{(z-w)^2}\,
\widetilde{\Sigma}_{C,\Delta}^\pm(w)+
\frac1{z-w}\,\partial_w\widetilde{\Sigma}_{C,\Delta}^\pm(w)+\dots \ ,
\label{TSigmaC}\nopg
T(z)\,\widetilde{\Sigma}_{D,\Delta}^\pm(w)&=&\frac{\Delta_C+\Delta}{(z-w)^2}\,
\widetilde{\Sigma}_{D,\Delta}^\pm(w)+
\frac1{(z-w)^2}\,\widetilde{\Sigma}_{C,\Delta}^\pm(w)+\frac1{z-w}\,\partial_w
\widetilde{\Sigma}_{D,\Delta}^\pm(w)+
\dots \ , \non&&\label{TSigmaD}\nopg
G(z)\,\widetilde{\Sigma}_{C,\Delta}^+(w)&=&\frac12\,\frac1{(z-w)^{3/2}}\,
\widetilde{\Sigma}_{C,\Delta}^-(w)+
\dots \ , \label{GSigmaC+}\nopg
G(z)\,\widetilde{\Sigma}_{C,\Delta}^-(w)&=&\frac12\,\left(\Delta-\frac{\hat
    c}{16}\right)\,\frac1{(z-w)^{3/2}}\,\widetilde{\Sigma}_{C,\Delta}^+(w)+
\dots \ , \label{GSigmaC-}\nopg
G(z)\,\widetilde{\Sigma}_{D,\Delta}^+(w)&=&\frac12\,\frac1{(z-w)^{3/2}}\,
\widetilde{\Sigma}_{D,\Delta}^-(w)+\dots \ , \label{GSigmaD+}\nopg
G(z)\,\widetilde{\Sigma}_{D,\Delta}^-(w)&=&\frac12\,\left(\Delta-\frac{\hat
    c}{16}\right)\,\frac1{(z-w)^{3/2}}\,
\widetilde{\Sigma}_{D,\Delta}^+(w)+\dots \ .
\label{GSigmaD-}\eea
The operator product expansions (\ref{TSigmaC}) and (\ref{TSigmaD})
yield a pair of ordinary, bosonic logarithmic conformal algebras,
while (\ref{GSigmaC+})--(\ref{GSigmaD-}) show that both
$\widetilde{\Sigma}_{C,\Delta}^\pm(z)$ and
$\widetilde{\Sigma}_{D,\Delta}^\pm(z)$ twist the fermionic supercurrent
$G(z)$ in exactly the same way that the original spin fields
$\Sigma_\Delta^\pm(z)$ do. In particular, the set of degenerate spin fields
$\widetilde{\Sigma}_{C,\Delta}^\pm(z)$,
$\widetilde{\Sigma}_{D,\Delta}^\pm(z)$ generate a pair of reducible but
indecomposable representations (\ref{highestwtdef}) of the R algebra,
of the {\it same} shifted weight $\Delta_C+\Delta$. The corresponding
excited highest-weight states $|C,\Delta\rangle^\pm_{\rm R}$,
$|D,\Delta\rangle^\pm_{\rm R}$ of the mutually
orthogonal degenerate Jordan blocks for the action of the Virasoro
operator $L_0$ on ${\cal H}_{\rm R}$ are created from the NS ground
state through the application of the logarithmic spin operators as
\bea
\widetilde{\Sigma}_{C,\Delta}^\pm(0)|0\rangle&=&|C,\Delta
\rangle^\pm_{\rm R} \ , \non\widetilde{\Sigma}_{D,\Delta}^\pm(0)
|0\rangle&=&|D,\Delta\rangle^\pm_{\rm R} \ ,
\label{SigmaCDpmdef}\eea
with
\bea
L_0|C,\Delta\rangle^\pm_{\rm R}&=&(\Delta_C+\Delta)|C,\Delta
\rangle^\pm_{\rm R} \ , \non L_0|D,\Delta\rangle^\pm_{\rm R}&=&
(\Delta_C+\Delta)|D,\Delta\rangle^\pm_{\rm R}+
|C,\Delta\rangle^\pm_{\rm R} \ .
\label{L0CDhpm}\eea

In the following we will be primarily interested in the spin fields associated
with the Ramond ground state $|\Delta\rangle^{~}_{\rm R}$ which is defined
by the condition $G_0|\Delta\rangle_{\rm R}^{~}=0$. This lifts the
degeneracy of the highest weight representation which by (\ref{G02L0})
necessarily has dimension $\Delta=\hat c/16$, corresponding to the
lowest energy in a unitary theory whereby $G_0^2\geq0$. In this case, the
Ramond state $G_0|\Delta\rangle^{~}_{\rm R}$ is a null vector and the R
sector contains a single copy of the logarithmic superconformal
algebra, as in the NS sector. We will return to the issue of
logarithmic null vectors within this context in
section~\ref{nullvectors}. The spin field
$\Sigma_{\hat c/16}^-(z)$ is then an irrelevant operator and may be
set to zero, while the other spin field will be simply denoted by
$\Sigma(z)\equiv\Sigma_{\hat c/16}^+(z)$. The spin field $\Sigma(z)$
corresponds to the unique supersymmetric ground state $|\frac{\hat
  c}{16}\rangle_{\rm R}^{~}$ of the Ramond system, with supersymmetry
generator $G_0$, in the logarithmic superconformal field
theory. Similarly, we may set $\widetilde{\Sigma}_{C,\hat
  c/16}^-(z)=\widetilde{\Sigma}_{D,\hat c/16}^-(z)=0$, and we denote the
remaining excited spin fields simply by
$\widetilde{\Sigma}_C(z)\equiv\widetilde{\Sigma}_{C,\hat
  c/16}^+(z)$ and $\widetilde{\Sigma}_D(z)\equiv\widetilde{\Sigma}^+_{D,\hat
c/16}(z)$.

\newsection{Correlation Functions\label{Correlators}}

Carrying on with an abstract logarithmic superconformal algebra, we
shall now describe the structure of logarithmic correlation functions
in both the NS and R sectors. In particular, we will determine all
two-point correlators involving the various logarithmic operators.

\subsection{Ward Identities and Neveu-Schwarz Correlation
  Functions\label{Ward}}

In the NS sector, we define the correlator of any periodic operator $\sf O$ as
its vacuum expectation value
\beq
\langle{\sf O}\rangle^{~}_{\rm NS}=\langle0|{\sf O}|0\rangle \ .
\label{NCcorrelator}\eeq
Such correlators of logarithmic operators, and their descendants, may be
derived as follows. Consider a collection of Jordan blocks in
the superconformal field theory of rank~2, weight $\Delta_{C_i}$, and
spanning logarithmic superfields $\scC_i(\scz)$, $\scD_i(\scz)$. Then, in
the standard way, we may deduce from the operator product expansions
(\ref{CDsuperOPE}) the superconformal Ward identities

\vbox{\bea
&&\Bigl\langle\scT(\scz)\,\scC_n(\scz_n)\cdots\scC_{n+k}(\scz_{n+k})\,
\scD_m(\scw_m)\cdots\scD_{m+l}(\scw_{m+l})\Bigr\rangle_{\rm
 NS}\non&&~~=~\left(\,\sum_{i=n}^{n+k}\left[\frac12\,\frac1{z-z_i-\theta\,
\theta_i}\,\deriv_{\scz_i}+\frac{\theta-\theta_i}{z-z_i-\theta\,\theta_i}
\,\partial_{z_i}+\frac{\Delta_{C_i}\,(\theta-\theta_i)}
{(z-z_i-\theta\,\theta_i)^2}\right]\right.\non&&~~~~~~+\left.
\sum_{i=m}^{m+l}\left[\frac12\,\frac1{z-w_i-\theta\,\zeta_i}
\,\deriv_{\scw_i}+\frac{\theta-\zeta_i}{z-w_i-\theta\,\zeta_i}
\,\partial_{w_i}+\frac{\Delta_{C_i}\,(\theta-\zeta_i)}
{(z-w_i-\theta\,\zeta_i)^2}\right]\right)
\non&&~~~~~~\times\,\Bigl\langle\scC_n(\scz_n)\cdots\scC_{n+k}(\scz_{n+k})\,
\scD_m(\scw_m)\cdots\scD_{m+l}(\scw_{m+l})\Bigr\rangle_{\rm
  NS}\non&&~~~~~~+\,\sum_{i=m}^{m+l}\frac{\theta-\zeta_i}
{(z-w_i-\theta\,\zeta_i)^2}
\,\Bigl\langle\scC_n(\scz_n)\cdots\scC_{n+k}(\scz_{n+k})\Bigr.\non&&~~~~~~
\times\Bigl.\scD_m(\scw_m)\cdots\scD_{i-1}(\scw_{i-1})\,
\scC_i(\scw_i)\,\scD_{i+1}(\scw_{i+1})\cdots
\scD_{m+l}(\scw_{m+l})\Bigr\rangle_{\rm NS} \ ,
\label{susyWardids}\eea}
\noindent
where the supercoordinates in (\ref{susyWardids}) are $\scz=(z,\theta)$,
$\scz_i=(z_i,\theta_i)$ and $\scw_i=(w_i,\zeta_i)$. These identities
can be used to derive correlation functions of descendants of the logarithmic
operators in terms of those involving the original superfields
$\scC_i$ and $\scD_i$. Notice, in particular, that the Ward identity
connects amplitudes of the descendants of $\scD_i$ with amplitudes
involving the primary superfields~$\scC_i$.

By expanding the super
energy-momentum tensor into modes using (\ref{TGmodes}) we may equate
the coefficients on both sides of (\ref{susyWardids}) corresponding to
the actions of the $OSp(2,1)$ generators $L_0$, $L_{\pm1}$ and
$G_{\pm1/2}$. By using global superconformal invariance of the vacuum
state $|0\rangle$, we then arrive at a set of superfield differential equations
\bea
0&=&\left(\,\sum_{i=n}^{n+k}\deriv_{\scz_i}+\sum_{i=m}^{m+l}\deriv_{\scw_i}
\right)\Bigl\langle\scC_n(\scz_n)\cdots\scC_{n+k}(\scz_{n+k})\,
\scD_m(\scw_m)\cdots\scD_{m+l}(\scw_{m+l})\Bigr\rangle_{\rm NS} \ , \non&&
{~~}^{~~}_{~~}\non
0&=&\left(\,\sum_{i=n}^{n+k}\Bigl[z_i\,\deriv_{\scz_i}+\theta_i\,
\partial_{\theta_i}+2\Delta_{C_i}\Bigr]+\sum_{i=m}^{m+l}\Bigl[w_i\,
\deriv_{\scw_i}+\zeta_i\,\partial_{\zeta_i}+2\Delta_{C_i}
\Bigr]\right)\non&&\times\,\Bigl\langle\scC_n(\scz_n)\cdots
\scC_{n+k}(\scz_{n+k})\,\scD_m(\scw_m)\cdots\scD_{m+l}(\scw_{m+l})
\Bigr\rangle_{\rm NS}\non&&+\,2\,\sum_{i=m}^{m+l}\Bigl\langle
\scC_n(\scz_n)\cdots\scC_{n+k}(\scz_{n+k})\Bigr.\non&&\times\Bigl.
\scD_m(\scw_m)\cdots\scD_{i-1}(\scw_{i-1})\,
\scC_i(\scw_i)\,\scD_{i+1}(\scw_{i+1})\cdots
\scD_{m+l}(\scw_{m+l})\Bigr\rangle_{\rm NS} \ ,\nonumber
\eea

\vbox{\bea
0&=&\left(\,\sum_{i=n}^{n+k}
\left[z_i^2\,\deriv_{\scz_i}+z_i\,(\theta_i\,\partial_{\theta_i}+2
\Delta_{C_i})\right]+\sum_{i=m}^{m+l}
\left[w_i^2\,\deriv_{\scw_i}+w_i\,(\zeta_i\,\partial_{\zeta_i}+2
\Delta_{C_i})\right]\right)\non&&\times\,\Bigl\langle\scC_n(\scz_n)\cdots
\scC_{n+k}(\scz_{n+k})\,\scD_m(\scw_m)\cdots\scD_{m+l}(\scw_{m+l})
\Bigr\rangle_{\rm NS}\non&&+\,2\,\sum_{i=m}^{m+l}w_i\,\Bigl\langle
\scC_n(\scz_n)\cdots\scC_{n+k}(\scz_{n+k})\Bigr.\non&&\times\Bigl.
\scD_m(\scw_m)\cdots\scD_{i-1}(\scw_{i-1})\,
\scC_i(\scw_i)\,\scD_{i+1}(\scw_{i+1})\cdots
\scD_{m+l}(\scw_{m+l})\Bigr\rangle_{\rm NS} \ .
\label{globalsusyWard}\eea}
\noindent
These equations can be used to
determine the general structure of the logarithmic correlators.

For the two-point correlation functions of the logarithmic superfields
one finds~\cite{KAG}
\bea
\Bigl\langle\scC(\scz_1)\,\scC(\scz_2)\Bigr\rangle^{~}_{\rm NS}&=&0 \
, \label{CC}\nopg
\Bigl\langle\scC(\scz_1)\,\scD(\scz_2)\Bigr\rangle^{~}_{\rm
  NS}&=&\Bigl\langle\scD(\scz_1)\,\scC(\scz_2)\Bigr\rangle^{~}_{\rm
  NS}~=~\frac b
{(\scz_{12})^{2\Delta_C}} \ , \label{CD}\nopg \Bigl\langle\scD(\scz_1)\,
\scD(\scz_2)\Bigr
\rangle^{~}_{\rm NS}&=&\frac1{(\scz_{12})^{2\Delta_C}}\,\Bigl(-2b\ln\scz_{12}+d
\Bigl) \ ,
\label{DD}\eea
where the constant $b$ is fixed by the leading logarithmic divergence of the
conformal blocks of the theory (equivalently by the normalization of
the $D$ operator), and the integration constant $d$ can be changed
by the field redefinitions
$\scD(\scz)\mapsto\scD(\scz)+\lambda\,\scC(\scz)$ which are induced by
the scale transformations $z\mapsto\e^\lambda\,z$. In
particular, the equality of two-point functions in (\ref{CD})
immediately implies that the conformal dimension $\Delta_C$ of the
logarithmic pair is necessarily an integer~\cite{ckt}. For the three-point
functions one gets~\cite{KAG}
\bea
\Bigl\langle\scC(\scz_1)\,\scC(\scz_2)\,\scC(\scz_3)\Bigr
\rangle^{~}_{\rm NS}&=&0 \ , \label{CCC}\nopg
\Bigl\langle\scC(\scz_1)\,\scC(\scz_2)\,\scD(\scz_3)\Bigr
\rangle^{~}_{\rm
  NS}&=&\frac1{(\scz_{12})^{\Delta_C}\,(\scz_{13})^{\Delta_C}
\,(\scz_{23})^{\Delta_C}}\,\Bigl(b_1+\beta_1\,\theta_{123}\Bigr) \ ,
\label{CCD}\nopg\Bigl\langle\scC(\scz_1)\,\scD(\scz_2)\,\scD(\scz_3)\Bigr
\rangle^{~}_{\rm
  NS}&=&\frac1{(\scz_{12})^{\Delta_C}\,(\scz_{13})^{\Delta_C}
\,(\scz_{23})^{\Delta_C}}\,\Bigl(b_2+\beta_2\,\theta_{123}-
2(b_1+\beta_1\,\theta_{123})\ln\scz_{23}\Bigr) \ ,\non&&
\label{CDD}\nopg\Bigl\langle\scD(\scz_1)\,\scD(\scz_2)\,\scD(\scz_3)\Bigr
\rangle^{~}_{\rm
  NS}&=&\frac1{(\scz_{12})^{\Delta_C}\,(\scz_{13})^{\Delta_C}
\,(\scz_{23})^{\Delta_C}}\,\Bigl[b_3+\beta_3\,\theta_{123}\Bigr.\non&&
-\,(b_2+\beta_2\,\theta_{123})\ln\scz_{12}\,\scz_{13}\,\scz_{23}
+(b_1+\beta_1\,\theta_{123})\Bigl(2\ln\scz_{12}\ln\scz_{13}\Bigr.\non&&+\left.
\left.2\ln\scz_{12}\ln\scz_{23}+2\ln\scz_{13}\ln\scz_{23}-\ln^2\scz_{12}-
\ln^2\scz_{13}-\ln^2\scz_{23}\right)\right] \ , \non
\label{DDD}\eea
where $b_i$ and $\beta_i$ are undetermined Grassmann even and odd
constants, respectively, and we have generally defined
\beq
\theta_{ijk}=\frac1{\sqrt{\scz_{ij}\,\scz_{jk}\,\scz_{ki}}}\,\Bigl(
\theta_i\,\scz_{jk}+\theta_j\,\scz_{ki}+\theta_k\,\scz_{ij}+
\theta_i\,\theta_j\,\theta_k\Bigr) \ .
\label{thetaijk}\eeq
The remaining three-point correlation functions can be obtained via
cyclic permutation of the superfields in (\ref{CCD}) and (\ref{CDD}). The
general form of the four-point functions may also be found in~\cite{KAG}.

\subsection{Ramond Correlation Functions\label{RCorrelators}}

In the R sector, we define the correlator of any operator $\sf O$ to
be its normalized expectation value in the supersymmetric Ramond ground state,
\beq
\langle{\sf O}\rangle^{~}_{\rm R}=\frac{\langle0|\Sigma(\infty)\,
{\sf O}\,\Sigma(0)|0\rangle}{\langle0|\Sigma(\infty)\,\Sigma(0)|0
\rangle} \ ,
\label{Rcorrelator}\eeq
where we have used the standard asymptotic out-state definition
\beq
\langle0|\Sigma(\infty)=\lim_{z\to\infty}\,\langle0|\Sigma(z)\,z^{\hat
  c/8}
\label{outstatedef}\eeq
and the fact that the spin field $\Sigma(z)$ is a primary field of the
ordinary Virasoro algebra of dimension $\Delta=\hat c/16$. In
particular, the two-point function of the (appropriately normalized)
spin operator is given by
\beq
\langle0|\Sigma(z)\,\Sigma(w)|0\rangle=
\frac1{(z-w)^{\hat c/8}} \ .
\label{Sigma2pt}\eeq
Since $\Sigma(z)$ does not act on the bosonic fields $C(z)$ and
$D(z)$, their R sector correlation functions coincide with those of
the NS sector, i.e. with those of an ordinary logarithmic conformal
field theory. In particular, for the two-point functions we
find~\cite{gurarie,ckt}
\bea
\Bigl\langle C(z)\,C(w)\Bigr\rangle^{~}_{\rm R}&=&0 \
, \non \Bigl\langle C(z)\,D(w)\Bigr\rangle^{~}_{\rm
  R}&=&\Bigl\langle D(z)\,C(w)\Bigr\rangle^{~}_{\rm R}~=~\frac b
{(z-w)^{2\Delta_C}} \ , \non \Bigl\langle D(z)\,D(w)\Bigr
\rangle^{~}_{\rm R}&=&\frac{d-2b\ln(z-w)}{(z-w)^{2\Delta_C}} \ .
\label{DDR}\eea

For the correlation functions of the fermionic fields, we proceed as
follows. Let us introduce the function
\beq
g_C^{~}(z,w|z_1,z_2)=\frac{\langle0|\Sigma(z_1)\,
\chi_C^{~}(z)\,\chi_C^{~}(w)\,\Sigma(z_2)|0\rangle}{\langle0|
\Sigma(z_1)\,\Sigma(z_2)|0\rangle} \ .
\label{g1zw}\eeq
All fields appearing in (\ref{g1zw}) behave as ordinary
primary fields under the action of the Virasoro algebra. The Green's
function (\ref{g1zw}) can therefore be evaluated using standard
conformal field theoretic methods~\cite{DFMS}. It obeys the asymptotic
conditions

\vbox{\bea
g_C^{~}(z,w|z_1,z_2)&\simeq&0+\dots ~~~~ {\rm as}~~z\to w \ ,
\label{g1ztow}\nopg
&\simeq&\frac{(z_1-z_2)^{\hat c/8}}{\sqrt{z-z_1}}\,
\langle0|\widetilde{\Sigma}_C(z_1)\,\chi_C^{~}(w)\,\Sigma(z_2)|0\rangle
+\dots ~~~~ {\rm as}~~z\to z_1 \ , \non&&\label{g1ztoz1}\nopg
&\simeq&\frac{(z_1-z_2)^{\hat c/8}}{\sqrt{z-z_2}}\,
\langle0|\Sigma(z_1)\,\chi^{~}_C(w)\,\widetilde{\Sigma}_C(z_2)|0\rangle
+\dots ~~~~ {\rm as}~~z\to z_2 \ .\non&&
\label{g1ztoz2}\eea}
\noindent
The first condition (\ref{g1ztow}) arises from the fact that the short
distance behaviour of the quantum field theory is independent of the
global boundary conditions, so that in the limit $z\to w$ the function
(\ref{g1zw}) should coincide with the corresponding Neveu-Schwarz
two-point function determined in (\ref{CC}), i.e. $\langle
\chi_C^{~}(z)\,\chi_C^{~}(w)\rangle^{~}_{\rm NS}=0$. The local
monodromy conditions (\ref{g1ztoz1}) and (\ref{g1ztoz2}) follow from
the operator product expansions (\ref{chiCDSigma}). In addition, by
Fermi statistics the Green's function (\ref{g1zw}) must be
antisymmetric under the exchange of its arguments $z$ and $w$,
\beq
g^{~}_C(z,w|z_1,z_2)=-g_C^{~}(w,z|z_1,z_2) \ .
\label{gCantisym}\eeq

By translation invariance, the conditions (\ref{g1ztow}) and
(\ref{gCantisym}) are solved by any
odd analytic function $f$ of $z-w$. Since the correlators appearing in
(\ref{g1ztoz1}) and (\ref{g1ztoz2}) involve only ordinary, primary conformal
fields, global conformal invariance dictates that the function
$f(z-w)$ must multiply a quantity which is a
function only of the $SL(2,\complex)$-invariant
anharmonic ratio $x$ of the four points of
$g^{~}_C(z,w|z_1,z_2)$ given by
\beq
x=\frac{(z-z_1)(w-z_2)}{(z-z_2)(w-z_1)} \ .
\label{crossratio}\eeq
By conformal invariance, the odd analytic function $f(z-w)$
is therefore identically $0$, and hence
\beq
g^{~}_C(z,w|z_1,z_2)=0 \ .
\label{gC0}\eeq

Using this result we can determine a number of correlation
functions. Setting $z_1=\infty$ and $z_2=0$ gives the Ramond correlator
\beq
\Bigl\langle\chi^{~}_C(z)\,\chi^{~}_C(w)\Bigr\rangle^{~}_{\rm R}=0 \ .
\label{chiCCR}\eeq
{}From (\ref{g1ztoz2}) and (\ref{gC0}) we obtain in
addition the vanishing mixed correlator
\beq
\langle0|\Sigma(z_1)\,\chi^{~}_C(z_2)\,\widetilde{\Sigma}_C(z_3)|0\rangle=0 \ .
\label{mixedNScorr0}\eeq
Fusing together the fields $\Sigma(z_1)$ and $\chi^{~}_C(z_2)$ in
(\ref{mixedNScorr0}) using (\ref{chiCDSigma}) then gives the conjugate
spin-spin correlator
\beq
\langle0|\widetilde{\Sigma}_C(z)\,\widetilde{\Sigma}_C(w)|0\rangle=0 \ .
\label{PiPi}\eeq
The vanishing of the $\widetilde{\Sigma}_C\widetilde{\Sigma}_C$
correlation function is consistent with
the fact that the excited spin field $\widetilde{\Sigma}_C(z)$ obeys
the logarithmic conformal algebra
(\ref{TSigmaC},\ref{TSigmaD})~\cite{gurarie,ckt}.

Next, let us consider the function
\beq
g_D^{~}(z,w|z_1,z_2)=\frac{\langle0|\Sigma(z_1)\,
\chi_C^{~}(z)\,\chi_D^{~}(w)\,\Sigma(z_2)|0\rangle}{\langle0|
\Sigma(z_1)\,\Sigma(z_2)|0\rangle} \ .
\label{gDzw}\eeq
The action of the Virasoro algebra in (\ref{gDzw}) does not
produce any additional terms from the logarithmic mixing of the
fermionic field $\chi^{~}_D(w)$, because of the vanishing property
(\ref{gC0}). Therefore, this function can also be evaluated as if the
theory were an ordinary conformal field theory~\cite{DFMS}. Using
(\ref{chiCDSigma}) and (\ref{CD}) the asymptotic conditions
(\ref{g1ztow})--(\ref{g1ztoz2}) are now replaced with
\bea
g_D^{~}(z,w|z_1,z_2)&\simeq&\frac{2\Delta_C\,b}{(z-w)^{2\Delta_C+1}}+\dots
{}~~~~ {\rm as}~~z\to w \ , \label{gDztow}\nopg
&\simeq&\frac{(z_1-z_2)^{\hat c/8}}{\sqrt{z-z_1}}\,
\langle0|\widetilde{\Sigma}_C(z_1)\,\chi_D^{~}(w)\,\Sigma(z_2)|0\rangle
+\dots ~~~~ {\rm as}~~z\to z_1 \ , \non&&\label{gDztoz1}\nopg
&\simeq&\frac{(z_1-z_2)^{\hat c/8}}{\sqrt{z-z_2}}\,
\langle0|\Sigma(z_1)\,\chi^{~}_D(w)\,\widetilde{\Sigma}_C(z_2)|0\rangle
+\dots ~~~~ {\rm as}~~z\to z_2 \ , \non&&\label{gDztoz2}\nopg
&\simeq&\frac{(z_1-z_2)^{\hat c/8}}{\sqrt{w-z_1}}\,
\langle0|\widetilde{\Sigma}_D(z_1)\,\chi_C^{~}(z)\,\Sigma(z_2)|0\rangle
+\dots ~~~~ {\rm as}~~w\to z_1 \ , \non&&\label{gDwtoz1}\nopg
&\simeq&\frac{(z_1-z_2)^{\hat c/8}}{\sqrt{w-z_2}}\,
\langle0|\Sigma(z_1)\,\chi^{~}_C(z)\,\widetilde{\Sigma}_D(z_2)|0\rangle
+\dots ~~~~ {\rm as}~~w\to z_2 \ .\non&&
\label{gDwtoz2}\eea

Again, from (\ref{mixedNScorr0}) it follows that the correlators in
(\ref{gDztoz1})--(\ref{gDwtoz2}) involving a single logarithmic operator
can be treated as an ordinary conformal correlator for primary
fields. In particular, we can treat (\ref{gDzw}) as a correlator for
two identical conformal fermion fields of dimension
$\Delta_C+\frac12$ and require it to be antisymmetric under exchange
of $z$ and $w$, as in (\ref{gCantisym}). This property follows from
the fact that the local NS correlator (\ref{gDztow}) is antisymmetric
in $z$ and $w$ and this feature should extend globally in the quantum
field theory. Again, by $SL(2,\complex)$-invariance the quantity
$(z-w)^{-2\Delta_C-1}\,g^{~}_D(z,w|z_1,z_2)$ is a function only of the
anharmonic ratio (\ref{crossratio}). The precise dependence on $x$ is
uniquely determined by the boundary conditions
(\ref{gDztow})--(\ref{gDwtoz2}) and the antisymmetry of $g^{~}_D$, and
we find\footnote{\baselineskip=12pt To show explicitly that
  (\ref{gDfinal}) is the unique function of $z$ and $w$ with the
  desired properties, we write it as
$$
g^{~}_D(z,w|z_1,z_2)=\frac1{\sqrt{(z-z_1)(z-z_2)(w-z_1)(w-z_2)}}\,
\frac{\Delta_C\,b}{(z-w)^{2\Delta_C+1}}\,\Bigl((z-z_1)(w-z_2)+
(z-z_2)(w-z_1)\Bigr) \ .
$$
The first factor here gives the correct behaviour for $g^{~}_D$ as
$z,w\to z_1,z_2$, while the second factor is the required pole at $z=w$ of
order $2\Delta_C+1$. The third factor is then chosen so that the
residue of the pole is $2\Delta_C\,b$ and such that it cancels the
lower order poles arising from the first factor in the limit $z\to w$,
and by further requiring that the overall combination be antisymmetric in $z$
and $w$.}
\beq
g^{~}_D(z,w|z_1,z_2)=\frac{\Delta_C\,b}{(z-w)^{2\Delta_C+1}}\,
\left(\,\sqrt{\frac{(z-z_1)(w-z_2)}{(z-z_2)(w-z_1)}}+
\sqrt{\frac{(z-z_2)(w-z_1)}{(z-z_1)(w-z_2)}}~\right) \ .
\label{gDfinal}\eeq

By taking various limits of (\ref{gDfinal}) we can generate another set
of correlation functions for Ramond sector operators. In the
simultaneous limit $z_1\to\infty$ and $z_2\to0$, the function
(\ref{gDfinal}) yields the Ramond two-point correlators
\beq
\Bigl\langle\chi^{~}_C(z)\,\chi^{~}_D(w)\Bigr\rangle^{~}_{\rm R}=
-\,\Bigl\langle\chi^{~}_D(z)\,\chi^{~}_C(w)\Bigr\rangle^{~}_{\rm R}=
\frac{\Delta_C\,b}{(z-w)^{2\Delta_C+1}}\,\left(\,\sqrt{\frac zw}+
\sqrt{\frac wz}~\right) \ .
\label{chiCDR}\eeq
Note that the term in parantheses has branch cuts at $z=0,\infty$ and
$w=0,\infty$, yielding the antiperiodic boundary conditions on the
spinor fields as they circle around the origin in the complex plane
and across the cut
connecting the spin operators $\Sigma(0)$ and $\Sigma(\infty)$ in
(\ref{Rcorrelator}). Taking the limits $z\to z_1,z_2$ and $w\to
z_1,z_2$ in (\ref{gDfinal}) and comparing with
(\ref{gDztoz1})--(\ref{gDwtoz2}) yields the correlation functions
\bea
\langle0|\widetilde{\Sigma}_C(z_1)\,\chi^{~}_D(z_2)\,\Sigma(z_3)|0\rangle
&=&\frac{\ii\Delta_C\,b}{(z_1-z_2)^{2\Delta_C+1/2}\,(z_1-z_3)^{\hat
    c/8-1/2}\,\sqrt{z_2-z_3}}
\non&=&-\,\langle0|\widetilde{\Sigma}_D(z_1)\,
\chi^{~}_C(z_2)\,\Sigma(z_3)|0\rangle \ .
\label{SigmaCchiDS}\eeq
Fusing $\Sigma(z_3)$ with $\chi^{~}_D(z_2)$ and $\chi^{~}_C(z_2)$ in
(\ref{SigmaCchiDS}) using (\ref{chiCDSigma}) then yields the spin-spin
correlators
\beq
\langle0|\widetilde{\Sigma}_C(z)\,\widetilde{\Sigma}_D(w)|0
\rangle=-\,\langle0|\widetilde{\Sigma}_D(z)\,\widetilde{\Sigma}_C(w)|0
\rangle=\frac{\ii\Delta_C\,b}{(z-w)^{2\Delta_C+\hat c/8}} \ .
\label{spinspinCD}\eeq
Note that the logarithmic pair
$\widetilde{\Sigma}_C,\widetilde{\Sigma}_D$ does not have the
canonical two-point functions of a logarithmic conformal field theory
(see~(\ref{DDR})). This is because the excited spin fields of the
theory are not bosonic fields, but are rather given by non-local
operators which interpolate between different sectors of the quantum
Hilbert space and which satisfy, in addition to the logarithmic
algebra, a supersymmetry algebra. In fact, their correlators are
almost identical in form to the correlation functions of the
logarithmic superpartners~$\chi^{~}_C,\chi^{~}_D$~\cite{MavSz}.

Finally, we need to compute the $DD$ type correlators. The above
techniques do not directly apply because Green's functions with two or
more logarithmic operator insertions will not transform covariantly
under the action of the Virasoro algebra. However, we may obtain the
$DD$ type correlators from the mixed $CD$ type ones above by the
following trick~\cite{KAG,MavSz,gezel}. We regard $\Delta_C$ as a continuous
weight and note that the logarithmic superconformal algebra can be
simply obtained by writing down the standard conformal operator
product expansions for the $C$ type operators, and then
differentiating them with respect to $\Delta_C$ to obtain the $D$ type
ones with the formal identifications $D=\partial C/\partial\Delta_C$,
$\chi^{~}_D=\partial\chi^{~}_C/\partial\Delta_C$ and
$\widetilde{\Sigma}_D=\partial\widetilde{\Sigma}_C/\partial\Delta_C$.
Since the basic spin fields $\Sigma(z)$ do not depend on the conformal
dimension $\Delta_C$, we can differentiate the correlation functions
(\ref{chiCDR})--(\ref{spinspinCD}) to get the
desired Green's functions. In doing so we regard the parameter $b$ as
an analytic function of the weight $\Delta_C$ and define $d=\partial
b/\partial\Delta_C$. In this way we arrive at the correlators
\bea
\Bigl\langle\chi^{~}_D(z)\,\chi^{~}_D(w)\Bigr\rangle^{~}_{\rm R}&=&
\frac{b+\Delta_C\Bigl(d-2b\ln(z-w)\Bigr)}{(z-w)^{2\Delta_C+1}}\,
\left(\,\sqrt{\frac zw}+\sqrt{\frac wz}~\right) \ , \non
\langle0|\widetilde{\Sigma}_D(z_1)\,\chi^{~}_D(z_2)\,\Sigma(z_3)|0\rangle
&=&-\frac{b+\Delta_C\Bigl(d-2b\ln(z_1-z_2)\Bigr)}
{\ii(z_1-z_2)^{2\Delta_C+1/2}\,(z_1-z_3)^{\hat
    c/8-1/2}\,\sqrt{z_2-z_3}} \ , \non
\langle0|\widetilde{\Sigma}_D(z)\,\widetilde{\Sigma}_D(w)|0
\rangle&=&-\frac{b+\Delta_C\Bigl(d-2b\ln(z-w)\Bigr)}
{\ii(z-w)^{2\Delta_C+\hat c/8}} \ .
\label{DDR3corrs}\eea

In a completely analogous way, we may easily determine the vanishing
two-point correlation functions
\beq
\Bigl\langle\phi(z)\,\chi^{~}_{\phi'}(w)\Bigr\rangle^{~}_{\rm R}~=~0~=~
\langle0|\widetilde{\Sigma}_{\phi'}(z_1)\,\phi(z_2)\,\Sigma(z_3)
|0\rangle \ ,
\label{othervanish}\eeq
where $\phi$ and $\phi'$ label either of the two fields $C$ or $D$.
The present technique unfortunately does not directly determine higher order
correlation functions of the fields. As they will not be required in
what follows, we will not pursue this issue in this paper.

\newsection{Null Vectors, Hidden Symmetries and Spin Models\label{nullvectors}}

It has been suggested~\cite{MavSz} that, in the limit $\Delta_C=0$,
the fermionic field $\chi^{~}_C(z)$ in (\ref{CDsuperfields}) may be a
null field, since its two-point correlation functions with all other
logarithmic fields vanish for zero conformal dimension. Furthermore,
the logarithmic scaling violations in the fermionic two-point
functions involving the field $\chi^{~}_D(z)$ disappear in this
limit. While this latter property is certainly true for all Green's
functions of the conformal field theory, a quick examination of the
three-point correlators (\ref{CCC}) and (\ref{CCD}) shows that
$\chi^{~}_C(z)$ is {\it not} a null field if $\beta_1\neq0$. The
situation is completely analogous to what happens generically to its
superpartner $C(z)$. Since the primary field $C(z)$ creates a
zero-norm state, and since $\Delta_C\in\zed$, there is a new hidden
continuous symmetry in the theory~\cite{ckt} generated by the conserved
holomorphic current $C(z)$, which is a symmetric tensor of rank
$\Delta_C$. For $\Delta_C=0$, the extra couplings of the $\chi^{~}_C$
field for $\beta_1\neq0$ show that it corresponds to a non-trivial,
dynamical fermionic symmetry of the logarithmic superconformal field
theory. In fact, in the R~sector the structure of these continuous
symmetries is even richer, given that the excited spin field
$\widetilde{\Sigma}_C(z)$ also creates a zero-norm state in the
logarithmic superconformal field theory, and that it has vanishing
two-point functions for $\Delta_C=0$. In $c\neq0$ theories where the
bosonic energy-momentum tensor $T(z)$ has a logarithmic partner, the
identity operator $I$ generates a Jordan cell with
$\Delta_C=0$~\cite{KogNich1} and the zero-norm state is the vacuum,
$\langle0|0\rangle=0$. In this case, of course, the fermion field
$\chi^{~}_I(z)=0$ is trivially a null field, and its
partner $\chi^{~}_D(z)$ is an ordinary, non-logarithmic primary
field of the Virasoro algebra of conformal dimension
$\frac12$. Similarly, in this case $\widetilde{\Sigma}_C(z)=0$, while
$\widetilde{\Sigma}_D(z)$ is an ordinary, non-logarithmic twist field
of weight $\hat c/16$.

In the Ramond sector, there are natural ways to generate null
states for any $\Delta_C$. One way is to build the representation of
the Ramond algebra from the supersymmetric ground state $|\frac{\hat
  c}{16}\rangle^{~}_{\rm R}$ as described at the end of
section~\ref{HighestWeight}. Another way is to introduce the fermion
parity operator $\Gamma=(-1)^F$, where $F$ is the fermion number operator
of the superconformal field theory. The operator $\Gamma$ commutes
with integer spin fields and anticommutes with half-integer spin
fields. It defines an inner automorphism $\pi^{~}_\Gamma:{\cal
  C}\to{\cal C}$ of the maximally extended chiral symmetry algebra
$\cal C$ of the superconformal field theory, such that there is an
exact sequence of vector spaces
\beq
0~\longrightarrow~{\cal C}^+~\longrightarrow~{\cal C}~\longrightarrow~
{\cal C}^-~\longrightarrow~0 \ , ~~
\pi^{~}_\Gamma({\cal C}^\pm)=\pm\,{\cal C}^\pm \ .
\label{calCGamma}\eeq
Under the operator-state correspondence, this determines a fermion
parity grading of the Hilbert space of states as
\beq
{\cal H}={\cal H}^+\oplus{\cal H}^- \ , ~~ \Gamma\,{\cal H}^\pm=\pm\,
{\cal H}^\pm \ .
\label{calHGammagrading}\eeq
Since $G_0$ reverses chirality, the paired Ramond ground states have
opposite chirality,
\beq
\Gamma\,\Sigma_\Delta^\pm(0)|0\rangle=\pm\,\Sigma_\Delta^\pm(0)|0\rangle \ .
\label{oppchirality}\eeq
The opposite chirality spin fields $\Sigma_\Delta^\pm(z)$ are non-local
with respect to each other
(c.f.~(\ref{GSigma+OPE}) and (\ref{GSigma-OPE})). In a unitary theory,
whereby $G_0^2\geq0$, all $\Delta=\hat c/16$ states are chirally asymmetric
highest-weight states, since the state $G_0|\Delta\rangle^{~}_{\rm R}$ is
then a null vector in the Hilbert space. On the other hand, the
orthogonal projection $\frac12\,(1+\Gamma):{\cal H}\to{\cal H}^+$
onto states of even fermion parity $\Gamma=1$ eliminates
the spin field $\Sigma_\Delta^-(z)$ and gives a local field theory which is
customarily refered to as a ``spin model''~\cite{FQS}. The fields of
the spin model live in the local chiral algebra ${\cal C}^+$. This
projection eliminates $G(z)$ and the other half-integer weight
fields. When combined with the projection onto $G_0=0$ it gives the
``GSO projection'' which will be important in the D-brane applications
of the next section.

The main significance of the chiral subalgebra restriction
$\frac12\,(1+\pi^{~}_\Gamma):{\cal C}\to{\cal C}^+$
is that the fermionic fields of the superconformal field theory can be
reconstructed from the $\Gamma=1$ spin fields $\Sigma(z)$, at least in
the examples that we consider in this paper. In an analogous way, the
logarithmic superpartners $\chi^{~}_C(z)$ and $\chi^{~}_D(z)$ can be
reconstructed from the $\Gamma=1$ excited spin fields
$\widetilde{\Sigma}_C(z)$ and $\widetilde{\Sigma}_D(z)$. By
supersymmetry, this yields the bosonic partners $C(z)$ and $D(z)$, and
so in this way the spin model determines the entire logarithmic
superconformal field theory. In fact, the spin field $\Sigma(z)$ can
be uniquely constructed from the underlying chiral current algebra
generated by currents which are formed by the primary fermionic fields
of the theory~\cite{FMS1}. The fermionic current algebra will thereby
completely determine the entire logarithmic superconformal field theory.

\newsection{The Recoil Problem in Superstring Theory\label{RecoilProb}}

In the remainder of this paper we will consider some concrete models
to illustrate the above formalism explicitly. These examples
will also serve to describe some of the basic constructions of
logarithmic spin operators and will illustrate the applicability of
the superconformal logarithmic formalism. In this section we will
discuss the logarithmic superconformal field theory that describes the
recoil of a D-particle in string theory~\cite{kmw,ms,MavSz}. This is
the simplest example in which to introduce some of the formalism that
will also be used in later sections. It also captures the essential
features of the general theory of the previous three sections in a
very simple setting.

\subsection{Supersymmetric Impulse Operators}

Consider the superconformal field theory defined by the classical worldsheet
action
\beq
S_{\rm D0}=\frac1{2\pi}\,\int\dd z~\dd\overline{z}~\dd\theta~\dd
\overline{\theta}~~\overline{{\cal D}_\scz}\,\scx^\mu\,\deriv_\scz
\scx_\mu-\frac1\pi\,\oint\dd\tau~\dd\vartheta~\Bigl(y_i\,
\scC_\epsilon+u_i\,\scD_\epsilon\Bigr)\,\deriv^{~}_{\!\perp}\scx^i \ ,
\label{SD0}\eeq
where $\scx^\mu(\scz,\overline{\scz}\,)=\scx^\mu(\scz)+
\scx^\mu(\overline{\scz}\,)$ with $\scx^\mu(\scz)$ the
chiral scalar superfield
\beq
\scx^\mu(\scz)=x^\mu(z)+\theta\,\psi^\mu(z) \ ,
\label{scxsuperfield}\eeq
whose Neveu-Schwarz two-point functions are given by
\beq
\Bigl\langle\scx^\mu(\scz_1)\,\scx^\nu(\scz_2)\Bigr\rangle^{~}_{\rm
  NS}=-\delta^{\mu\nu}\,\ln\scz_{12} \ .
\label{scx2ptfn}\eeq
Here $x^\mu$, $\mu=1,\dots,10$ are maps from the upper complex
half-plane $\complex_+$ into
ten dimensional {\it Euclidean} space $\real^{10}$, and $\psi^\mu$ are
their spin $\frac12$ fermionic superpartners that transform in the vector
representation of $SO(10)$ and each of which is a Majorana-Weyl spinor
in two-dimensions. We will identify the coordinate $x^{10}$ as the
Euclidean time, while $x^i$, $i=1,\dots,9$ lie along the spatial
directions in the target space of the open strings. As in the previous
section, we concentrate on the chiral sector of the worldsheet field
theory with superfields (\ref{scxsuperfield}). The chiral super
energy-momentum tensor is given by
\beq
\scT(\scz)=-\frac12\,\deriv_\scz\scx^\mu(\scz)\,\partial_z
\scx_\mu(\scz) \ .
\label{chiralscT}\eeq

The reasons for working with Euclidean spacetime signature are
technical. First of all, it is easier to deal with spinor
representations of the Euclidean group $SO(10)$ than with those of the
Lorentz group $SO(9,1)$. In the former case all of the $\psi^\mu$ are
treated on equal footing and one is free from the possible
complications arising from the time-like nature of $x^0$, which would
otherwise imply a special role for its superpartner
$\psi^0$~\cite{gsw}. Secondly, for the recoil problem, Euclidean target
spaces are necessary to ensure convergence of worldsheet correlation
functions among the logarithmic operators~\cite{kmw}. For calculational
definiteness and convenience of the worldsheet path integrals, we
shall therefore adopt a Euclidean signature convention in the
following.

The second term in the action (\ref{SD0}) is a marginal deformation of the free
$\hat c=10$ superconformal field theory by the vertex operator describing the
recoil, within an impulse approximation, of a non-relativistic D-brane
in target space due to its interaction with closed string scattering
states~\cite{km}--\cite{ms}. It is the appropriate operator to use
when regarding the branes as string solitons. The coordinate $\tau$
parametrizes the boundary of the upper half-plane, and $\vartheta$ is
a real Grassmann coordinate. The fields in this part of the action are
understood to be restricted to the worldsheet boundary. The coupling
constants $y_i$ and $u_i$ are interpreted as the initial position and
constant velocity of the D-particle, respectively, and the subscript
$\perp$ denotes differentiation in the direction normal to the boundary of
$\complex_+$. The recoil operators are given by chiral superfields
$\scC_\epsilon(\scz)$ and $\scD_\epsilon(\scz)$ whose components are defined
in terms of superpositions over tachyon vertex operators
$\e^{\ii qx^{10}(z)}$ in the time direction as~\cite{MavSz}
\bea
C_\epsilon(z)&=&\frac\epsilon{4\pi\ii}\,\int
\limits_{-\infty}^\infty\frac{\dd q}{q-\ii\epsilon}~
\e^{\ii qx^{10}(z)} \ , \non \chi^{~}_{C_\epsilon}(z)&=&
\ii\epsilon\,C_\epsilon(z)\otimes\psi^{10}(z) \ , \non
D_\epsilon(z)&=&-\frac1{2\pi}\,\int\limits_{-\infty}^\infty\frac{\dd
  q}{(q-\ii\epsilon)^2}~\e^{\ii qx^{10}(z)} \ , \non
\chi^{~}_{D_\epsilon}(z)&=&\ii\left(\epsilon\,D_\epsilon(z)-
\frac2\epsilon\,C_\epsilon(z)\right)\otimes\psi^{10}(z) \ .
\label{susyrecoilops}\eea
Here and in the following, singular operator products taken at
coincident points are always understood to be normal ordered according
to the prescription
\beq
O(z)\,O'(z)\equiv\oint\limits_{w=z}\frac{\dd w}{2\pi\ii}~
\frac{O(w)\,O'(z)}{w-z} \ .
\label{normalordering}\eeq

The target space regularization parameter $\epsilon\to0^+$ is related
to the worldsheet ultraviolet cutoff $\Lambda\to0^+$ by
\beq
\frac1{\epsilon^2}=-\ln\Lambda \ .
\label{epsilonLambda}\eeq
In this limit, careful computations~\cite{kmw,MavSz} establish that,
to leading orders in $\epsilon$, the superfield recoil operators
(\ref{susyrecoilops}) satisfy the relations (\ref{CDsuperOPE}) and
(\ref{CC})--(\ref{DD}) of the $N=1$ logarithmic superconformal algebra
in the NS sector of the worldsheet field theory, with
\bea
\Delta_{C_\epsilon}&=&-\frac{\epsilon^2}4 \ , \non
b&=&\frac{\pi^{3/2}}4 \ , \non d&=&\frac{\pi^{3/2}}{2\epsilon^2} \ .
\label{SLCFTrecoilconsts}\eea
In the following we will describe how to properly incorporate the
Ramond sector of this system.

\subsection{Spin Fields\label{logspinrecoil}}

We will now construct the operators $\Sigma$ which create cuts in the
fields $\psi^{10}$ appearing in the superpartners of the recoil operators
(\ref{susyrecoilops}) and are thereby responsible for
changing their boundary conditions as one circumnavigates the cut~\cite{gsw}.
In fact, one needs $\Sigma(z)$ in the neighborhood of the fields $\psi^{10}$
but this is readily done in bosonized form~\cite{fermspin}, as we
shall now discuss, by means of a boson translation operator which
relates $\Sigma (z)$ to $\Sigma (0)$. Bosonization of the free fermion
system defined by (\ref{SD0}) allows us to express in a local-looking
form the non-local effects of the spin operators. In what follows we
shall only require the bosonization of the spinor field appearing in
(\ref{susyrecoilops}).

In the Euclidean version of the target space theory there are ten
fermion fields $\psi^\mu$ which we can treat on equal footing. Given
the pair of right-moving NSR fermion fields $\psi^9$, $\psi^{10}$
corresponding to the light-cone of the recoiling D0-brane system, we
may form complex Dirac fermion fields
\beq
\psi^\pm(z)=\psi^9(z)\pm\ii\psi^{10}(z) \ .
\label{psipmdef}\eeq
The worldsheet kinetic energy in (\ref{SD0}) associated to this pair is
of the form
\beq
\int\dd^2z~\left(\psi^9\,\overline{\partial_z}\,\psi^9+\psi^{10}\,
\overline{\partial_z}\,\psi^{10}
\right)=\int\dd^2z~\psi^+\,\overline{\partial_z}\,\psi^- \ .
\label{nsraction}\eeq
{}From the corresponding equations of motion and (\ref{scx2ptfn}) it
follows that the field
\beq
j(z)=\psi^+(z)\,\psi^-(z)
\label{Jpsi}\eeq
is a conserved $U(1)$ fermion number current which is a primary field
of the Virasoro algebra of dimension 1 and which generates a $U(1)$
current algebra at level 1. Its presence allows the
introduction of spin fields, and hence twisted sectors in the quantum
Hilbert space, through the bosonization formulas
\bea
j(z)&=&2\ii\,\partial_z\phi(z) \ , \non
\psi^\pm(z)&=&\sqrt2\,\e^{\pm\ii\phi(z)} \ ,
\label{bosonformulas}\eea
where $\phi(z)$ is a free, real, compact chiral scalar field,
i.e. its two-point function is
\beq
\langle0|\phi(z)\,\phi(w)|0\rangle=-\ln(z-w) \ .
\label{freebos2ptfn}\eeq
In this representation all fields are taken to act in the NS
sector.

The holomorphic part of the Sugarawa energy-momentum tensor
corresponding to the worldsheet action (\ref{nsraction}) is given in
bosonized form by
\beq
T_\kappa(z)=-\frac12\,\partial_z\phi(z)\,\partial_z\phi(z)+\frac{\ii\kappa}2\,
\partial_z^2\phi(z) \ ,
\label{stressboson}\eeq
where the constant $\kappa$ is arbitrary because the second term in
(\ref{stressboson}) is identically conserved for all $\kappa$. This
energy-momentum tensor derives from the Coulomb gas model defined by
the Liouville action
\beq
S_\kappa=\frac1{4\pi}\,\int\dd z~\dd\overline{z}~\sqrt g\,\left(
\partial_z\phi\,\overline{\partial_z}\,\phi+\frac{\ii\kappa}2\,
R^{(2)}\,\phi\right) \ ,
\label{Skappa}\eeq
where $g(z,\overline{z}\,)$ and $R^{(2)}(z,\overline{z}\,)$ are the
metric and curvature of the worldsheet. The topological curvature term
in (\ref{Skappa}) provides a deficit term to the central charge
$c_\kappa$ of the free boson field $\phi(z)$,
\beq
c_\kappa=1-3\kappa^2 \ ,
\label{ckappa}\eeq
and it also induces a vacuum charge at infinity (the singular point of
the metric on the Riemann sphere). In particular, the primary
field $\e^{\ii q\phi(z)}$ has dimension
\beq
\Delta_{q,\kappa}=\frac q2\,\Bigl(q-\kappa\Bigr) \ .
\label{hqkappa}\eeq
What fixes $\kappa$ here, and thereby lifts the ambiguity, is the
charge conjugation symmetry $\psi^{10}(z)\mapsto-\psi^{10}(z)$ of the
NSR model (\ref{nsraction}), which interchanges the two Dirac fields
$\psi^\pm(z)$ and hence acts on the free boson field as
$\phi(z)\mapsto-\phi(z)$. This symmetry implies that $\kappa=0$ in
(\ref{stressboson}).

Let us now consider the tachyon vertex operators corresponding to the
free boson,
\begin{equation}
           \Sigma _q(z)=\e^{\ii q\phi(z)} \ ,
\label{cutoper}\end{equation}
which have conformal dimension $\Delta_{q,0}=q^2/2$.
In bosonized language the pair of Dirac
fermion fields corresponds to the operators (\ref{cutoper}) at $q=\pm
1$, $\psi^\pm(z)=\sqrt2\,\Sigma_{\pm 1}(z)$. On the other hand, the operators
(\ref{cutoper}) at $q=\pm\,\frac12$ introduce a branch cut in the
field $\psi^{10}(z)$. To see this, we note the standard free
field formula for multi-point correlators of tachyon vertex operators,
\bea
\langle0|\Sigma_{q_1}(z_1)\cdots\Sigma_{q_n}(z_n)|0\rangle&=&
\prod_{k=1}^n\,\prod_{l=1}^n\e^{-q_kq_l\langle0|\phi(z_k)\,\phi(z_l)
|0\rangle/2}\non&=&\Lambda^{\bigl(\sum_lq_l\bigr)^2/2}\,
\prod_{k<l}(z_k-z_l)^{q_kq_l} \ ,
\label{tachyoncorr}\eeq
where we have regulated the coincidence limit of the two-point
function (\ref{freebos2ptfn}) by the short-distance cutoff
$\Lambda\to0^+$. In particular, the correlator (\ref{tachyoncorr})
vanishes unless
\beq
\sum_{l=1}^nq_l=0 \ ,
\label{selectionrule}\eeq
which is a consequence of the
continuous $U(1)$ symmetry generated by the current (\ref{Jpsi}) which
acts by global translations of the fields $\phi(z_l)$. From the
general result (\ref{tachyoncorr}) we may infer the three-point
correlation functions
\beq
\langle0|\Sigma_{\pm1/2}(z_1)\,\Sigma_{\pm1/2}(z_2)\,\Sigma_{\mp1}(z_3)
|0\rangle=\frac{(z_1 - z_2)^{1/4}}{\sqrt{(z_1 - z_3)(z_2 - z_3)}} \ .
\label{cuteq}\eeq
The correlator (\ref{cuteq}) has square root branch points at
$z_3=z_1$ and $z_3=z_2$. This implies that the elementary fermion fields
$\psi^\pm(z_3)$ are double-valued in the fields of the operators
$\Sigma_{\mp1/2}(z_1)$, respectively.

It follows that the spin operators for the recoil problem are given by
\beq
\Sigma^+_{1/8}(z)=\sqrt2\,\cos\frac{\phi(z)}2
\label{spinrecoilpm}\eeq
and they have weight $\Delta=\Delta_{\pm1/2}=\frac18$. They create
branch cuts in the fermionic fields
\beq
\psi^{10}(z)=\sqrt2\,\sin\phi(z) \ .
\label{psi10sinphi}\eeq
Note that the spin operators need only be inserted at the origin
$z=0$, because it is there that they are required to change the
boundary conditions on the fermion fields. These operators are all
understood as acting on the NS vacuum state $|0\rangle$, thereby
creating highest weight states in the Ramond sector. The spin fields
$\Sigma^\pm_{1/8}(0)$ may be extended to operators
$\Sigma^\pm_{1/8}(z)$ in the neighbourhood of $\psi^{10}(z)$ via application
of the boson translation operator $\e^{z\,\partial_z}=\e^{z\,L_{-1}}$.

Using the operator product expansions
\beq
\Sigma_q(z)\,\Sigma_{q'}(w)=(z-w)^{qq'}\,\Sigma_{q+q'}(w)\Bigl(1+
\ii q\,(z-w)\,\partial_w\phi(w)\Bigr)+\dots \ , ~~ qq'<-1
\label{SigmaqqprimeOPE}\eeq
and (\ref{chiralscT}), it is straightforward to check that the term of
order $(z-w)^{-3/2}$ in the operator product
$G(z)\,\Sigma_{1/8}^+(w)$ vanishes, and hence that
\beq
\Sigma_{1/8}^-(z)=0 \ .
\label{Sigma18minus0}\eeq
This means that the spin field $\Sigma(z)=\Sigma_{1/8}^+(z)$
corresponds to the supersymmetric ground state
$|\frac18\rangle^{~}_{\rm R}$ in the Ramond sector of the system,
associated with superconformal central charge $\hat c=2$. By using
the selection rule (\ref{selectionrule}) and the factorization of
bosonic and fermionic correlation functions in the free superconformal
field theory determined by (\ref{SD0}), it is straightforward to
verify both the NS two-point functions (\ref{CC})--(\ref{DD}) and the
spin-spin two-point function as normalized in (\ref{Sigma2pt}). The
central charge $\hat c=2$ is the one pertinent to the recoil operators
because in the bosonized representation they only refer to two of the
ten superconformal fields present in the total action (\ref{SD0}).

Using (\ref{SigmaqqprimeOPE}) one can also easily derive
the excited logarithmic spin operators of dimension
$\Delta_{C_\epsilon}+\frac18$, which along with (\ref{chiCDSigma}) and
(\ref{susyrecoilops}) yields
\bea
\widetilde{\Sigma}_{C_\epsilon}(z)&=&\ii\epsilon\,C_\epsilon(z)\otimes
\sin\frac{\phi(z)}2 \ , \non\widetilde{\Sigma}_{D_\epsilon}(z)&=&
\ii\left(\epsilon\,D_\epsilon(z)-
\frac2\epsilon\,C_\epsilon(z)\right)\otimes\sin\frac{\phi(z)}2 \ .
\label{Sigmaexcitedrecoil}\eea
The corresponding logarithmic operator product expansions
(\ref{TSigmaC}) and (\ref{TSigmaD}) are straightforward consequences
of the factorization of the bosonic and fermionic sectors in the recoil
problem. Because of this same factorization property, all of the two-point
correlation functions of section~\ref{RCorrelators} may be easily
derived. The basic identities are given by (\ref{cuteq}) and the
four-point function
\bea
\langle0|\Sigma(z_1)\,\psi^{10}(z)\,\psi^{10}(w)\,\Sigma(z_2)|0\rangle&=&
\frac1{2(z_1-z_2)^{1/4}\,(z-w)}\non&&\times\,
\left(\,\sqrt{\frac{(z_1-z)(w-z_2)}{(z_1-w)(z-z_2)}}+
\sqrt{\frac{(z_1-w)(z-z_2)}{(z_1-z)(w-z_2)}}~\right) \ , \non&&
\label{4ptfnrecoil}\eea
where we have again used the selection rule
(\ref{selectionrule}). Thus, by using bosonization techniques it is
straightforward to describe the $N=1$ supersymmetric extension of the
logarithmic operators of the recoil problem in both the NS and R
sectors of the worldsheet superconformal field theory.

\subsection{Fermionic Vertex Operators for the Recoil Problem}

As a simple application of the above formalism, we will now construct
the appropriate spacetime vertex operators which create recoil states of the
D-branes. The crucial point is that one can now build states
that are consistent with the target space supersymmetry of Type~II
superstring theory, which thereby completes the program of
constructing recoil operators in string theory. Spacetime
supersymmetry necessitates vertex operators which describe the
excitations of fermionic states in target space. Such supersymmetric
operators were constructed in~\cite{MavSz} from a target space
perspective. Here we shall construct fermionic states for the recoil
problem from a worldsheet perspective by using appropriate
combinations of the spin operators (\ref{cutoper}). We have already
seen how this arises above, in that the Ramond state
$G_0|\frac18\rangle^{~}_{\rm R}$ is a null vector and
one recovers a single logarithmic superconformal algebra among the
physical states, as in the NS sector. This construction relies heavily
on the Euclidean signature of the spacetime, and yields states that
transform in an appropriate spinor representation of the Euclidean
group.

The recoil operators (\ref{susyrecoilops}) are all built as
appropriate superpositions of the off-shell tachyon vertex operators
$\e^{\ii qx^{10}(z)}$. It is well-known how to construct the boson and
fermion emission operators which create corresponding tachyon ground
states from the NS vacuum state $|0\rangle$~\cite{gsw}. In the bosonic
sector the vertex operator is $[G_r,\e^{\ii
  qx^{10}(z)}]=q~\e^{\ii qx^{10}(z)}\otimes\psi^{10}(z)$, where the fermion
field $\psi^{10}(z)$ has the periodic mode expansion
\beq
\psi^{10}(z)=\frac1{\sqrt2}~\sum_{n=-\infty}^\infty\psi_{n+1/2}^{10}
{}~z^{-n-1}
\label{psi10permode}\eeq
appropriate to the NS sector, with $(\psi^{10}_r)^\dag=\psi^{10}_{-r}$,
$\{\psi_r^{10},\psi_s^{10}\}=\delta_{r+s,0}$, and
$\psi^{10}_{n+1/2}|0\rangle=0~~\forall n\geq0$. By construction, the
corresponding recoil operators are of course just the fermionic operators
$\chi^{~}_{C_\epsilon}(z)$ and $\chi^{~}_{D_\epsilon}(z)$ in
(\ref{susyrecoilops}). The emission of a fermion by a spinor
$u_\alpha$ is described by the vertex operator
$\e^{\ii qx^{10}(z)}\otimes\overline{u}^{\,\alpha}(q)\,\Sigma_\alpha(z)$,
where $\alpha=\pm\,\frac12$ are regarded as spinor indices of the
two-dimensional Euclidean group $SO(2)$ and $u(q)$ is a two-component
off-shell Dirac spinor.

The recoil emission vertex
operators are therefore given by the chiral superfields
\bea
\scV_{C_\epsilon}(\scz)&=&\Xi_{C_\epsilon}(z)+\theta\,
\chi^{~}_{C_\epsilon}(z) \ , \non \scV_{D_\epsilon}(\scz)&=&
\Xi_{D_\epsilon}(z)+\theta\,\chi^{~}_{D_\epsilon}(z) \ ,
\label{recoilemission}\eea
where the boson emission operators are
\bea
\chi^{~}_{C_\epsilon}(z)&=&\frac{\epsilon^2}{4\pi}\,\int
\limits_{-\infty}^\infty\frac{\dd q}{q-\ii\epsilon}~\e^{\ii
  qx^{10}(z)}
\otimes\psi^{10}(z) \ , \non\chi^{~}_{D_\epsilon}(z)&=&-\frac1{2\pi}\,
\int\limits_{-\infty}^\infty\frac{\dd q~q}{(q-\ii\epsilon)^2}~
\e^{\ii qx^{10}(z)}\otimes\psi^{10}(z) \ ,
\label{bosemission}\eea
while the emission operators for the fermionic recoil states
are\footnote{\baselineskip=12pt Strictly speaking, the spin operators
  in these relations should include a non-trivial cocycle~\cite{KLLSW} for the
  lattice of charges $\alpha$ in the exponentials of the bosonized
  representation, which also depend on the fields
  $\Sigma_\alpha(z)$. The cocycle factor is defined on the weight
  lattice of the spinor representation of the Euclidean group, and it
  ensures that the vertex has the correct spinor transformation
  properties. Its inclusion becomes especially important in the generalization
  of these results to higher-dimensional branes. To avoid clutter in
  the formulas, we do not write these extra factors explicitly.}
\bea
\Xi_{C_\epsilon}(z)&=&\frac\epsilon{4\pi\ii}\,\int
\limits_{-\infty}^\infty\frac{\dd q}{q-\ii\epsilon}~
\e^{\ii qx^{10}(z)}\otimes\mu(z)\otimes
\overline{u}^{\,\alpha}(q)\,\Sigma_\alpha(z) \ , \non
\Xi_{D_\epsilon}(z)&=&-\frac1{2\pi}\,\int\limits_{-\infty}^\infty\frac{\dd
  q}{(q-\ii\epsilon)^2}~\e^{\ii qx^{10}(z)}\otimes\mu(z)\otimes
\overline{u}^{\,\alpha}(q)\,\Sigma_{\alpha}(z) \ .
\label{fermemission}\eea
Here $\mu(z)$ is an appropriate auxilliary ghost spin operator of
conformal dimension $-\frac18$~\cite{FMS1}. For example, it can be taken to
be a plane wave $\mu(z)=\e^{\ii k_ix^i(z)}$ in the
directions $x^i$ transverse to the $(x^9,x^{10})$ light cone, with
$k^2=-\frac14$. In the physical conformal limit $\epsilon\to0^+$, the
superfields (\ref{recoilemission}) then have vanishing superconformal
dimension.

The spinor $u(q)$ in (\ref{fermemission}) is not constrained by any
on-shell equations such as the Dirac equation which would normally
guarantee that the corresponding states respect spacetime
supersymmetry. It can be partially
restricted by implementing the GSO truncation of the superstring
spectrum. The fermion chirality operator $\Gamma$ acts on the
operators (\ref{cutoper}) as
\beq
\Gamma\,\Sigma_{q+(1-\lambda)/2}(z)\,\Gamma^{-1}=(-1)^{q-\lambda+1}
\,\Sigma_{q+(1-\lambda)/2}(z)
\label{GammaSigmaq}\eeq
for $q\in\zed$. This is only consistent with the operator product
expansions in the combined superconformal field theory including ghost
fields, because the action of $\Gamma$ on the fields of (\ref{SD0})
alone is not an automorphism of the local algebra of spin
fields~\cite{FMS1}. Then the action of the chirality operator can be
extended to the spin fields with the $\Gamma=1$ projection giving a
local field theory. The chiral $\Gamma=1$ projection requires that
$u_\alpha(q)$ be a right-handed Dirac spinor, after which the operators
(\ref{fermemission}) become local fermionic fields. Then the vertex operators
(\ref{recoilemission})--(\ref{fermemission}) describe the appropriate
supersymmetric states for the recoil problem. The relevant spacetime
supersymmetry generator $Q_\alpha$ is given by the contour integral of the
fermionic vertex corresponding to the basic tachyon operator $\e^{\ii
  qx^{10}(z)}$ at zero momentum,
\beq
Q_\alpha=\oint\limits_{z=0}\frac{\dd z}{2\pi}~\partial_zx^{10}(z)
\otimes z^{1/4}\,\mu\left(\frac1z\right)
\otimes\varepsilon_\alpha^{~\beta}\,\Sigma_\beta(z)\, \ .
\label{Qalphadef}\eeq
The integrand of (\ref{Qalphadef}), which involves the adjoint ghost
field $\mu^\dag(z)$, is a BRST invariant conformal field
of dimension~1. From the various operator product expansions above it
follows that the supercharge (\ref{Qalphadef}) relates the two vertices
(\ref{bosemission}) and (\ref{fermemission}) through the anticommutators
\beq
\Bigl\{Q_\alpha\,,\,\e^{\ii qx^{10}(z)}\otimes\mu(z)\otimes\Sigma_\beta(z)
\Bigr\}=-\ii\delta_{\alpha\beta}~\e^{\ii qx^{10}(z)}\otimes\psi^{10}(z) \ .
\label{QalphaSUSY}\eeq
Notice, however, that the target space supersymmetry alluded to here
refers only to the fields which live on the worldline of the
D-particle, or more precisely on the corresponding light-cone. The
full target space supersymmetry is of course broken by the motion of
the D-brane~\cite{MavSz}.

\newsection{The $\widehat{c}=-2$ Model}

By far the best understood example of a logarithmic conformal field
theory is the $c=-2$ model. This model has various realizations in
terms of ghost fields, symplectic fermions, and twist
fields~\cite{Kausch1,Saleur1}. In this section we shall describe the
superconformal extension of this logarithmic conformal field theory
through the local triplet model~\cite{GabKausch1}, and discuss some of
the supersymmetric generalizations of the extended algebras that arise
in the bosonic case.

\subsection{Superconformal Symplectic Fermions\label{SymplFerm}}

We begin with the standard superconformal ghost system which has
classical action~\cite{FMS1}
\beq
S_{\rm gh}=\frac1{2\pi}\,\int\dd z~\dd\overline{z}~\dd\theta~
\dd\overline{\theta}~
\left(\scb\,\overline{\deriv_\scz}\,\scc+\overline{\scb}\,\deriv_\scz
\overline{\scc}\right) \ ,
\label{Sgh}\eeq
where the chiral parts of the superfields
\bea
\scb(\scz)&=&\beta(z)+\theta\,b(z) \ , \non
\scc(\scz)&=&c(z)+\theta\,\gamma(z)
\label{bcsuperfields}\eea
have superconformal dimensions $\frac12$ and 0, respectively. The
fermionic $(b,c)$ system has spin $(1,0)$ and central charge $-2$,
while the bosonic $(\beta,\gamma)$ system has spin $(\frac12,\frac12)$
and central charge $-1$. The superconformal central charge of the combined
system is therefore $\hat c=-2$. The non-vanishing Neveu-Schwarz
two-point functions are
\beq
\Bigl\langle\scb(\scz_1)\,\scc(\scz_2)\Bigr\rangle^{~}_{\rm NS}=
\Bigl\langle\scc(\scz_1)\,\scb(\scz_2)\Bigr\rangle^{~}_{\rm NS}=
\frac{\theta_{12}}{\scz_{12}} \ ,
\label{bcNS2ptfns}\eeq
and the chiral super energy-momentum tensor is given by
\beq
\scT(\scz)=\frac12\,\deriv_\scz\scc(\scz)\,\deriv_\scz\scb(\scz)-
\frac12\,\partial_z\scc(\scz)\,\scb(\scz) \ .
\label{superEMbc}\eeq

Logarithmic behaviour in the bosonic part of this model is well-known
to arise from extending the normal fermionic fields $(b,c)$ to fields
with extra zero modes included~\cite{Kausch1}. This introduces the chiral,
two-component scalar symplectic fermion field $\chi^\pm(z)$ through
\bea
b(z)&=&\partial_z\chi^-(z) \ , \non c(z)&=&\chi^+(z) \ .
\label{bcchipm}\eea
The action then has a global $SL(2,\real)$ symmetry acting by rotations
of $\chi^\pm$. The Grassmann fields $\chi^\pm(z)$ have mode expansions
\beq
\chi^\pm(z)=\chi_0^\pm+\tilde\chi_0^\pm\,\ln z+\sum_{n\neq0}
\frac{\chi_n^\pm}n~z^{-n}
\label{chipmmodeexp}\eeq
with the non-vanishing anticommutation relations
\bea
\left\{\chi_n^\pm\,,\,\chi_m^\mp\right\}&=&\mp\,n\,\delta_{n+m,0}
\ , \non\left\{\chi_0^\pm\,,\,\tilde\chi_0^\pm\right\}&=&\pm\,1
\label{chianticommrels}\eea
and operator product expansions
\beq
\chi^+(z)\,\chi^-(w)=-\chi^-(z)\,\chi^+(w)=\ln(z-w)+\dots \ .
\label{chipm2ptfn}\eeq
The symplectic, $SL(2,\complex)$-invariant Fock vacuum
$|0\rangle$ is defined by
$\tilde\chi_0^\pm|0\rangle=\chi_n^\pm|0\rangle=0~~\forall n>0$. The
symplectic fermion differs from the fermionic ghost system through
the treatment of the $b(z)$ zero mode. The ghost system may be
identified with a non-logarithmic subsector of the symplectic fermion
model, whose chiral algebra ${\cal C}_\chi$ is defined by including the
extra zero mode $\chi_0^-$ in out-states of correlation functions as
\bea
&&\Bigl\langle b(z_1)\cdots b(z_n)\,c(w_1)\cdots
c(w_m)\Bigr\rangle_{{\cal C}_\chi}\non&&~~~~~~~
\equiv~\langle\chi_0^-|\partial_{z_1}\chi^-(z_1)\cdots
\partial_{z_n}\chi^-(z_n)\,\chi^+(w_1)\cdots\chi^+(w_m)
|0\rangle\non&&~~~~~~~=~\delta_{n+1,m}\,
\prod_{i<i'}(z_i-z_{i'})\,\prod_{j<j'}(w_j-w_{j'})\,
\prod_{i=1}^n\,\prod_{j=1}^m\frac1{z_i-w_j} \ ,
\label{bc0modecorrs}\eea
where $|\chi_0^\pm\rangle=\chi_0^\pm|0\rangle$.

There are several consequences of the presence of these extra
non-trivial zero modes. First of all, they couple the holomorphic and
antiholomorphic parts of the theory and thereby create a full non-chiral
algebra~\cite{GabKausch1}. Nevertheless, we shall continue to
concentrate on the chiral sector of the symplectic fermion field
theory. Secondly, they are directly responsible for the logarithmic
structure of the theory. Because of (\ref{bc0modecorrs}), in ${\cal
  C}_\chi$ the vacuum state has zero norm, $\langle0|0\rangle=0$, while
$\langle0|\chi^-(z)\,\chi^+(z)|0\rangle=1$. The field
\beq
\omega(z)=\chi^-(z)\,\chi^+(z)
\label{omegadef}\eeq
has the operator product expansions
\bea
T(z)\,\omega(w)&=&\frac1{(z-w)^2}+\frac1{z-w}\,\partial_w\omega(w)+\dots
\ , \non\omega(z)\,\omega(w)&=&-\ln^2(z-w)-2\ln(z-w)\,\omega(w)+\dots
\ ,
\label{omegaOPEs}\eea
and hence the two-point function
\beq
\langle0|\omega(z)\,\omega(w)|0\rangle=-2\ln(z-w) \ .
\label{omega2ptfn}\eeq
It follows that the field $\omega(z)$ is the logarithmic partner of
the identity operator $I$, i.e. the pair of operators $C(z)=I$,
$D(z)=\omega(z)$ generate a logarithmic conformal algebra with
\bea
\Delta_I&=&0 \ , \non b&=&1 \ , \non d&=&0 \ .
\label{sympllogpars}\eea
In terms of highest weight representations, there is another, degenerate
(non-invariant) vacuum state $|\omega\rangle=\omega(0)|0\rangle$ which is
conjugate to the $SL(2,\complex)$-invariant one $|0\rangle$, with
$|0\rangle=\tilde\chi_0^-\,\tilde\chi_0^+|\omega\rangle=L_0|\omega\rangle$.

The superconformal extension of the symplectic fermion system is
straightforward. Using the fermionic supercurrent in (\ref{superEMbc})
we easily derive the superpartners of the scalar Grassmann fields
$\chi^\pm(z)$ to be the bosonic spin $\frac12$ fields $\psi^\pm(z)$,
where
\bea
\psi^+(z)&=&\gamma(z) \ , \non\psi^-(z)&=&-\beta(z) \ .
\label{psipmgammabeta}\eea
As expected from Bose statistics, there are no special roles played by
the zero modes in this sector. Combining these operators into the
chiral scalar superfields
\beq
\scX^\pm(\scz)=\chi^\pm(z)\pm\theta\,\psi^\pm(z)
\label{Xpmchipsi}\eeq
with the non-vanishing operator product expansion
\beq
\scX^-(\scz_1)\,\scX^+(\scz_2)=-\ln\scz_{12}+\dots \ ,
\label{scXOPE}\eeq
we may write the superconformal ghost fields as
\bea
\scb(\scz)&=&\deriv_\scz\scX^-(\scz) \ , \non\scc(\scz)&=&\scX^+(\scz)
\ .
\label{scbsccscX}\eea
The chiral action and super energy-momentum tensor then assume the
standard free scalar superfield forms
\bea
S_{\rm gh}&=&\frac1\pi\,\int\dd^2z~\dd^2\theta~\deriv_\scz\scX^-\,
\overline{\deriv_\scz}\,\scX^+ \ , \non\scT(\scz)&=&\frac12\,
\deriv_\scz\scX^+(\scz)\,\partial_z\scX^-(\scz)-\frac12\,
\partial_z\scX^+(\scz)\,\deriv_\scz\scX^-(\scz) \ .
\label{SghscTsympl}\eeq

The superpartners to the logarithmic operators $I$ and $\omega(z)$ in
(\ref{omegadef}) may also be easily determined by using
(\ref{GCDchiexpl}) and the supercurrent
\beq
G(z)=\frac12\,\Bigl(\partial_z\chi^+(z)\otimes\psi^-(z)+\partial_z
\chi^-(z)\otimes\psi^+(z)\Bigr)
\label{Gzexpl}\eeq
to get
\bea
\chi^{~}_I(z)&=&0 \ , \non\chi^{~}_\omega(z)&=&
\chi^+(z)\otimes\psi^-(z)+\chi^-(z)\otimes\psi^+(z) \ .
\label{Iomegasuperpart}\eea
The operator product expansions (\ref{CDsuperOPE}) are straightforward
to derive by using (\ref{scXOPE}), (\ref{SghscTsympl}) and
factorization of the bosonic and fermionic sectors of the free field
theory. The NS correlation functions (\ref{CC})--(\ref{DD}) are likewise
straightforward consequences of factorization and the
two-point functions (\ref{bcNS2ptfns}). Note that $\chi^{~}_\omega(z)$ is an
ordinary primary field of the Virasoro algebra of weight $\frac12$
which corresponds to a supersymmetric state in the NS sector. This is
a consequence of the fact that its partner
$\chi^{~}_C(z)$ in this case is (trivially) a null field, as we
generally anticipated in section~\ref{nullvectors} for the case of
zero dimension superconformal logarithmic operators.

\subsection{Spin Fields\label{SymplSpin}}

To deal with the R sector of the symplectic fermion superconformal
field theory, we proceed via bosonization in complete analogy with
section~\ref{logspinrecoil}. Again, there is the conserved $U(1)$
ghost number current (\ref{Jpsi}). The bosonization proceeds in
exactly the same way as before, except for some crucial sign changes
owing to the Bose statistics of the fields $\psi^\pm(z)$ in the
present case. The bosonization of the $U(1)$ current has the same form
as in (\ref{bosonformulas}), but the chiral scalar
field $\phi(z)$ is now non-compact and so has kinetic energy of
opposite sign to that of
section~\ref{logspinrecoil}, i.e. its two-point function is now
\beq
\langle0|\phi(z)\,\phi(w)|0\rangle=\ln(z-w) \ .
\label{boswrongsign}\eeq
As a consequence, the central extension of the $U(1)$ current algebra
is now equal to $-1$. Due to ghost conjugation symmetry, the Sugawara
energy-momentum tensor is given by (\ref{stressboson}) with $\kappa=0$
and an overall sign change. The tachyon vertex operators
(\ref{cutoper}) now have conformal dimension $-\Delta_{q,0}=-q^2/2$,
and the relations (\ref{tachyoncorr}) and (\ref{SigmaqqprimeOPE}) are
modified to
\bea
\Sigma_q(z)\,\Sigma_{q'}(w)&=&(z-w)^{-qq'}\,\Sigma_{q+q'}(w)
\Bigl(1+\ii q\,(z-w)\,\partial_w\phi(w)\Bigr)+\dots \ ,
{}~~ qq'>1 \ , \non
\langle0|\prod_{l=1}^n\Sigma_{q_l}(z_l)|0\rangle&=&
\delta^{~}_{q_1+\dots+q_n\,,\,0}~\prod_{k<l}(z_k-z_l)^{-q_kq_l} \ .
\label{Sigmaqchanges}\eea

However, the actual bosonization of the fields $\psi^\pm(z)$ is a bit
more involved, because in the present case the $U(1)$ current algebra
does not yield their complete dynamics~\cite{FMS1}. By Bose
statistics, the central charge of the conformal algebra generated by
the Sugarawa energy-momentum tensor $\frac18\,j(z)\,j(z)$ is $c=1$. To
get the required central charge $-1$, we need to add the generators of
a conformal algebra with $c=-2$. This is also clear from the fact that
the soliton fields $\e^{\pm\ii\phi(z)}$ are fermionic, while $\psi^\pm(z)$ are
bosonic. To this end, we introduce an auxilliary $c=-2$ ghost system
$(\eta,\xi)$, which comprises a pair of conjugate free fermion fields
of conformal weights $(1,0)$ with the non-vanishing two-point functions
\beq
\langle0|\eta(z)\,\xi(w)|0\rangle=\langle0|\xi(z)\,\eta(w)|0\rangle=
\frac1{z-w} \ .
\label{etaxi2ptfn}\eeq
These auxilliary fields commute with the original $c=-2$ $(b,c)$
system and also with the scalar field $\phi(z)$. Thus, the Ramond
sector of the symplectic fermion superconformal field theory involves
not only a non-compact $U(1)$ current algebra, but also a fermionic
$c=-2$ ghost system $(\eta,\xi)$.

The Bose fields $\psi^\pm(z)$ can now each be written in terms of a product of
two fermionic fields as
\bea
\psi^+(z)&=&\sqrt2~\e^{\ii\phi(z)}\otimes\eta(z) \ , \non
\psi^-(z)&=&-\sqrt2~\e^{-\ii\phi(z)}\otimes\partial_z\xi(z) \ .
\label{psipmbosetaxi}\eea
Note that the zero mode of $\xi(z)$ does not contribute and so the
auxilliary ghost system does not yield any additional logarithmic
behaviour. In fact, the superconformal symplectic fermion field theory
is only defined on the subalgebra ${\cal C}_\psi$ of the full chiral
algebra of the $(\eta,\xi)$ system which is generated by the fields $\eta(z)$
and $\partial_z\xi(z)$~\cite{Kausch1,FMS1}. In ${\cal C}_\psi$ it is
consistent to fix the zero-mode of the $\eta$ field to 0, i.e. $\oint_{z=0}\dd
  z~\eta(z)=0$. This lifts the two-fold degeneracy of the
vacuum state which would otherwise lead to extra logarithmic behaviour. The
operator products of the chiral algebra ${\cal C}_\psi$ are then
obtained by defining its correlation functions as
\bea
&&\Bigl\langle\eta(z_1)\cdots\eta(z_n)\,\partial_{w_1}\xi(w_1)\cdots
\partial_{w_m}\xi(w_m)\Bigr\rangle_{{\cal C}_\psi}\non&&~~~~~~~
\equiv~\langle\xi^{~}_0|\eta(z_1)\cdots\eta(z_n)\,\partial_{w_1}\xi(w_1)
\cdots\partial_{w_n}\xi(w_m)|0\rangle \ ,
\label{calCpsicorr}\eea
where $|\xi^{~}_0\rangle=\frac1{2\pi\ii}\,\oint_{z=0}\dd
  z~\xi(z)|0\rangle$.

As before, we can easily show that the operators $\Sigma_{\mp1/2}(z)$
create branch cuts in the fields $\psi^\pm(z)$. However, because of
the auxilliary $c=-2$ $(\eta,\partial_z\xi)$ system that is present,
there is no longer a unique spin field~\cite{Kausch1,Saleur1,LMRS},
but rather an infinite family of them labelled by a continuous
parameter $\lambda\in[0,1]$ which interpolates between Neveu-Schwarz
and Ramond boundary conditions on the fermion fields $\eta(z)$ and
$\xi(z)$. This possibility arises because the conserved $U(1)$ ghost
number current $\eta(z)\,\xi(z)$ of the $(\eta,\xi)$ system can itself
be bosonized, yielding twisted sectors of the corresponding Hilbert
space. In ${\cal C}_\psi$, the bosonization formulas are given by
\bea
\eta(z)&=&\frac1{\sqrt2}~\e^{-\ii\varphi(z)} \ , \non\partial_z\xi(z)&=&
\frac1{\sqrt2}~\e^{2\ii\varphi(z)} \ ,
\non\eta(z)\,\xi(z)&=&\frac\ii4\,\partial_z\varphi(z) \ ,
\label{etaxivarphibos}\eea
where $\varphi(z)$ is a compact Coulomb gas field of central charge deficit
$\kappa=1$. The twist fields in the $\lambda$-twisted sector of the
$(\eta,\xi)$ Hilbert space are defined by
\beq
\mu^{~}_\lambda(z)=\e^{\ii\lambda\varphi(z)/2} \ ,
\label{mulambdaphi1}\eeq
and they generate the local monodromy conditions
\bea
\eta(z)\,\mu^{~}_\lambda(w)&=&\frac1{\sqrt2}\,\Bigl(z-w\Bigr)^{-\lambda/2}\,
\mu^{~}_{\lambda-2}(w)+\dots \ , \non\partial_z\xi(z)\,\mu^{~}_\lambda(w)&=&
\frac1{\sqrt2}\,\Bigl(z-w\Bigr)^\lambda\,\mu^{~}_{\lambda+4}(w)+\dots \ .
\label{etaximonodromy}\eea

{}From (\ref{Sigmaqchanges}) and (\ref{etaximonodromy}) it follows that
the spin operators which generate $\zed_2$-twists in the fields
(\ref{psipmbosetaxi}) are given by
\beq
\Sigma^{(\lambda)}(z)=\Sigma_{-1/8}^+(z)=2\,\cos\Bigl(\lambda-1
\Bigr)\,\frac{\phi(z)}2\otimes\mu^{~}_\lambda(z) \ .
\label{spinsymplectic}\eeq
For each $\lambda$, they have weight
\beq
\Delta=-\Delta_{\frac{\lambda-1}2\,,\,0}+\Delta_{\frac\lambda2\,,\,1}=-\frac18
\label{hlambdaweight18}\eeq
and therefore correspond to the Ramond ground state $|\frac{\hat
  c}{16}\rangle^{~}_{\rm R}$ for $\hat c=-2$. Thus, in contrast to the
model of the previous section, the superconformal symplectic fermion
system possesses an infinite family of supersymmetric ground states
which are labelled by their $U(1)$ charge $1-\lambda$ in the non-compact
current algebra generated by (\ref{Jpsi}). This charge partly labels
  the inequivalent representations of the
underlying superconformal algebra. Using (\ref{chiCDSigma}),
(\ref{Iomegasuperpart}), (\ref{Sigmaqchanges}) and
(\ref{etaximonodromy}), we may easily compute the corresponding family
of excited logarithmic spin fields of dimension $-\frac18$ to be
\bea
\widetilde{\Sigma}^{(\lambda)}_I(z)&=&0 \ , \non
\widetilde{\Sigma}^{(\lambda)}_\omega(z)&=&
\chi^-(z)\otimes\Sigma_{-(\lambda-3)/2}(z)\otimes
\mu^{~}_{\lambda-2}(z)-\chi^+(z)\otimes
\Sigma_{(\lambda-3)/2}(z)\otimes z^{3\lambda/2}\,\mu^{~}_{\lambda+4}(z)
\ . \non&&
\label{logspinsympl}\eeq
Again, the complete logarithmic superconformal algebra of
sections~\ref{SLCFTgen} and \ref{Correlators} may be easily verified
explicitly in the present case by using factorization of the various
free field sectors and the correlation functions given above.

\subsection{The $sl(2,\realbm)$ Symmetry Algebra}

The generators of the $SL(2,\real)$ symmetry of the ordinary
symplectic fermion action are given by the conformal dimension 1
primary currents~\cite{Kausch1}
\bea
J^0(z)&=&\frac13\,\Bigl(\chi^+(z)\,\partial_z\chi^-(z)+
\chi^-(z)\,\partial_z\chi^+(z)\Bigr) \ , \non
J^\pm(z)&=&\frac23\,\chi^\pm(z)\,\partial_z\chi^\pm(z) \ .
\label{SL2Rgens}\eea
By Fermi statistics, these are the complete set of bosonic primary
fields with $\Delta=1$. Together with the logarithmic operator
$\omega(z)$, they generate a logarithmic extension of an $sl(2,\real)$
Kac-Moody type symmetry algebra~\cite{KogNich1}, which is very similar
to an ordinary affine Lie algebra. The relations are given by the
operator product expansions (\ref{omegaOPEs}) and
\bea
J^a(z)\,J^b(w)&=&\frac29\,g^{ab}\,\left(\frac{\ln(z-w)+
\omega(w)+1}{(z-w)^2}+\frac{1/2}{z-w}\,\partial_w\omega(w)\right)+
\frac{f^{ab}_{~~c}}{z-w}\,J^c(w)+\dots \ , \non
J^a(z)\,\omega(w)&=&\ln(z-w)\,J^a(w)+\dots \ ,
\label{SL2Rlogalg}\eea
where $g^{ab}$ and $f^{ab}_{~~c}$ are the metric and structure
constants of the $sl(2,\real)$ Lie algebra with, in the chosen basis,
the non-vanishing components $g^{+-}=-2g^{00}=-2$ and
$f^{+-}_{~~~3}=\mp\,2f^{3\pm}_{~~~\pm}=-2$. Up to the $\ln(z-w)$
  terms and the appearence of the logarithmic operator
  (\ref{omegadef}), the relations (\ref{SL2Rlogalg}) are just those of
  a standard $sl(2,\real)$ affine Lie algebra. This algebra is related to the
  $W$-algebra of type $W(2,3^3)$ that arises from combining the
  $sl(2,\real)$ and Virasoro algebras and is normally used to classify
  the triplet theory~\cite{GabKausch1,Kausch1}. The corresponding generators
  are obtained from (\ref{SL2Rgens}) by the substitutions
  $\chi^\pm(z)\mapsto\partial_z\chi^\pm(z)$~\cite{Kausch1}. The fields
  $J^a(z)$ then become the $W_3$-triplet and the logarithmic operator
  $\omega(z)$ becomes the bosonic energy-momentum tensor~$T(z)$. This
  eliminates the logarithms in (\ref{SL2Rlogalg}) (effectively
  replacing $\ln z$ with $1/z^2$).

As a simple application of the above formalism, let us now examine the
supersymmetric enhancement of this logarithmically extended symmetry
algebra. As is usual in a supersymmetric current algebra, we regard
the bosonic sector current $J^a(z)$ as the highest component of a
weight $\frac12$ holomorphic superfield
\beq
\scJ^a(\scz)=j^a(z)+\theta\,J^a(z) \ .
\label{scurrent}\eeq
Using the supercurrent (\ref{Gzexpl}), the superpartners $j^a(z)$ with
$J^a(z)=\chi_{j^a}^{~}(z)$ of the currents (\ref{SL2Rgens}) are easily
calculated to be the dimension $\frac12$ primary fields
\bea
j^0(z)&=&\frac13\,\Bigl(\chi^-(z)\otimes\psi^+(z)-\chi^+(z)\otimes
\psi^-(z)\Bigr) \ , \non j^\pm(z)&=&\pm\,\frac23\,\chi^\pm(z)
\otimes\psi^\pm(z) \ .
\label{SL2Rsupergens}\eea

Together with $\omega(z)$, its superpartner $\chi^{~}_\omega(z)$ in
(\ref{Iomegasuperpart}), and the bosonic $U(1)$ ghost number current
(\ref{Jpsi}), the fermionic currents (\ref{SL2Rsupergens}) generate an
algebra whose non-vanishing operator product expansions are given by
(\ref{omegaOPEs}) and
\bea
j^a(z)\,j^b(z)&=&\frac29\,g^{ab}\,\frac{\ln(z-w)+\omega(w)+1}{z-w}+
\frac29\,d^{ab}\,\ln(z-w)\,j(w)+\dots \ , \non j^a(z)\,\omega(w)&=&
\ln(z-w)\,j^a(w)+\dots \ , \non j^0(z)\,j(w)&=&\frac{1/3}{z-w}\,
\chi^{~}_\omega(w)+\dots \ , \non j^\pm(z)\,j(w)&=&\pm\,\frac1{z-w}
\,j^\pm(w)+\dots \ , \non j(z)\,j(w)&=&-\frac1{(z-w)^2}+\dots \ ,
\non j^0(z)\,\chi^{~}_\omega(w)&=&-\frac23\,\ln(z-w)\,j(w)+\dots \ ,
\non\omega(z)\,\chi_\omega^{~}(w)&=&\ln(z-w)\,\chi^{~}_\omega(w)+\dots
\ , \non j(z)\,\chi^{~}_\omega(w)&=&-\frac3{z-w}\,j^0(w)+\dots \ ,
\non \chi^{~}_\omega(z)\,\chi^{~}_\omega(w)&=&\frac{2\,\omega(w)-2
\ln(z-w)}{z-w}+\dots \ ,
\label{SL2Rsusylogalg}\eea
where $d^{ab}$ is an $sl(2,\real)$ tensor whose only non-vanishing
components, in the given basis, are $d^{\pm\mp}=\pm\,1$. In
addition, there are mixed correlations between these operators and the
original currents (\ref{SL2Rgens}) given by the non-vanishing operator
product expansions
\bea
J^a(z)\,j^b(w)&=&\frac{1/3}{z-w}\,\left(f^{ab}_{~~c}\,j^c(w)+
\frac{g^{ab}}3\,\chi^{~}_\omega(w)\right)+\dots \ , \non
J^a(z)\,\chi^{~}_\omega(w)&=&\frac1{z-w}\,j^a(w)+\dots \ .
\label{SL2Radditional}\eea
Again, up to the $\ln(z-w)$ terms and the appearence of the
logarithmic superpartner $\chi^{~}_\omega(z)$, these are just the
usual relations of the $N=1$ superconformal extension of an
$sl(2,\real)$ Kac-Moody algebra.

Thus the currents (\ref{SL2Rgens}) and (\ref{SL2Rsupergens}) define a
logarithmic extension of the $N=1$ super Kac-Moody symmetry algebra of
the superconformal symplectic fermion model. In the bosonic case, the
extended $W$-algebra determines the irreducible representations of the
triplet theory and furnishes it with an appropriate definition of
rationality~\cite{GabKausch1}. It is therefore the building block of the $c=-2$
model. Under the extension $\chi^\pm(z)\mapsto\partial_z\chi^\pm(z)$
(with $\psi^\pm(z)$ unchanged), the operator $\chi^{~}_\omega(z)$
becomes the fermionic supercurrent (\ref{Gzexpl}), consistently with the
transformations of the bosonic sector and the operator product expansions
(\ref{SL2Rsusylogalg},\ref{SL2Radditional}). We may thereby conclude
that an $N=1$ superconformal extension of the $W$-algebra of type
$W(2,3^3)$ yields the algebraic characterization of the $\hat c=-2$,
superconformal symplectic fermion model. This enhancement of the
$W$-symmetry in the supersymmetric case is reminescent of the direct product
extension~\cite{wdoubl} of the $W_{1+\infty}$ symmetry algebra in
the twisted $N=2$ superconformal extension of the $SL(2,\real)/U(1)$
coset current algebra~\cite{twist}.

\subsection{Fusion Rules in the $\zedbm_2$ Orbifold Model}

The original $(b,c)$ ghost system of the $c=-2$ model has its own
conserved $U(1)$ current $c(z)\,b(z)$, and as a consequence it
possesses itself both twisted and untwisted sectors. The logarithmic
behaviour described above originates from the
$\zed_2$-twisted sector of the theory~\cite{Kausch1}, in which the symplectic
fermion fields $\chi^\pm(z)$ have anti-periodic boundary conditions
and half-integer weighted mode expansions (\ref{chipmmodeexp}). From
the $U(1)$ current it is possible to construct, in the usual way,
$\zed_2$ spin operators $\mu(z)$ of dimension $-\frac18$ which act on
the symplectic fermion fields through the operator products
\beq
\mu(z)\,\partial_w\chi^\pm(w)=\frac1{\sqrt{z-w}}\,\nu^\pm(w)+\dots \
,
\label{muchipm}\eeq
where $\nu^\pm(z)$ are excited twist fields of dimension
$\frac38$. The twist field $\mu(z)$ corresponds to the unique ground
state $|\mu\rangle=\mu(0)|0\rangle$ of weight $-\frac18$ in the
twisted sector, while $\nu^\pm(z)$ correspond to the doublet of
excited states $|\nu^\pm\rangle=\chi^\pm_{-1/2}|\mu\rangle$ of
dimension $\frac38$. In the twisted sector, the zero modes in
(\ref{chipmmodeexp}) are naturally absent, and the logarithmic
operator $\omega(z)$ can be split into two twist fields through the
fusion relation~\cite{gurarie,Saleur1}
\beq
\mu(z)\,\mu(w)=(z-w)^{1/4}\,\Bigl(\omega(w)+\ln(z-w)\Bigr)+\dots \ .
\label{mufusion}\eeq
In the language of highest weight representations, the degenerate
vacua of the untwisted sector are given by direct products of twisted
sector states as~\cite{GabKausch1}
\bea
|\omega\rangle&=&|\mu\rangle\otimes|\mu\rangle \ , \non
|0\rangle&=&-\frac14\,|\mu\rangle\otimes|\mu\rangle+L_{-1}|\mu\rangle
\otimes|\mu\rangle \ .
\label{tensortwisted}\eea
The complete set of fusion relations among the
twist fields may be summarized succinctly in terms of appropriately
defined product representations of the Virasoro algebra. For any
quasi-primary or primary chiral field $O(z)$, let $[O]$ denote the irreducible
representation built on the ground state $O(0)|0\rangle$. From
(\ref{muchipm}) and (\ref{mufusion}) we then have the non-trivial
fusion rules~\cite{GabKausch1}
\bea
[\mu]\times[\mu]&=&[\omega] \ , \non{} [\mu]\times[\nu^\pm]&=&
[\chi_{-1}^\pm\,\omega] \ , \non{} [\nu^\pm]\times[\nu^\mp]&=&
[\omega] \ ,
\label{munufusions}\eea
where the right-hand sides of (\ref{munufusions}) are indecomposable
representations of the Virasoro algebra generated by the logarithmic
pairs of states $(|0\rangle\,,\,|\omega\rangle)$ and
$(\chi_{-1}^\pm|0\rangle\,,\,\chi_{-1}^\pm|\omega\rangle)$.

Let us now describe the appropriate superconformal extension of the
$\zed_2$ orbifold conformal field theory. Supersymmetry of the
symplectic fermion model severely constrains the possible
fields. Namely, whatever twist is imposed on the fields $\chi^\pm(z)$,
a compensating twist should be put in their superpartners $\psi^\pm(z)$
such that the fermionic supercurrent (\ref{Gzexpl}) is
single-valued~\cite{DFMS}. As we have seen above, $\zed_2$ twists in the
superpartner fields $\psi^\pm(z)$ are generated by spin operators
$\Sigma(z)$ of dimension $-\frac18$ through the operator products
\beq
\Sigma(z)\,\psi^\pm(w)=\frac1{\sqrt{z-w}}\,\sigma^\pm(w)+\dots \ ,
\label{Sigmapsipm}\eeq
where the twist fields $\sigma^\pm(z)$ also have weight
$-\frac18$. The bosonization formulas\footnote{\baselineskip=12pt As
  before, the proper bosonization relations must include a non-trivial
  cocycle on the $U(1)$ charge lattice, which in the present case will
  also depend on the field $\mu(z)$.} and (\ref{Sigmapsipm})
imply the non-trivial fusion rules
\beq
[\Sigma]\times[\Sigma]&=&[I] \ , \non{} [\Sigma]\times[\sigma^\pm]&=&
[\psi^\pm] \ , \non{} [\sigma^\pm]\times[\sigma^\mp]&=&[I] \ .
\label{Sigmapsifusions}\eea
In (\ref{Sigmapsifusions}) we have used supersymmetry to keep only
descendants which are compatible with the operator product expansion
convention in (\ref{mufusion}). The right-hand sides of
(\ref{Sigmapsifusions}) are highest-weight Virasoro modules.

{}From the operator product expansions (\ref{muchipm}) and
(\ref{Sigmapsipm}) it follows that the lowest dimension spin-twist field
$\Omega(z)$ which leaves (\ref{Gzexpl}) single-valued is given by
\beq
\Omega(z)=\mu(z)\otimes\Sigma(z)
\label{Omegadef}\eeq
and it has conformal dimension $-\frac14$. Supersymmetry implies that
$\Omega(z)$ must have a superpartner. Since it is the spin-twist field of
lowest possible dimension, it must be the lowest component of a
superfield
\beq
\scOm(\scz)=\Omega(z)+\theta\,\chi^{~}_\Omega(z) \ .
\label{Wsuperfield}\eeq
The weight $\frac14$ superpartner $\chi^{~}_\Omega(z)$ of $\Omega(z)$
is easily found by applying the supercurrent (\ref{Gzexpl}) to
(\ref{Omegadef}), and using the operator product expansions
(\ref{muchipm}) and (\ref{Sigmapsipm}) to get
\beq
\chi^{~}_\Omega(z)=\nu^+(z)\otimes\sigma^-(z)+\nu^-(z)\otimes
\sigma^+(z) \ .
\label{chiOmega}\eeq
Using (\ref{munufusions}) and
(\ref{Sigmapsifusions})--(\ref{chiOmega}) we may then compute the
fusion relations generated by the spin-twist superfield and we find
\beq
[\scOm]\times[\scOm]=\Bigl([\omega]\otimes[I]\Bigr)
\oplus\Bigl([\chi_{-1}^+\,\omega]\otimes[\psi^-]\Bigr)
\oplus\Bigl([\chi_{-1}^-\,\omega]\otimes[\psi^+]\Bigr) \ .
\label{Omegafusion}\eeq
This provides the superconformal extension of the first fusion
relation in (\ref{munufusions}) and it agrees with the expressions for
the logarithmic superfields obtained in (\ref{Iomegasuperpart}). It
demonstrates explicitly how modules over the superconformal algebra in
this case split into indecomposable representations of the ordinary
Virasoro algebra. The fusion rule (\ref{Omegafusion})
thereby yields an alternative description of the superconformal
symplectic fermion system in terms of fusion relations, whereby the
logarithmic operators are descendants of non-integer weight primary
fields and are therefore no longer really required in the
theory. The Ramond sector in this language is incorporated by
introducing, in the usual way, the appropriate spin operators for the
fermionic partners $\chi^{~}_\Omega(z)$ in (\ref{chiOmega}). Along with the
construction of the previous subsection, this formalism leads to the
representation theoretic classification of this
logarithmic superconformal field theory. We will not explore this
classification any further in this paper.

\newsection{Supersymmetric Wess-Zumino-Witten Models}

As our final example, in this section we will describe explicitly the
appearance and properties of logarithmic conformal algebras in $N=1$
supersymmetric WZW models~\cite{coulomb,swzw}, including
the Ramond sector. For definiteness, we shall concentrate on
the compact $su(2)_k$ and non-compact $sl(2,\real)_k$ Kac-Moody algebras,
but a lot of the discussion can be extended to more general current
algebras based on generic simple Lie groups. Our analysis will be
facilitated enormously by a Coulomb gas representation of these
models~\cite{terao}, which allows for a free field
representation of the superconformal dynamics. In particular, in this
formalism the fermion sector of the supersymmetric WZW model
completely decouples from the bosonic sector, thereby yielding free
fermion fields which enable a direct construction of the spin operators
in analogy with the other examples studied in this paper.
As before, the spin fields are completely determined by the chiral, fermionic
current algebra of the superconformal field theory.

\subsection{Supersymmetric $su(2)_k$ Current Algebras}

Consider the $N=1$ supersymmetric extension of a level $k\in\zed_+$ Kac-Moody
algebra based on the $SU(2)$ group, which is generated by a triplet of
supercurrents $\scJ^a(\scz)$ as in (\ref{scurrent}) that are holomorphic,
weight $\frac12$ superfields transforming in the adjoint
representation of $SU(2)$. Here $j^a(z)$ are spin $\frac12$ fermionic
currents and $J^a(z)$ are bosonic currents of dimension~1 which
generate an ordinary Kac-Moody algebra. The operator product
expansions of the super Kac-Moody algebra are given by
\begin{equation}
\scJ^a(\scz _1)\,\scJ^b(\scz _2)=\frac{k\, g^{ab}}{2\scz_{12}}
+\frac{ \theta_{12}}{\scz_{12}}\, f^{ab}_{~~c}\,\scJ^c(\scz
_2)+\dots \ .
\label{skma}\end{equation}
In an orthonormal basis, the metric and structure constants of $su(2)$
are given by $g^{ab}=\delta^{ab}$ and
$f^{ab}_{~~c}=\ii\varepsilon^{ab}_{~~c}$. In analogy with the bosonic
case, the super energy-momentum tensor may be represented in terms of
the Kac-Moody supercurrents via a Sugarawa construction as~\cite{coulomb,swzw}
\begin{equation}
\scT(\scz)=\frac{1}{k}\, g_{ab}\,\deriv_\scz\scJ^a(\scz)\,\scJ^b(\scz)
+\frac{2 }{3k^2}\, f_{abc}\,\scJ^a(\scz)\,\scJ^b(\scz)\,
\scJ^c(\scz) \ ,
\label{superstresssug}\end{equation}
and from (\ref{skma}) it follows that (\ref{superstresssug}) generates
a super Virasoro algebra (\ref{TOPE}) of superconformal central charge
\beq
\hat c=\frac{3k-4}k \ .
\label{hatcKMk}\eeq

The bosonic part of the supersymmetric WZW model
at level $k$ is just the ordinary WZW model at level $k-2$, and the
spectrum of the supersymmetric field theory is the same as that of the
WZW model at level $k-2$ plus free Majorana fermion fields in the adjoint
representation of the $SU(2)$ group~\cite{coulomb,swzw}. The fermion
decoupling can be seen explicitly in terms of the Kac-Moody currents
by defining the modified currents
\begin{equation}
{\widehat J}^{\,a}(z)= J^a (z) + J_{\rm f}^a (z) \ ,
\label{modified}\end{equation}
where the fermionic currents $J_{\rm f}^a(z)$ are given in terms of the
original fermion currents $j^a(z)$ in (\ref{scurrent}) by
\begin{equation}
J_{\rm f}^a(z) = -\frac1k\, f^a_{~bc}\,j^b(z)\,j^c(z) \ .
\label{fermioncurr}\end{equation}
By using (\ref{skma}), one finds that the modified currents have the
operator product expansions
\bea
j^a(z)\,j^b(w)&=&\frac{k\, g^{ab}}{2(z-w)}+\dots \ , \nonumber \\
j^a(z)\,\widehat{J}^{\,b}(w)&=&{\widehat J}^{\,a}(z)\,j^b(w)~=~0+\dots
\ , \nonumber \\{\widehat J}^{\,a}(z)\,{\widehat
  J}^{\,b}(w)&=&\frac{(k-2)\, g^{ab}}{2(z-w)^2}+
\frac{  f^{ab}_{~~c}}{z-w}\,{\widehat J}^{\,c}(w)+\dots \ ,
\label{modcurrope}\eea
from which it follows that the operators ${\widehat J}^{\,a}(z)$
generate a bosonic Kac-Moody algebra at level $k-2$ and are independent
of the fermionic currents $j^a(z)$. The fields $j^a(z)$ pertain to the
supersymmetric partners in the current algebra, which thereby decouple
from the rest of the quantum field theory.

In the supersymmetric $SU(2)_k$ WZW model, the primary superfields
\beq
\scV_{j,m}(\scz)=V_{j,m}(z)+\theta\,W_{j,m}(z)
\label{scVjm}\eeq
are labelled by the usual angular momentum quantum numbers
$j\in\zed_+/2$, $m\in\{-j,-j+1,\dots,j-1,j\}$ of the $su(2)$ Lie
algebra~\cite{coulomb}. They have superconformal dimension
\beq
\Delta_{j}=\frac{j(j+1)}k
\label{Deltajm}\eeq
and obey
\bea
\scJ^3(\scz_1)\,\scV_{j,m}(\scz_2)&=&\frac{\theta_{12}}{\scz_{12}}\,
m\,\scV_{j,m}(\scz_2)+\dots \ ,
\non\scJ^\pm(\scz_1)\,\scV_{j,m}(\scz_2)&=&\frac{\theta_{12}}{\scz_{12}}
\,\Bigl(j\mp m\Bigr)\,\scV_{j,m\pm1}(\scz_2)+\dots \ ,
\label{scJscVjm}\eea
where $\scJ^\pm(\scz)=\scJ^1(\scz)\pm\ii\scJ^2(\scz)$. The mode
expansion of (\ref{superstresssug}) yields null state constraints
which, along with the superconformal Ward identities of
section~\ref{Ward}, leads to the supersymmetric version of the
Knizhnik-Zamolodchikov equations~\cite{swzw}
\beq
\left[\deriv_{\scz_l}-\frac2k\,\sum_{l'\neq
    l}\frac{\theta_{ll'}}{\scz_{ll'}}\, g_{ab}\,t_{(l)}^a\otimes
t_{(l')}^b\right]\Bigl\langle\scV_{j_1,m_1}(\scz_1)\cdots\scV_{j_n,m_n}(\scz_n)
\Bigr\rangle_{\rm NS}=0 \ , ~~ l=1,\dots,n \ ,
\label{SUSYKZeqn}\eeq
where $t^a_{(l)}$ are the generators of $SU(2)$ in the spin-$j_l$
representation with
\beq
\scJ^a(\scz_1)\,\scV_{j_l,m_l}(\scz_2)=\frac{\theta_{12}}{\scz_{12}}\,
t_{(l)}^a\,\scV_{j_l,m_l}(\scz_2)+\dots \ .
\label{Japrimarydef}\eeq
In particular, from (\ref{Japrimarydef}) we may infer the Ward
identities for the supersymmetric Kac-Moody algebra in the form
\beq
\Bigl\langle\scJ^a(\scz_0)\,\scV_{j_1,m_1}(\scz_1)\cdots
\scV_{j_n,m_n}(\scz_n)\Bigr\rangle^{~}_{\rm NS}
=\sum_{l=1}^n\frac{\theta_{0l}}{\scz_{0l}}
\,t_{(l)}^a\,\Bigl\langle\scV_{j_1,m_1}(\scz_1)\cdots
\scV_{j_n,m_n}(\scz_n)\Bigr\rangle^{~}_{\rm NS} \ .
\label{SUSYWardKM}\eeq

The prototypical example for logarithmic behaviour in this class of
superconformal field theories is provided by the spin $\frac12$
primary superfields at level $k=2$. The decoupled bosonic sector then
admits the operator product expansions~\cite{CKLT}
\bea
V_{\frac12,\alpha_1}(z_1)\,V_{\frac12,\alpha_2}(z_2)&=&
\frac1{(z_1-z_2)^{3/4}}\,\Bigl[I_{\alpha_1\alpha_2^\vee}\Bigr.\non
&&-\left.(z_1-z_2)\,t^a_{\alpha_1\alpha_2^\vee}\,\Bigl(D_a(z_2)+
\ln(z_1-z_2)\,C_a(z_2)\Bigr)\right]+\dots \ , \non&&
\label{spin12OPElogs}\eea
where
\beq
t^a=\frac{\sigma^a}2
\label{spin12gens}\eeq
are the spin $\frac12$ generators (with $\sigma^a$ the usual
$2\times2$ Pauli matrices) satisfying
\beq
\left[t^a\,,\,t^b\right]=\ii\varepsilon^{ab}_{~~c}\,t^c \ , ~~
\Tr\left(t^a\,t^b\right)=\frac12\,\delta^{ab} \ ,
\label{spin12gensrels}\eeq
and $\alpha=\pm\,\frac12$ are the fundamental weights of the
chiral $SU(2)$ group with $\alpha^\vee$ denoting
the weight conjugate to $\alpha$. From (\ref{SUSYKZeqn}) it follows
that the operators $C_a(z)$ and $D_a(z)$ have the non-vanishing NS
two-point functions~\cite{CKLT}
\bea
\Bigl\langle C_a(z)\,D_b(w)\Bigr\rangle^{~}_{\rm NS}&=&
\frac{\delta_{ab}}{(z-w)^2} \ , \non\Bigl\langle D_a(z)\,
D_b(w)\Bigr\rangle^{~}_{\rm NS}&=&\frac{2\,\delta_{ab}}{(z-w)^2}\,
\Bigl(1-\ln4(z-w)\Bigr) \ ,
\label{KCD2pt}\eea
and we thereby obtain a logarithmic superconformal algebra with
\bea
\Delta_{C_{a}}&=&1 \ , \nn\\b&=&1 \ , \nn\\d&=&2-4\ln2 \ .
\label{KMlogpars}\eea
The indecomposability of the corresponding affine $su(2)_2$ super Lie
algebra representation is implied by the Sugarawa construction
(\ref{superstresssug}). This logarithmic behaviour arises from the
fact, discussed above, that the bosonic sector of the supersymmetric
$SU(2)_2$ WZW model is just the ordinary $SU(2)$ WZW model at level
$k=0$, which is a well-known example of a $c=0$ theory which is both
logarithmic and unitary~\cite{km,CKLT}, and therefore contains no
negative dimension operators. Taking the product of the bosonic
logarithmic operators with the corresponding free fermion fields then
yields their superpartners, and hence the logarithmic behaviour
described above.

At the level of the fusion relations (\ref{spin12OPElogs}), one cannot
determine the precise form of the logarithmic pair. We will
find these operators explicitly later on by another technique. The
basic idea is that although the classical WZW action vanishes in the
bosonic $SU(2)_0$ model, the effective {\it quantum} action is
non-vanishing~\cite{Gerasimov} and yields a deformation of the classical
energy-momentum tensor which embeds the $c=0$ theory into a twisted
$N=2$ superconformal algebra~\cite{twist}. The BRST supercharge of the
corresponding topological field theory is the zero mode of one of the
fermionic supercurrents of this $N=2$ algebra, which is used as a
deformation field. The BRST operator can be represented in terms of a
fermionic screening operator for the WZW model, from which the
logarithmic operators may be explicitly constructed.

\subsection{Coulomb Gas Representation}

To facilitate our analysis, we shall now construct a free field
representation of the current algebra in the superfield
formalism~\cite{terao} and use it to construct highest-weight
representations of the supersymmetric $su(2)_k$ algebra in the Fock
space of the free fields. For this, we introduce three free, real
scalar superfields
\beq
{\sc\Phi}^a(\scz)=\phi^a(z)-\ii^a\,\theta\,\psi^a(z) \ , ~~ a=1,2,3 \ ,
\label{Phiasczdef}\eeq
which have the NS two-point functions
\beq
\Bigl\langle\sc\Phi^a(\scz_1)\,\sc\Phi^b(\scz_2)\Bigr\rangle_{\rm NS}
=(-1)^a\,\delta^{ab}\,\ln\scz_{12} \ .
\label{scPhi2ptfn}\eeq
For simplicity, here and in the remainder of this paper we shall work in an
orthonormal basis of the $su(2)$ Lie algebra.

The superspace currents $\scJ^a(\scz)$ can be expressed as
\begin{eqnarray}
\scJ^3(\scz)&=&\ii\,\sqrt{\frac{k}{2}}\,\deriv_\scz\sc\Phi^3(\scz) \ ,
\nonumber \\\scJ^{\pm}(\scz)&=&\sqrt{\frac{k}{2}}\,\Bigl(
\deriv_\scz\sc\Phi^2(\scz)\mp\deriv_\scz\sc\Phi^1(\scz)\Bigr)
{}~\e^{\pm\ii\,\sqrt{\frac{2}{k}}\,\bigl(\sc\Phi^2(\scz) +
\sc\Phi^3(\scz)\bigr)} \ .
\label{scJCoulomb}\end{eqnarray}
The Sugarawa operator (\ref{superstresssug}) can also
be expressed in terms of the superfields (\ref{Phiasczdef}) as
\begin{equation}
\scT(\scz)=\frac{1}{2}\,\sum_{a=1}^3(-1)^a\,\partial_z\sc\Phi^a(\scz)
\,\deriv_\scz\sc\Phi^a(\scz)+\frac{\ii}{\sqrt{2k}}\,\partial_z
\deriv_\scz\sc\Phi^1(\scz) \ ,
\label{scTCoulomb}\end{equation}
which in component form yields an expression for the bosonic energy-momentum
tensor that is familiar from the usual Coulomb gas representation
\begin{eqnarray}
T(z)&=&\frac{1}{2}\,\sum_{a=1}^3\Bigl((-1)^a\,\partial_z\phi^a(z)\,
\partial_z\phi^a(z)-\psi^a(z)\,\partial_z\psi^a(z)\Bigr)+
\frac{\ii}{\sqrt{2k}}\,\partial_z^2\phi^1(z) \ , \label{tcomp}\\
G(z)&=&-\frac12\,\sum_{a=1}^3\ii^a\,\partial_z\phi^a(z)\otimes\psi^a(z)+
\frac1{\sqrt{2k}}\,\partial_z\psi^1(z) \ .
\label{tgcomp} \end{eqnarray}
It is straightforward to check that this Coulomb gas representation
correctly reproduces the superconformal central charge
(\ref{hatcKMk}). The last term in (\ref{tcomp}) yields a background
charge $-1/\sqrt k$ at infinity in the Coulomb gas model for the
$\phi^1$ field, i.e. a central charge deficit
\beq
\kappa=\sqrt{\frac2k}
\label{kappasqrt2k}\eeq
in the notation of~(\ref{stressboson},\ref{Skappa}).

Let us now exhibit explicitly the decoupling of the fermionic partners
from the bosonic Kac-Moody algebra at level $k-2$, represented by the
operator product expansions (\ref{modcurrope}), in the Coulomb gas
representation. For this, we need to decouple the fermion fields
$\psi^a(z)$ from the scalar fields $\phi^a(z)$ in the expressions for
the modified fermionic currents of the previous subsection. We begin
by bosonizing the complex fermion fields
\beq
\psi^\pm(z)\equiv\psi^2(z)\mp\ii\psi^1(z)=\sqrt{2}~\e^{\pm\ii\phi^4(z)}
\label{psipmphi4}\eeq
by means of a fourth scalar boson field $\phi^4(z)$ with the two-point
function
\beq
\langle0|\phi^4(z)\,\phi^4(w)|0\rangle=-\ln(z-w) \ ,
\eeq
just as we did in section~\ref{logspinrecoil}. We now define an
$SO(2,1)$ transformation of the four boson fields
$\phi^i(z)$, $i=1,\dots,4$ by
\begin{eqnarray}
\varphi^1(z)&=&\phi^1(z) \ , \nonumber \\
\varphi^2(z)&=&\sqrt{\frac{2}{k-2}}\,\phi^2(z)+\sqrt{\frac{2}{k-2}}\,
\phi^4(z) \ , \nonumber \\
\varphi^3(z)&=&-\frac{2}{\sqrt{k(k-2)}}\,\phi^2(z)+\sqrt{\frac{k-2}{k}}\,
\phi^3(z)-\sqrt{\frac{2}{k-2}}\,\phi^4(z) \ , \nonumber \\
\varphi^4(z)&=&\sqrt{\frac{2}{k}}\,\phi^2(z)+\sqrt{\frac{2}{k}}\,
\phi^3(z)+\phi^4(z) \ ,
\label{newbosons}\end{eqnarray}
with the corresponding two-point functions
\beq
\langle0|\varphi^i(z)\,\varphi^j(w)|0\rangle=-(-1)^{\delta_{i,2}}\,
\delta^{ij}\,\ln(z-w) \ .
\label{SO212ptfn}\eeq
Note that (\ref{newbosons}) is only defined for $k>2$. With this
transformation, the modified currents (\ref{modified}) decouple from
the boson field $\varphi^4(z)$, and hence the fermion fields, and are
given explicitly by
\begin{eqnarray}
{\widehat J}^{\,3}(z)&=&\ii\,\sqrt{\frac{k-2}{2}}\,\partial _z
\varphi^3(z) \ , \nonumber \\
{\widehat J}^{\,\pm}(z)&=&\sqrt{\frac{k-2}{2}}\,
\left(\partial_z\varphi^2(z)\mp\sqrt{\frac{k}{k-2}}\,
\partial_z\varphi^1(z)\right)~\e^{\pm\ii\,\sqrt{\frac{2}{k-2}}\,
\bigl(\varphi^2(z)+\varphi^3(z)\bigr)} \ .
\label{modifiedCoulomb}\end{eqnarray}
It is straightforward to check directly that the operators
(\ref{modifiedCoulomb}) generate a bosonic $su(2)_{k-2}$ Kac-Moody
algebra. Indeed, (\ref{modifiedCoulomb}) is just the Nemeschansky
representation of the ordinary $su(2)$ current algebra at level
$k-2$~\cite{nemesh}.

On the other hand, we can fermionize back the boson field $\varphi^4(z)$
by introducing new free fermion fields ${\tilde \psi}^a(z)$ through
\bea
{\tilde\psi}^\pm(z)&=&\sqrt{2}~\e^{\pm\ii\varphi^4(z)} \ , \nn\\
{\tilde\psi}^3(z)&=&\psi^3(z) \ ,
\label{newfermions}\eea
and express the fermionic currents $j^a(z)$ solely in terms of three
free Majorana fermion fields as
\beq
j^a(z)=\sqrt{\frac{k}{2}}\,{\tilde \psi}^a(z) \ .
\label{fercurrent}\eeq
It follows that the supersymmetric $su(2)_k$ affine Lie algebra
decomposes into bosonic algebras as $su(2)_{k-2}\oplus su(2)_2$, where
$su(2)_2$ is the current algebra associated with the three
real fermion fields $\tilde\psi^a(z)$. Let us note from
(\ref{modifiedCoulomb}) that the bosonic $su(2)_{k-2}$ currents
formally vanish identically at $k=2$, and the supercurrents can be written as
$\scJ^a(\scz)=\tilde\psi^a(z)+\theta\,\varepsilon^a_{~bc}\,\tilde
\psi^b(z)\,\tilde\psi^c(z)$.
Thus the minimal level $k=2$ supersymmetric Kac-Moody algebra is
described solely in terms of three Majorana fermion fields, or
equivalently by means of a free boson field and a free fermion field,
i.e. a single scalar superfield. For $k>2$ the affine algebra is
determined instead by three scalar superfields.

Finally, let us describe how to evaluate correlation functions of
primary operators in the above free field representation. The vertex
operators for the primary superfields (\ref{scVjm}) can be
written in terms of the scalar superfields (\ref{Phiasczdef}) as
\beq
\scV_{j,m}(\scz)=\e^{\ii\,\sqrt{\frac2k}\,\bigl[-j\,\sc\Phi^1(\scz)
+m\bigl(\sc\Phi^2(\scz)+\sc\Phi^3(\scz)\bigr)\bigr]} \ .
\label{scVjmvertexop}\eeq
For $k>2$, the bosonic component of (\ref{scVjmvertexop}) is
mapped under the $SO(2,1)$ transformation (\ref{newbosons}) to
\beq
V_{j,m}(z)=\e^{\ii\bigl[-j\,\sqrt{\frac2k}\,\varphi^1(z)+
m\,\sqrt{\frac2{k-2}}\,\bigl(\varphi^2(z)+\varphi^3(z)\bigr)\bigr]} \ ,
\label{primary}\eeq
which, as expected, coincides with the primary fields of the
bosonic level $k-2$ Kac-Moody algebra for the spin $j$ highest weight
representation. The superpartner of (\ref{primary}) in this decoupling
free field representation may be determined in the NS algebra through
the supersymmetry transformation $W_{j,m}(z)=[G_{-1/2},V_{j,m}(z)]$.

To compute correlation functions of the vertex operators
(\ref{scVjmvertexop}), we need to insert appropriate Feigin-Fuks
operators~\cite{FF1} which commute with the Kac-Moody supercurrents
and have vanishing superconformal dimension. As such, their insertion
into correlators of primary superfields does not affect their
conformal or affine Ward identities, but merely serve to ``screen'' out extra
charges such that the correlators satisfy the appropriate neutrality
conditions required of (non-vanishing) singlet solutions to the
Knizhnik-Zamolodchikov equations (\ref{SUSYKZeqn}). Screening
operators are not unique, and they are provided by supercontour
integrals of the form~\cite{screen}
\begin{equation}
Q=\oint\limits_{z=0}\frac{\dd z}{2\pi\ii}~
\int\dd\theta~\scQ(\scz)=
\oint\limits_{z=0}\frac{\dd z}{2\pi\ii}~E(z) \ ,
\label{QWZW}\end{equation}
where $\scQ(\scz)=\rho(z)+\theta\,E(z)$. A screening operator for the
compact, $SU(2)_k$ supersymmetric WZW model may be constructed to
conveniently have only ${\sc\Phi}^1$-charge and is given by
\beq
\scQ(\scz)=\deriv_\scz\sc\Phi^2(\scz)~\e^{\ii\,\sqrt{\frac{2}{k}}\,
\sc\Phi^1(\scz)} \ .
\label{scQscPhi1}\eeq
In terms of the decoupling free field representation
(\ref{newbosons}), the weight~1 fermionic component of
(\ref{scQscPhi1}) may be written as (up to normalization for $k>2$)
\beq
E(z)=\partial_z\varphi^2(z)~
\e^{\ii\,\sqrt{\frac{2}{k}}\,\varphi^1(z)} \ .
\label{screening}\eeq
That the scalar field $\varphi^1(z)$ is appropriate for the screening
operator here is clear from the representation (\ref{tcomp}) of the
energy-momentum tensor, whereby the $\partial_z^2\varphi^1(z)$ term is
responsible for inducing a vacuum charge
$\e^{-\ii\,\sqrt{\frac2k}\,\varphi^1(\infty)}$ (see
(\ref{Skappa})). Indeed, one can check explicitly that the
first order pole in the operator product expansion of
$E(z)\,\widehat{J}^{\,a}(w)$ vanishes.

The insertion of (\ref{screening}) into primary state correlators rids
us of the background neutralizing constraint $\sum_lq_l=\kappa$ required of the
$\phi^1$ tachyon vertex operators, which generalizes
(\ref{selectionrule}) to the case of a non-vanishing central charge
deficit $\kappa$. To calculate correlation functions explicitly in the
Coulomb gas representation, we insert screening charges (\ref{QWZW})
and define vertex operators which are dual to (\ref{scVjmvertexop}) by
\beq
\tilde\scV_{j,m}(\scz)=\e^{\ii\,\sqrt{\frac2k}\,\bigl[(j+1)\,\sc\Phi^1(\scz)
+m\bigl(\sc\Phi^2(\scz)+\sc\Phi^3(\scz)\bigr)\bigr]} \ .
\label{scVjmvertexopdual}\eeq
Then the NS correlators in the free field representation are defined
by
\bea
\Bigl\langle\scV_{j_1,m_1}(\scz_1)\cdots\scV_{j_n,m_n}(\scz_n)\Bigr
\rangle^{~}_{\rm NS}&=&\delta^{~}_{m_1+\dots+m_n\,,\,0}~\langle0|
\scV_{j_1,m_1}(\scz_1)\cdots\scV_{j_{n-1},m_{n-1}}(\scz_{n-1})\non&&
\times\,\tilde\scV_{j_n,m_n}(\scz_n)~Q^{j_1+\dots+j_{n-1}-j_n}|0\rangle \ ,
\label{NSfreefield}\eea
where the charge neutrality constraint $\sum_lm_l=0$ follows from the
Ward identity (\ref{SUSYWardKM}) for the super affine $SU(2)$ symmetry.

\subsection{Emergence of Chiral Logarithmic Operators\label{LogConstruct}}

We will now proceed to an explicit construction of logarithmic
operators in the supersymmetric $SU(2)_k$ WZW model for appropriate
choices of the level $k\in\zed_+$. The construction is motivated by the
manner in which logarithmic behaviour arises in fusion relations among the
chiral primary superfields (\ref{scVjmvertexop}). For instance, at
level $k=2$ for the spin $j'=\frac12$ representation, it follows from
(\ref{spin12OPElogs}) that the logarithmic operators transform as a
conjugate representation of $SU(2)$ and have dimension
$\Delta_{1}=2/k=1$. This suggests that generally, for certain
values of $k$, logarithmic operators arise in the conjugate spin-$2j'$
representation in the Clebsch-Gordan decomposition of the fusion
product of two spin-$j'$ primary fields according to the affine
$SU(2)$ fusion rules
\beq
[V_{j'}]\times[V_{j'}]=[I]\oplus[V_1]\oplus[V_2]\oplus\cdots
\oplus[V_{2j'}] \ .
\label{CGjdecomp}\eeq
In the following we will show that this is indeed the case for the
spin~$\frac12$ representation, which necessitates the level number
$k=2$. For other spins $j'$ we have no general proof that this is the case, but
nevertheless logarithmic operators {\it can} be built for certain
$k$. The precise values of $j'$ and $k$ will be determined by demanding that
the logarithmic pair determine a marginal deformation of the original
superconformal field theory, and thereby generate an isomorphic
superconformal algebra.

We will first present a very simple version of this construction, and
then afterwards show more precisely that it amounts to a deformation
of the superconformal algebra. For this, we note that logarithmic
behaviour in a conformal field theory arises when two (primary)
operators, one of negative norm and the other of positive norm, have
weights which become degenerate~\cite{Cardy1}. Their other quantum
numbers must also coincide, so that they form a Jordan cell structure
with respect to the full, maximally extended chiral symmetry
algebra. Motivated by these remarks, we begin with the chiral primary
fields (\ref{primary}) written in terms of the original scalar
fields~$\phi^a(z)$,
\beq
V_{j,m}(z)=\e^{\ii\,\sqrt{\frac2k}\,\bigl[-j\,\phi^1(z)+m\bigl(
\phi^2(z)+\phi^3(z)\bigr)\bigr]} \ .
\label{Vjmphi}\eeq
It is important to note that, as follows from (\ref{tcomp}), with
respect to the relative signs in the respective kinetic energies, the
field $\phi^2(z)$ is time-like, while $\phi^1(z)$ and $\phi^3(z)$ are
space-like. Furthermore, $\phi^2(z)$ and $\phi^3(z)$ are free fields,
while $\phi^1(z)$ is a Coulomb gas field with central charge deficit
(\ref{kappasqrt2k}).

We now consider the state with $m=0$ and {\it deform} the operator
(\ref{Vjmphi}) by giving the $\phi^2$ field a coefficient
$\sqrt{\frac2k}\,(m+\epsilon)$, where $\epsilon\to0^+$ is an
infinitesimally small, but non-zero, number. Thus we define the field
\beq
V_{j,\epsilon}(z)=\e^{\ii\,\sqrt{\frac2k}\,\bigl(-j\,\phi^1(z)+
\epsilon\,\phi^2(z)\bigr)} \ .
\label{Vjepsilon}\eeq
{}From (\ref{tcomp}) it follows that the operator (\ref{Vjepsilon}) has
conformal dimension
\beq
\Delta_{j}(\epsilon)=\frac{j(j+1)}k-\frac{\epsilon^2}k \ ,
\label{Deltajepsilon}\eeq
where the minus sign in the second term is due to the time-like
signature of the free field $\phi^2(z)$. We will only consider those
states deformed in this way whose spin $j$, for a given level $k$, is
subject to the constraint
\beq
j(j+1)=k \ .
\label{jkconstr}\eeq
In that case, the conformal dimension (\ref{Deltajepsilon}) deviates
from unity by an infinitesimally negative amount, thereby making the
operator (\ref{Vjepsilon}) {\it relevant} in a renormalization group
sense. In the limit $\epsilon\to0^+$, the operator (\ref{Vjepsilon})
thus yields a marginal deformation of the superconformal field theory and
an isomorphic superconformal algebra, i.e. because of the time-like
signature of the field $\phi^2(z)$, deforming its momentum does {\it
  not} produce off-shell quantities.

Before moving on with the construction, let us pause to briefly
discuss what the implications of this deformation will be. The unitary
and topological (or rational) constraint $k\in\zed_+$ implies that the
only spin-$j$ representations which can be used in the ensuing
construction, i.e. those which obey (\ref{jkconstr}), are the integral
ones $j=1,2,3,4,\dots$. This is consistent with the unitarity of the
corresponding $SU(2)$ representation and the fact that the deformation
requires states of magnetic quantum number $m=0$. It will restrict
logarithmic behaviour in supersymmetric $su(2)_k$ current algebras to
those with level coefficient $k=2,6,12,20,\dots$,
respectively. Logarithmic behaviour can occur for other discrete
values of $k$ not in this list~\cite{nicholsmult}, but in that case
the operators will not yield a marginal deformation of the
superconformal algebra. Notice that only the case $k=2$ violates the
ground state unitarity condition $2j\leq k-2$, once again illustrating
the special features of this level number. Fractional level numbers
are also possible~\cite{Gab1,nicholsmult}, but in that
case the deformation will not apply. Finally, note that $k\in\zed_-$
implies by (\ref{jkconstr}) that also $j\in\zed_-$, leading to
non-unitary, continuous representations of the $su(2)_k$ current
algebra. Such values are most naturally thought of in terms of
non-compact, $sl(2,\real)_{-k}$ WZW models, which we shall briefly
discuss in section~\ref{sl2RAlgs}.

With the restriction (\ref{jkconstr}), let us now define the primary
operator
\beq
C_\epsilon(z)=-\ii\epsilon\,V_{j,\epsilon}(z)=-\ii\epsilon
{}~\e^{\ii\,\sqrt{\frac2{j(j+1)}}\,\bigl(-j\,
\phi^1(z)+\epsilon\,\phi^2(z)\bigr)} \ .
\label{CepsilonWZW}\eeq
This field has a logarithmic partner $D_\epsilon(z)$ which may be
found either by computing the operator product of the energy-momentum
tensor (\ref{tcomp}) with (\ref{CepsilonWZW}), or by the formal
derivative rule~\cite{gezel} $D_\epsilon=\partial
C_\epsilon/\partial\Delta_{j}(\epsilon)=-\frac k{2\epsilon}\,\partial
C_\epsilon/\partial\epsilon$. Using either method we find
\beq
D_\epsilon(z)=-\,\sqrt{\frac{j(j+1)}2}\,\phi^2(z)~
\e^{\ii\,\sqrt{\frac2{j(j+1)}}\,\bigl(-j\,
\phi^1(z)+\epsilon\,\phi^2(z)\bigr)} \ .
\label{DepsilonWZW}\eeq
Since the operator (\ref{DepsilonWZW}) contains the free scalar field
$\phi^2(z)$ itself (and not merely exponentials or derivatives
thereof), it has logarithmic correlation functions. In writing
(\ref{DepsilonWZW}) we have found it convenient to exploit
the ambiguity in the definition of the $D$ operator to eliminate the
terms of order $1/\epsilon$ which diverge as $\epsilon\to0^+$ and are
due to the particular form of the operator (\ref{CepsilonWZW}). These
operators then obey the standard logarithmic conformal algebra
provided that one uses the free field prescription (\ref{NSfreefield})
with {\it connected} correlation functions $\langle O\,O'\rangle_{\rm
  conn}=\langle O\,O'\rangle-\langle
  O\rangle\,\langle O'\rangle$~\cite{grav}, and makes the
  identification~(\ref{epsilonLambda}).\footnote{\baselineskip=12pt There
  are interesting similarities between the deformation introduced here and the
  logarithmic recoil operators in (\ref{susyrecoilops}). Formally, the
  $x^{10}$ field is analogous to the $\phi^2$ field here. Setting
  $\epsilon=0^+$ in the Fourier integrals of (\ref{susyrecoilops})
  yields formally the Heaviside function $\Theta(x^{10})$, which has
  vanishing momentum $q=0$ and hence yields a unit weight deformation
  in the action (\ref{SD0}). This is similar to setting $m=0$ here. As
  follows from the expressions for the logarithmic operators which
  follow, the states are related formally by
  taking the $k\to\infty$ limit of the WZW model. From
  (\ref{jkconstr}) we then have also $j\to\infty$ with $k=j^2$, and
  the logarithmic operators appear to map into each other. It would be
  interesting to understand this formal correspondence better,
  particularly in light of the analysis of the next subsection.}

Under the operator-state correspondence, the state created by the
``puncture'' operator (\ref{DepsilonWZW}) is the logarithmic partner of
the highest-weight state $|j,0\rangle$ corresponding to the primary
field~$V_{j,0}(z)$. It is straightforward to see, at least at $k=2$, that
(\ref{CepsilonWZW},\ref{DepsilonWZW}) is the logarithmic pair that
arises through appropriate fusion products of primary operators, as in
(\ref{spin12OPElogs}). In terms of the fusion rules (\ref{CGjdecomp}),
the condition (\ref{jkconstr}) restricts the spin $j'$ used in the
operator product according to
\beq
j'\left(j'+\frac12\right)=\frac k4 \ .
\label{jprimeconstr}\eeq
Let $m'\in\{-j',-j'+1,\dots,j'-1,j'\}$. Using (\ref{Sigmaqchanges}),
with $m_1'=m'$, $m_2'=-m'+\epsilon$ and the constraint
(\ref{jprimeconstr}) we may compute in the limit $\epsilon\to0^+$ the
operator product expansion
\bea
\e^{\ii m_1'\,\sqrt{\frac2k}\,\phi^2(z)}~
\e^{\ii m_2'\,\sqrt{\frac2k}\,\phi^2(w)}&=&(z-w)^{2m'^{\,2}/k}\,
\left(1-\frac{2m'\,\epsilon}k\,\ln(z-w)\right)~\e^{\ii\epsilon\,
\sqrt{\frac2k}\,\phi^2(w)}\non&&+\,\dots \ .
\label{m1m2fusion}\eeq
This illustrates how, by deforming the momentum of the $\phi^2$ field,
logarithmic singularities will be produced at leading order $\epsilon$
in four-point correlation functions of the form
\beq
F_{j',m'\,;\,l',n'}(z_1,z_2\,;\,w_1,w_2)=\Bigl\langle V_{j',m'}(z_1)\,
V_{j',-m'}(z_2)\,V_{l',n'}(w_1)\,V_{l',-n'}(w_2)\Bigr\rangle^{~}_{\rm NS} \ .
\label{F4pointfn}\eeq
Using (\ref{NSfreefield}) we write (\ref{F4pointfn}) explicitly as
\beq
F_{j',m'\,;\,l',n'}(z_1,z_2\,;\,w_1,w_2)=
\langle0|V_{j',m'}(z_1)\,V_{j',-m'}(z_2)\,V_{l',n'}(w_1)\,
\tilde V_{l',-n'}(w_2)~Q^{2j'}|0\rangle \ .
\label{F4freefield}\eeq
Note that at $k=2$, the decoupled bosonic sector of the theory has
vanishing central charge, and (\ref{QWZW}) is the nilpotent BRST
operator of the theory~\cite{Felder1}, i.e. $Q^2=0$. Then (\ref{F4freefield})
vanishes unless $j'=\frac12$, which is consistent with the condition
(\ref{jprimeconstr}). For $k>2$, $Q$ is only nilpotent when acting on
appropriate Feigin-Fuks representations.

The logarithms arising in (\ref{F4pointfn}) are accounted for by
the fusion products
\beq
V_{j',m'}(z)\,V_{j',-m'}(w)=(z-w)^{j'/(2j'+1)}\,\Bigl(D_\epsilon(w)+
\ln(z-w)\,C_\epsilon(w)\Bigr)+\dots \ ,
\label{fusionaccom}\eeq
where the $C_\epsilon$ operator arises from (\ref{m1m2fusion}) while
the $D_\epsilon$ operator arises, at least for $k=2$, from the
insertion of the screening operator (\ref{screening}). This
illustrates how the logarithmic operators constructed above appear as
descendant states of non-integer weight primary fields, in the manner
described at the beginning of this subsection. Let us stress that
while we have only really proven the appearence of (\ref{DepsilonWZW})
in this manner for $k=2$, the construction of logarithmic operators
above holds for all $k$ obeying the constraint (\ref{jkconstr}). In
the next subsection we shall give another, similar explanation for the
appearence of the logarithmic pair~(\ref{CepsilonWZW},\ref{DepsilonWZW}).

Finally, to construct the superpartners of the logarithmic pair
(\ref{CepsilonWZW},\ref{DepsilonWZW}), we apply the supercurrent
(\ref{tgcomp}) to the primary field (\ref{Vjmphi}) to get its superpartner
\beq
W_{j,m}(z)=\sqrt{\frac2k}~\e^{\ii\,\sqrt{\frac2k}\,\bigl[-j\,\phi^1(z)+m
\bigl(\phi^2(z)+\phi^3(z)\bigr)\bigr]}\otimes\left[-j\,\psi^1(z)+m
\Bigl(\ii\psi^2(z)+\psi^3(z)\Bigr)\right] \ . \non
\label{Wjmphi}\eeq
The partner of the operator (\ref{CepsilonWZW}) is then obtained by
setting $m=0$ and shifting the coefficient of the $\psi^2$ field by
$\epsilon$ in (\ref{Wjmphi}) to give
\beq
\chi^{~}_{C_\epsilon}(z)=\ii\,\sqrt{\frac2{j(j+1)}}\,
C_\epsilon(z)\otimes\Bigl(-j\,\psi^1(z)+\ii\epsilon\,\psi^2(z)\Bigr) \ .
\label{chiCepsilonWZW}\eeq
Differentiating (\ref{chiCepsilonWZW}) then gives the superpartner of
(\ref{DepsilonWZW}) as
\beq
\chi^{~}_{D_\epsilon}(z)=\sqrt{\frac{j(j+1)}2}\,\left(D_\epsilon(z)+
\frac1\epsilon\,C_\epsilon(z)\right)\otimes\psi^2(z) \ ,
\label{chiDepsilonWZW}\eeq
where as before we have removed divergent terms of order
$1/\epsilon$. One can now easily verify the NS sector logarthmic
superconformal algebra with the parameters (\ref{KMlogpars}).

\subsection{Deforming the Current Algebra\label{Deforming}}

We shall now relate the deformation above to deformations of the
energy-momentum tensor~\cite{fuchs} and hence prove invariance of the
superconformal field theory with respect to the appearence of such
operators. This will provide a better understanding of
the origin of logarithmic operators in WZW models. Recall that the
logarithmic pair is constructed by deforming the operator
(\ref{Vjmphi}) via a shift in the coefficient of the $\phi^2$ field by
an infinitesimal amount. This clearly deforms the entire $su(2)_k$ current
algebra $\cal C$ of fields of the supersymmetric WZW model. We shall now
describe the nature of this deformation and how it leads to the
appearence of Jordan cells in a more precise way.

For this, we consider the mode expansion of the weight~1 fermionic
screening operator~(\ref{screening}),
\beq
E(z)=\frac{Q}z+\sum_{n\neq0}E_n~z^{-n-1} \ ,
\label{screeningmode}\eeq
with $Q$ the screening charge (\ref{QWZW}), and the vacuum module
conditions $Q|0\rangle=E_n|0\rangle=0~~\forall n\geq1$. We then
introduce the field $\nabla_E(z)$ defined by
\beq
\partial_z\nabla_E(z)=E(z) \ ,
\label{Pizdef}\eeq
which using (\ref{screeningmode}) has the mode decomposition
\beq
\nabla_E(z)=\chi^{~}_0+Q\,\ln z-\sum_{n\neq0}\frac{E_n}n~z^{-n}
\label{nablaEmode}\eeq
with $\chi^{~}_0$ an arbitrary fermionic zero mode. The operator product
expansion of $\nabla_E(z)$ with any field $O(z)$ of the current
algebra is given by
\bea
\nabla_E(z)\,O(w)&=&\ln(z-w)\,\Bigl[Q\,,\,O(w)\Bigr]\non&&+
\,\sum_{n=1}^\infty\frac1n\,\frac1{(z-w)^n}\,\sum_{l=1}^{n+1}(-1)^{n+l}
\,{n\choose{l-1}}\,w^{n-l+1}\,\Bigl[E_{l-1}\,,\,O(w)\Bigr]+\dots \ , \non&&
\label{PiOPE}\eea
which along with (\ref{normalordering}) can be used to define a
deformation $O^E(z)$ of the field $O(z)$ by
\beq
O^E(z)=O(z)+\chi^\dag_0\,\nabla_E(z)\,O(z) \ ,
\label{deformfield}\eeq
where $\chi^\dag_0$ is a Grassmann-valued mode which is conjugate
to $\chi^{~}_0$,
\beq
\left\{\chi^{~}_0\,,\,\chi^\dag_0\right\}=1 \ .
\label{chitildexirel}\eeq
The operation $O\mapsto O^E$ defines a graded outer derivation of the
corresponding operator product algebra~\cite{fuchs}.

Since $Q$ is a screening operator for the current algebra, it serves
as a BRST charge for the physical state space and $[Q,O(z)]=0$. Thus the
logarithms naturally disappear from (\ref{PiOPE}) and
(\ref{deformfield}). Since $E(z)$ is a weight~1 primary field, it satisfies
\beq
\sum_{l=1}^n(-1)^{n+l}\,{{n-1}\choose{l-1}}\,z^{n-l}\,
\Bigl[E_{l-1}\,,\,T(z)\Bigr]=\delta_{n,2}\,E(z) \ , ~~ n>0 \ ,
\label{weight1eq}\eeq
and hence the corresponding deformation of the energy-momentum tensor
is given by
\beq
T^E(z)=T(z)+\frac{\chi^\dag_0}z\,E(z) \ .
\label{TEzdef}\eeq
This ensures that $T^E(z)$ has the correct operator product to be an
energy-momentum tensor, i.e. to generate a Virasoro algebra.

Since logarithms do not appear in the transformed fields
(\ref{deformfield}), the algebra of the deformed operators $O^E(z)$ is
{\it isomorphic} to the original current algebra $\cal C$. In complete
analogy with the relationship between the fermionic ghost system and
the symplectic fermion in section~\ref{SymplFerm}, the WZW model is
identified with a subsector of a larger conformal field theory whose
chiral algebra ${\cal C}_E$ of fields contains the operator
(\ref{nablaEmode}), so that the operation $O\mapsto O^E$ defines an
{\it inner} derivation of ${\cal C}_E$. This extension of $\cal C$ may
be realized explicitly by noting that the fermionic operators
$\chi^{~}_0$ and $\chi^\dag_0$, obeying the Clifford algebra
(\ref{chitildexirel}), force the vacuum state of the corresponding
extended physical state space ${\cal H}_E$ to be two-fold
degenerate. The two ground states are the $SL(2,\complex)$ invariant
vacuum $|0\rangle$, with $\chi^\dag_0|0\rangle=0$, and its conjugate
$Q^\dag\,\chi^{~}_0|0\rangle$, where the operator $Q^\dag$ is
conjugate to the fermionic screening charge,
i.e. $\{Q,Q^\dag\}=1$. From (\ref{chitildexirel}) and (\ref{TEzdef})
it then follows that the field $\nabla_E(z)$ corresponds to the state
$-\chi^{~}_0|0\rangle$. Furthermore, if $\cal H$ denotes the physical
state space of the original WZW model, then the new space of states
may be identified as ${\cal H}_E={\cal H}\oplus{\cal H}$, with
operators in ${\cal C}_E$ generically acting between the two copies of
$\cal H$. As in section~\ref{SymplFerm}, these definitions immediately
lead to logarithmic singularities in correlation functions of
operators in ${\cal C}_E$.

Since $E(z)$ is a {\it fermionic} screening operator, the charge $Q$
can serve as a differential in a complex. For $k=2$, $Q$ is the
nilpotent BRST operator, and the space of states $\cal H$ can be
defined as the cohomology of the screening operator acting in the
larger space ${\cal H}_E$~\cite{Felder1}. For $k>2$ we need to extend
${\cal H}_E$ further in order to guarantee nilpotency of $Q$. In any
case, working with operators over the appropriate extended state
space, from the above constructions it follows explicitly that
\bea
&&T^E(z)\,V_{j,m}^E(w)\non&&~~~~~~=~\sum_{n=1}^\infty\frac1{(z-w)^{n+2}}\,
\sum_{l=1}^{n+1}(-1)^{n+l+1}\,{n\choose{l-1}}\,w^{n-l+1}\,
\Bigl[E_{l-1}\,,\,V_{j,m}^E(w)\Bigr]\non&&~~~~~~~~~~+\,
\frac{\Delta_{j}}{(z-w)^2}\,
V_{j,m}^E(w)+\frac1{(z-w)^2}\,\Bigl[Q^E\,,\,
V_{j,m}^E(w)\Bigr]+\frac1{z-w}\,\partial_wV_{j,m}^E(w)+\dots \ , \non&&
\label{TEVjmE}\\&&T^E(z)\,\Bigl[Q^E\,,\,V_{j,m}^E(w)\Bigr]
\non&&~~~~~~=~\frac{\Delta_{j}}{(z-w)^2}\,\Bigl[Q^E
\,,\,V_{j,m}^E(w)\Bigr]+\frac1{z-w}\,\Bigl[Q^E\,,\,
\partial_wV_{j,m}^E(w)\Bigr]+\dots \ .
\label{TEQWZWE}\eea
Thus the Jordan cells, of energy $\Delta_{j}$, are spanned by
  $D_{j,m}(z)=V_{j,m}^E(z)$ and $C_{j,m}(z)=[Q^E,V_{j,m}^E(z)]\neq0$
  (note that $Q^E$ is {\it not} a
  screening operator in the deformed chiral symmetry
  algebra).

To understand the appearence of rank~2 Jordan blocks in this context
more clearly, let us consider the very special case of level $k=2$. In the
associated Virasoro representation provided by the Sugarawa
construction, the decoupled bosonic sector of the Kac-Moody algebra can then be
naturally embedded inside a topologically twisted $N=2$ superconformal
algebra~\cite{twist} such that the screening operator $E(z)$ becomes one of the
fermionic supercurrents. To see this, we note that the $N=2$ algebra
of central charge $c=1$ has three unitary irreducible representations
${\cal H}_\ell$, $\ell=0,1,2$, each of which decomposes as
\beq
{\cal H}_\ell=\bigoplus_{n=-\infty}^\infty{\cal F}_\ell^n
\label{calHlcalFln}\eeq
into Feigin-Fuks (free field Fock space) representations ${\cal
  F}_\ell^n$ of the $c=0$ Virasoro algebra~\cite{FF1}. This is precisely the
  central charge of the decoupled bosonic WZW model at $k=2$. The
  $N=2$ algebra contains a $U(1)$ current subalgebra, under which the
  Fock states of the module ${\cal F}_\ell^n$ all carry $U(1)$ charge
  $n+\ell/3$. The vacuum state $|0\rangle^{~}_\ell$ of ${\cal H}_\ell$
  satisfies the highest-weight conditions
  $E_n|0\rangle^{~}_\ell=0~~\forall n\geq\ell$ and
  $Q|0\rangle^{~}_0=0$. The screening operator $Q$, of $U(1)$ charge $-1$, then
  provides a proper BRST differential making each ${\cal H}_\ell$ into
  an entire Felder complex of Virasoro modules over $\cal C$,
\beq
\dots~\stackrel{Q}{\longrightarrow}~{\cal F}_\ell^n~
\stackrel{Q}{\longrightarrow}~{\cal F}_\ell^{n-1}~
\stackrel{Q}{\longrightarrow}~\dots \ , ~~ Q^2=0 \ .
\label{QFelder}\eeq
The corresponding irreducible Virasoro representations, in the Fock
spaces of the free scalar field which bosonizes the $U(1)$ current of
the $N=2$ algebra, are then furnished through the BRST cohomology
$\ker^{~}_{{\cal F}_\ell^n}Q/{\rm im}^{~}_{{\cal
    F}_\ell^{n+1}}Q$~\cite{Felder1}.

Using this embedding, we now define the extension ${\cal C}_E$ of the
chiral algebra $\cal C$ to be the extended $N=2$ vertex operator
algebra, constructed as explained above. Corresponding to ${\cal C}_E$
there are then three representations ${\cal L}_\ell$, $\ell=0,1,2$.
Because of (\ref{chitildexirel}), each ${\cal L}_\ell$ is an
indecomposable extension of ${\cal H}_\ell$ by itself, i.e. there is a
short exact sequence
\beq
0~\longrightarrow~{\cal H}_\ell~\longrightarrow~{\cal L}_\ell~
\longrightarrow~{\cal H}_\ell~\longrightarrow~0 \ .
\label{calLlexact}\eeq
As Virasoro representations, the decomposition
(\ref{calHlcalFln},\ref{QFelder}), along with (\ref{TEQWZWE}), implies
that (\ref{calLlexact}) is a direct sum of sequences
\beq
0~\longrightarrow~{\cal F}^n_\ell~\longrightarrow~{\cal L}^n_\ell~
\longrightarrow~{\cal F}^{n+1}_\ell~\longrightarrow~0 \ , ~~ n\in\zed \ ,
\label{calLlnexact}\eeq
where ${\cal L}_\ell=\bigoplus_{n\in\zeds}{\cal L}_\ell^n$~\cite{fuchs}. The
non-diagonalizable Virasoro modules ${\cal L}_\ell^n$ carry definite
$U(1)$ charge $n\in\zed$, and they are generated by
the original Kac-Moody currents of the $SU(2)_2$ WZW model along with the
extra zero modes $\chi^{~}_0$, $\chi^\dag_0$. In terms of the
operator product expansions (\ref{TEVjmE},\ref{TEQWZWE}), the $C$
operators generate the sub-modules ${\cal F}_\ell^n\subset{\cal
  L}_\ell^n$, while the descendants of the field $\nabla_E(z)$, by which
$\cal C$ is extended, live in the quotient space ${\cal L}_\ell^n/{\cal
  F}_\ell^n\cong{\cal F}_\ell^{n+1}$. This illustrates explicitly how the
Hamiltonian operator
\beq
L_0^E=L_0+\chi^\dag_0\,Q
\label{L0Eexpl}\eeq
for $k=2$ acquires rank~2 Jordan blocks. For $k>2$, we have not been
able to find such indecomposable representations of the current
algebra explicitly.

Finally, we relate the present Jordan block structure to the
logarithmic pair (\ref{CepsilonWZW},\ref{DepsilonWZW}). For this, we
write the fermionic screening operator (\ref{screening}) as
\beq
E(z)=\left(\partial_z\phi^2(z)+\sqrt{\frac2k}\,\partial_z\phi^4(z)
\right)~\e^{\ii\,\sqrt{\frac2k}\,
\phi^1(z)} \ .
\label{screeningphi}\eeq
The deformation operator satisfying (\ref{Pizdef}) is then given
explicitly in the Coulomb gas representation by
\bea
\nabla_E(z)&=&\chi^{~}_0+\left(\phi^2(z)+\sqrt{\frac2k}\,\phi^4(z)
\right)~\e^{\ii\,\sqrt{\frac2k}\,\phi^1(z)}\non&&-\,
\ii\,\sqrt{\frac2k}\,\int^z\dd w~\left(\phi^2(w)+\sqrt{\frac2k}\,
\phi^4(w)\right)\,\partial_w\phi^1(w)~
\e^{\ii\,\sqrt{\frac2k}\,\phi^1(w)} \ .
\label{PiECoulomb}\eea
As usual, to neutralize the background charge of the $\phi^1$ field, we
introduce the vertex operator dual to (\ref{Vjmphi}) at $m=0$,
\beq
\tilde V_{j,0}(z)=\e^{\ii\,\sqrt{\frac2k}\,(j+1)\,\phi^1(z)} \ .
\label{tildeVj0}\eeq
Using (\ref{PiECoulomb}) and the adjoint $\tilde V_{j,0}^\dag(z)$ of
(\ref{tildeVj0}), we now define the field
\beq
V_{j,\epsilon}^E(z)=\nabla_E(z)\cdot\Bigl(z^{2\Delta_{j+1}}
\,\tilde V_{j,0}(z^{-1})\Bigr) \ ,
\label{VjepsilonE}\eeq
with $\epsilon$ defined by (\ref{epsilonLambda}). Among other terms,
the field
$\phi^2(z)~\e^{\ii\,\sqrt{\frac2k}\,\bigl(-j\,\phi^1(z)+\epsilon\,
\phi^2(z)\bigr)}$ is present in the operator product
(\ref{VjepsilonE}), {\it provided} that the constraint
(\ref{jkconstr}) is satisfied. Up to normalization, this is just the
logarithmic operator (\ref{DepsilonWZW}). Similarly, it is
straightforward to check that $[Q^E,V_{j,\epsilon}^E(z)]$, with the
constraint (\ref{jkconstr}), contains the $C$ operator
(\ref{CepsilonWZW}).

The other operators arising in (\ref{TEVjmE}) and (\ref{VjepsilonE})
may be attributed to higher-order terms in the deformation parameter
$\epsilon\to0^+$, as they are absent in the logarithmic conformal
algebra of the operators (\ref{CepsilonWZW},\ref{DepsilonWZW}) to
leading order in $\epsilon$. While the Jordan blocks
(\ref{TEVjmE},\ref{TEQWZWE}) arise for generic values of $j$, $m$ and
$k$, it is only with the constraint (\ref{jkconstr}) that the
deformation leaves an isomorphic superconformal algebra and thus
naturally explains the appearence of logarithmic operators in fusion
relations within the original supersymmetric WZW model. In particular,
it is only for $k=2$ that we have been able to explicitly demonstrate
both the appearence of the indecomposable representations and of the
logarithmic operators through fusion relations. The analogy between
the logarithmic pair (\ref{CepsilonWZW},\ref{DepsilonWZW}) and that of
the recoil problem of section~\ref{RecoilProb} is very suggestive of
the fact that logarithmic operators are generally hidden within
(subtle) marginal deformations of the original conformal
algebra~\cite{fuchs}. This feature is further supported by the
similarity between (\ref{CepsilonWZW},\ref{DepsilonWZW}) and the
logarithmic operators which arise in gravitationally dressed conformal field
theories~\cite{liouv,lcftgeneral}.

\subsection{Spin Fields\label{SpinWZW}}

The Coulomb gas representation, and the ensuing decoupling of
fermions, allows the appropriate definition of the spin operator for the
Ramond sector of the model. We seek an operator that introduces cuts
in the free fermion fields $\psi^a(z)$, $a=1,2,3$. For the fields
$\psi^1(z)$ and $\psi^2(z)$ this is straightforward to construct using
the technique of the previous sections. Namely, from the bosonization
formula (\ref{psipmphi4}) it follows immediately that the $\psi^1$ and
$\psi^2$ fields are twisted by the standard spin operator
$\sqrt2\,\cos\frac{\phi^4(z)}2$ of dimension $\frac18$.

For the last unpaired Majorana fermion field $\psi^3(z)$, however, we
cannot use bosonization and must treat it as a
parafermion field, i.e. we must instead appeal to the standard,
non-local oscillator construction of its spin
field~\cite{Corrigan1}. For this, we introduce the positive and
negative frequency mode expansions of the operator
$\psi^3(z)=\psi_{(\lambda)}^3(z)_++\psi_{(\lambda)}^3(z)_-$ in the
Neveu-Schwarz ($\lambda=0$) and Ramond ($\lambda=1$) sectors by
\bea
\psi_{(\lambda)}^3(z)_+&=&\frac1{\sqrt2}\,\sum_{n<0}
\psi_{n+(1-\lambda)/2}^3~z^{-n-1+\lambda/2} \ , \non
\psi_{(\lambda)}^3(z)_-&=&\frac1{\sqrt2}\,\sum_{n\geq0}
\psi_{n+(1-\lambda)/2}^3~z^{-n-1+\lambda/2} \ ,
\label{psia3pm}\eea
with $(\psi_r^3)^\dag=\psi_{-r}^3$ and
\beq
\left\{\psi_r^3\,,\,\psi_s^3\right\}=\delta_{r+s,0} \ .
\label{psirsanticomm}\eeq
The ground states of the NS and R sectors obey the highest-weight
conditions
\beq
\psi^3_{n+1/2}|0\rangle=\psi^3_{n+1}|\Delta\rangle^{~}_{\rm R}=0~~~~
\forall n\geq0
\label{psi3vac}\eeq
with $\Delta=\hat c/16$. The $\zed_2$ twist field for $\psi^3(z)$ is
then given by
\beq
\Pi(z)=\e^{z\,L_{-1}^{\rm R}}\,\langle0|\e^{H(z)}~\e^{I(z)}|\Delta
\rangle_{\rm R}^{~} \ ,
\label{twist3def}\eeq
where
\beq
L_{-1}^{\rm R}=-\frac18~\sum_{n=-\infty}^\infty(2n+1)\,\psi_n^3\,
\psi^3_{-n-1}
\label{Virtranslpsi3}\eeq
is the Virasoro translation generator acting in the Ramond sector of
the Fock space defined by (\ref{psia3pm})--(\ref{psi3vac}). The field
\beq
H(z)=\oint\limits_{w=z}\frac{\dd w}{2\pi\ii}~\oint\limits_{w'=w}
\frac{\dd w'}{2\pi\ii}~\frac1{w-w'}\,\left(1-\sqrt{\frac{z-w}{z-w'}}~
\right)\,\psi_{(0)}^3(w)\,\psi^3_{(0)}(w')
\label{Hzpsi3NS}\eeq
is quadratic in the NS oscillators of $\psi^3(z)$, while the field
\beq
I(z)=\oint\limits_{w=z}\frac{\dd
  w}{2\pi\ii}~\frac{w-z+1}{\sqrt{w-z}}\,\psi^3_{(0)}(w-z)_+\,
\psi^3_{(1)}(w)_-
\label{Izpsi3NSR}\eeq
is bilinear in the NS and R oscillators.

The coefficients of the oscillator expansions of (\ref{Hzpsi3NS}) and
(\ref{Izpsi3NSR}) are complicated functions of
$z$~\cite{Corrigan1}. In this representation, zero modes do not appear
(in contrast to the bosonized representations), and the different vacua
are described instead by the different states $|0\rangle$ and
$|\frac{\hat c}{16}\rangle^{~}_{\rm R}$. Because the fermionic
currents in this case coincide with the fermion fields themselves (see
(\ref{fercurrent})), the twisting of the current algebra by the spin
field $\Pi(z)$ is determined by the R sector operator product expansion
\beq
\psi^3(z)\,\Pi(w)=\frac1{\sqrt{z-w}}\,\pi(w)+\dots \ ,
\label{psi3PiOPE}\eeq
where $\pi(z)$ is the corresponding excited twist field. The operator
(\ref{twist3def}) has scaling dimension $\frac1{16}$~\cite{Corrigan1}.

The desired total spin field for the set of free fermion fields
$\psi^a(z)$, $a=1,2,3$ is therefore given by
\beq
\Sigma(z)=\sqrt2\,\cos\frac{\phi^4(z)}2\otimes\Pi(z) \ ,
\label{Sigmaztotalpsia}\eeq
and it has dimension $\frac3{16}$. Finally, the excited logarithmic
spin fields defined by (\ref{chiCDSigma}) may be computed explicitly
by using (\ref{psipmphi4}), (\ref{chiCepsilonWZW}),
(\ref{chiDepsilonWZW}) and (\ref{Sigmaztotalpsia}) to get
\bea
\widetilde{\Sigma}_{C_\epsilon}(z)&=&2\ii\,\sqrt{\frac1{j(j+1)}}\,
C_\epsilon(z)\otimes\left(-\sqrt2\,j\,\sin\frac{\phi^4(z)}2
\otimes\Pi(z)+\ii\epsilon\,\Sigma(z)\right) \ , \non
\widetilde{\Sigma}_{D_\epsilon}(z)&=&\sqrt{j(j+1)}\,
\left(D_\epsilon(z)+\frac1\epsilon\,C_\epsilon(z)\right)\otimes
\Sigma(z)
\label{SigmaCDexcited}\eeq
in the limit $\epsilon\to0^+$, and they have dimension
$\frac{19}{16}$. With these explicit representations, it is now
straightforward to verify explicitly the relations of the R sector
logarithmic superconformal algebra derived earlier. Let us note that
at level $k=2$, whereby the supersymmetric $su(2)_2$ current algebra
is generated by a single scalar superfield, we may take $\phi^4(z)=0$
in (\ref{Sigmaztotalpsia}). Then the spin field $\Sigma(z)$ has
dimension $\frac1{16}$, and hence it creates the supersymmetric R
sector ground state $|\frac{\hat c}{16}\rangle_{\rm R}^{~}$ at
$k=2$. On the other hand, for $k>2$ there are three scalar superfield
generators and $\Sigma(z)$ has dimension $\frac3{16}$. Therefore, in
the WZW models with $k>2$, one has $\Sigma_{3/16}^-(z)\neq0$ and the
Ramond supersymmetry is {\it broken}. As indicated earlier, features
such as this single out the value $k=2$ as a very special point, and
indeed it provides the lowest-lying marginal logarithmic deformation
of section~\ref{LogConstruct}. We will see another very special
feature of this level in the next subsection.

\subsection{Extra $c=-2$ Sectors\label{Underlying}}

It has been suggested~\cite{nichols,lcftgeneral} that the occurence of
logarithmic behaviour in the bosonic $SU(2)_0$ WZW model is due to the
presence of an underlying $c=-2$ sector. We will now explore this
possibility within the present context. Let us concentrate on the
field $\phi^1(z)$ in the free field representation. From (\ref{tcomp})
its energy-momentum tensor is
\beq
T_{\phi^1}(z)=-\frac12\,\partial_z\phi^1(z)\,\partial_z\phi^1(z)+
\frac\ii{\sqrt{2k}}\,\partial_z^2\phi^1(z) \ ,
\label{Tphi1z}\eeq
and hence it has central charge
\beq
c_{\phi^1}=1-\frac6k \ .
\label{cphi1}\eeq
It follows that at $k=2$, the Coulomb gas field $\phi^1(z)$ represents
a bosonized symplectic fermion system with $c=c_{\phi^1}=-2$. We shall
examine how this $\phi^1$ sector of the field theory is related to the
logarithmic structures that we have unveiled above. As we will see,
the extra $c=-2$ sector which arises is responsible for the
isomorphism between the deformed algebra of fields and the original
superconformal algebra.

Explicitly, the bosonization formulas at $k=2$ are given by
\bea
\eta(z)&=&\frac1{\sqrt2}\,V_{1,0}(z)~=~
\frac1{\sqrt2}~\e^{-\ii\phi^1(z)} \ , \non\xi(z)&=&
\frac1{\sqrt2}\,\tilde V_{0,0}(z)~=~\frac1{\sqrt2}~\e^{\ii\phi^1(z)} \ ,
\non\eta(z)\,\xi(z)&=&-\frac\ii{2}\,\partial_z\phi^1(z) \ .
\label{etaxiphi1bos}\eea
The fermionic $c=-2$ ghost system
$(\eta,\xi)$ has weight $(1,0)$ and non-vanishing two-point
functions~(\ref{etaxi2ptfn}). Note that its bosonization is different
from that of section~\ref{SymplSpin}, because now all modes of the
field $\xi(z)$ will contribute. The energy-momentum tensor
(\ref{Tphi1z}) for the $\phi^1$ sector can then be written as
\beq
T_{\phi^1}(z)=-\eta(z)\,\partial_z\xi(z) \ ,
\label{Tphi1etaxi}\eeq
and the logarithmic operators of section~\ref{LogConstruct} may be
expressed as
\bea
C_\epsilon(z)&=&-\sqrt2\,\ii\epsilon\,\eta(z)\otimes\e^{\ii\epsilon\,\phi^2(z)}
\ , \non D_\epsilon(z)&=&-\sqrt2\,\eta(z)\otimes\phi^2(z)~\e^{\ii\epsilon\,
\phi^2(z)} \ .
\label{CDepsiloneta}\eea
{}From (\ref{CDepsiloneta}) it is clear that at $k=2$ the logarithmic
structure is {\it not} due to the presence of this underlying $c=-2$
sector in the theory, because the logarithmic behaviour of
(\ref{CDepsiloneta}) arises from the $\phi^2$ sector instead. Indeed,
in the setting of section~\ref{LogConstruct} the field $\phi^1(z)$
acts as a Liouville dressing for the operators constructed from the
field $\phi^2(z)$. Thus the $\phi^1$ sector in (\ref{CDepsiloneta})
ensures that the deformations are marginal and hence that the deformed
algebra is conformal. This is even more apparent upon examination of
the screening operator for the WZW model, which in this case yields
directly the purely fermionic BRST operator $Q$ that can be written in
terms of the $c=-2$ ghost fields as
\beq
E(z)=\xi(z)\otimes\Bigl(\partial_z\phi^2(z)+\partial_z\phi^4(z)\Bigr) \ .
\label{screeningxi}\eeq
Deforming the current algebra in the manner described in
section~\ref{Deforming} using the spin~0 ghost field in
(\ref{screeningxi}) will not produce the logarithmic deformation
(\ref{CDepsiloneta}) that is determined instead by the spin~1 ghost field.

The main reason for this discrepency here lies in the fact that the field
$\xi(z)$ is {\it not} a screening operator in the $c=-2$ model. Let us
introduce the mode expansions of the fermionic ghost fields,
\bea
\eta(z)&=&\eta^{~}_0-\sum_{n\neq0}\eta_n~z^{-n-1} \ , \non
\xi(z)&=&\xi^{~}_0+\sum_{n\neq0}\frac{\xi_n}n~z^{-n} \ ,
\label{etaximode}\eea
with the non-vanishing anticommutation relations
\bea
\{\eta_n,\xi_m\}&=&n\,\delta_{n+m,0} \ , ~~ n\neq0 \ , \non
\{\eta^{~}_0,\xi^{~}_0\}&=&1 \ ,
\label{etanxim}\eea
and the vacuum state annihilation conditions
$\eta_n|0\rangle=\xi_{n+1}|0\rangle=0~~\forall n\geq0$. From the
bosonization formulas (\ref{etaxiphi1bos}) it follows that a fermionic
screening operator for the $c=-2$ model is given by the zero mode of
the $\eta$ field,
\beq
\eta^{~}_0=\oint\limits_{z=0}\frac{\dd z}{2\pi\ii}~\eta(z) \ .
\label{eta0screening}\eeq
The presence of the extra screening operator $\eta(z)$ in fact opens
the door for the construction of yet {\it another} logarithmic
deformation ${\cal C}_\eta$ of the current algebra $\cal C$, with the
same properties as that constructed in section~\ref{Deforming} and in
parallel with the symplectic fermion model of
section~\ref{SymplFerm}~\cite{fuchs}.

For this, we note that the deformation field of ${\cal C}_\eta$ is the
operator
\beq
\nabla_\eta(z)=\chi^{~}_0+\eta^{~}_0\,\ln z+\sum_{n\neq0}\frac{\eta_n}n~
z^{-n} \ ,
\label{nablaetaz}\eeq
and the corresponding deformation of the $\xi$ field is given by
\beq
\xi^\eta(z)=\xi^{~}_0+\chi^\dag_0\,\ln z+\sum_{n\neq0}
\frac{\xi_n}n~z^{-n} \ .
\label{xietaz}\eeq
The extra zero mode $\chi^\dag_0$ generated for the field $\xi^\eta(z)$,
along with the anticommutation relations (\ref{chitildexirel}) and
(\ref{etanxim}), leads to the desired logarithmic correlation
\beq
\nabla_\eta(z)\,\xi^\eta(w)=\ln(z-w)+\dots \ .
\label{nablaetaxietaOPE}\eeq
Moreover, the deformed energy-momentum tensor is given by
\beq
T_{\phi^1}^\eta(z)=-\eta(z)\,\partial_z\xi(z)+\frac{\chi^\dag_0}z\,
\eta(z)=-\eta(z)\,\partial_z\xi^\eta(z) \ ,
\label{Tphi1etaz}\eeq
and its form invariance in terms of the transformed fields is a direct
consequence of the fact~\cite{fuchs} that the operator products are
preserved by the deformation. These are of course just the standard
relations of the symplectic fermion field theory that we studied in
section~\ref{SymplFerm}, with the identifications
$(\nabla_\eta\,,\,\xi^\eta)=(-\chi^-\,,\,\chi^+)$.

The indecomposable module extensions in ${\cal C}_\eta$ are determined
by the Felder complex $\eta^{~}_0:{\cal F}^n\to{\cal F}^{n+1}$ with Feigin-Fuks
representations ${\cal F}^n$ consisting of Fock space states of fixed
$(\eta,\xi)$ ghost number $n\in\zed$~\cite{fuchs}. In the mapping from
the $c=0$ model to the $N=2$ superconformal field theory that we discussed in
section~\ref{Deforming}, this is the same $U(1)$ current that is used
to grade the Felder complex (\ref{QFelder})~\cite{Gerasimov}. However,
we stress again that the screenings in the two sectors are completely
different, and hence so are the logarithmic deformations. For the symplectic
fermion, the logarithmic pair in the $\phi^1$ sector for $n=0$ is given by
$(C\,,\,D)=(I\,,\,\nabla_\eta\,\xi^\eta)$, and it generates a Jordan cell
spanned by the states $(|0\rangle\,,\,\xi^{~}_0\,\chi_0^{~}|0\rangle)$. Note
that this Jordan block has spin~0, and so the logarithmic deformation
does not preserve the conformal algebra. This is in contrast to the
spin~1 blocks generated by (\ref{CDepsiloneta}) that arise from
primary field fusion relations and have $U(1)$ ghost charge $n=1$.

It follows from this analysis that there is another, independent
logarithmic sector of the supersymmetric $SU(2)_2$ WZW
model. It is not the same as that which comes about from logarithmic
singularities in the WZW conformal blocks. The full current algebra
has much more symmetry and a somewhat richer logarithmic
structure. Nevertheless, several striking similarities between the
triplet model and the WZW model seem to arise due to this underlying ghost
system~\cite{nichols,lcftgeneral}. Of course, this analysis breaks
down at higher levels $k>2$. On the other hand, for generic values of
the level $k\in\zed_+$, there exist alternative (but equivalent) free
field representations of the Kac-Moody currents
which make manifest an underlying $c=-2$ structure. The bosonic currents can
be represented by either the three scalar fields $\varphi^a(z)$
studied above, or they can be ``debosonized'' into ghost
fields~$(\beta,\gamma)$~\cite{fuchs,nichols,FMS1}. These ghost fields
constitute a first order {\it bosonic} system and a current such that
the $SU(2)_{k-2}$ currents become those of the Wakimoto
construction~\cite{wakimoto}. The ghost system $(\beta,\gamma)$ may
then be bosonized in terms of a symplectic fermion system
$(\eta,\xi)$, as in section~\ref{SymplSpin}, which can be used to
construct a logarithmic conformal algebra by deforming the fields as
above. In this way one can obtain logarithmic deformations, at least
in a certain sector of the quantum field theory, for generic
values of $k\in\zed_+$ in supersymmetric $SU(2)_k$ WZW
models. However, it should be stressed again that such deformations
only lead to isomorphic superconformal algebras when the
highest-weight constraint (\ref{jkconstr}) is obeyed.

\subsection{Supersymmetric $sl(2,\realbm)_k$ Current Algebras~\label{sl2RAlgs}}

For completeness we shall now briefly describe the situation in the
case of non-compact current algebras, which can be treated in a similar manner.
Formally, one obtains the non-compact case from the compact one essentially
by flipping the sign $k\mapsto-k$ of the level number $k\in\zed_+$, as
a sort of analytic continuation of the field theory~\cite{witten}.
In supersymmetric $sl(2,\real)_k$ WZW models one can again demonstrate
the decoupling of the supersymmetric fermionic partners, and
hence an equivalence of the bosonic sector at level $k-2$.
It therefore suffices for our purposes again to concentrate
on bosonic $sl(2,\real)_{k-2}$ affine algebras.

The bosonic currents in this case, for an arbitrary level $k\in\zed_+$,
can be represented by means of free scalar fields.
A specific representation may be given in which the two currents
$J^+(z)$ and $J^0(z)$ are represented in terms of two scalar fields
$\phi(z)$ and $\varphi(z)$, while the third current $J^-(z)$
depends explicitly on the energy-momentum tensor $T(z)$~\cite{fuchs}.
The field $T(z)$ can itself be bosonised by means of a scalar current
$\partial_zu(z)$, and the entire system debosonized to ghost fields
$(\beta,\gamma)$ plus a current in such a way that the Wakimoto
representation~\cite{wakimoto} for the currents arises. The
logarithmic deformations then appear through a fermionic screening charge
as before~\cite{fuchs}, and are given by the zero mode $\eta^{~}_0$
of the $c=-2$ system $(\eta,\xi)$ that is used in the bosonization of the ghost
system $(\beta,\gamma)$. The crucial difference from the previous analysis is
that now the operator $\eta^{~}_0$ {\it coincides} exactly with the screening
operator of the $sl(2,\real)_{k-2}$ current algebra for any
$k\in\zed_+$. Thus the logarithmic behaviour in this class of WZW
models is indeed due entirely to symplectic fermions and occurs for
generic values of the level coefficient. In particular, the bosonic part of the
current algebra has central charge $c=-2$ at $k=2$, and hence
logarithmic behaviour arises directly at this special level number due
to the presence of an underlying $c=-2$ sector of the quantum field theory.

For completeness, in order to compare with the corresponding
expression (\ref{screening}) for the compact current algebras, we note
that a fermionic screening operator in the present case is given by
\begin{equation}
Q'=\oint\limits_{z=0}\frac{\dd z}{2\pi\ii}~\e^{\sqrt{\frac{k-2}{2}}\,
\phi(z)}\,{\cal V}_{1,2}(z) \ ,
\label{Qprime}\end{equation}
where ${\cal V}_{r,s}(z)$ are Virasoro primary fields of conformal
dimension
\beq
\Delta_{r,s} = \frac{r^2 - 1}{4k}+\frac{s^2 - 1}{4}\,k+
\frac{1 - rs}{2} \ .
\label{Deltars}\eeq
Notice that the exponential in (\ref{Qprime}) is real-valued and
amounts to the analytic continuation $k-2\mapsto-(k-2)$ in the passage
from the compact $su(2)$ current algebra to the non-compact
$sl(2,\real)$ case. The $SL(2,\real)$ primary fields are given
by~\cite{fuchs}
\begin{equation}
V_{r,s}(z)=\e^{j(r,s)\,\sqrt{\frac{k-2}{2}}\,\bigl(\varphi(z)-
\phi(z)\bigr)}\,{\cal V}_{r,s}(z) \ ,
\end{equation}
where
\beq
j(r,s)=\frac{r-1}{2} - \frac{s-1}{2}\,k \ .
\label{jrsdef}\eeq
Using these primary states one can now proceed in an analogous way to
the compact case above and demonstrate explicitly the appearance of
logarithmic deformations for generic values of the level $k$. The
indecomposable extensions of the corresponding highest-weight
representations $[V_{r,s}]$ are constructed explicitly
in~\cite{fuchs}, as are the corresponding states which generate the Jordan
cells. The appearence of logarithmic operators through fusion
relations of spin $j=-\frac12$ primary fields is described
in~\cite{giribet}. These models are essentially equivalent to
Liouville theory, which at $c=1$ is known to contain logarithmic
operators~\cite{liouv}. As indicated earlier, the logarithmic
operators in these theories at generic $k$~\cite{lcftgeneral,giribet} are
strikingly similar to the ones we have found in
section~\ref{LogConstruct} above. These non-compact current algebras
also describe string propagation in an $AdS_3$ background~\cite{Mald1}.

\subsection{Supersymmetric Coset Models}

Finally, let us mention some brief points about coset constructions,
although we will not deal with these models explicitly here. The
analysis of such models is complicated by the fact that the decoupling
of fermionic partners is not complete. The equivalence of coset models
over $G/H$ with a gauged $N=1$ supersymmetric WZW model, in which a
diagonal subgroup $H\subset G$ is gauged, has been analysed
in~\cite{figueroa}. There it is demonstrated, by means of
appropriate fermionic currents as above, that the fermion fields
in the adjoint representation of the subgroup $H\subset G$ decouple.
One is then left with a theory of coset fermion fields living in $G/H$ which
are {\it gauged}, and also an ordinary WZW model in which
the diagonal $H$ symmetry is gauged. In the limit $H=G$ there
are no coset fermion fields and the gauged supersymmetric WZW model
reduces to an ordinary WZW model in the sense of complete decoupling
of fermion fields, as we have seen above. The detailed analysis of
such models represents an interesting arena in which to search for
logarithmic operators. They are also relevant to the standard coset
constructions of $N=2$ superconformal field theories. They might
therefore lead to logarithmic $N=2$ superconformal algebras and, in
the string context, there may even be the possibility of breaking
$N=2$ worldsheet supersymmetry, in an analogous way that $N=1$
supersymmetry was broken in the $k>2$ WZW models (see
section~\ref{SpinWZW}).

These theories, combined with those of the previous subsection, also
represent interesting physical situations where logarithmic behaviour
arises. For example, logarithmic operators may arise in
two-dimensional black holes which can be described as exact gauged WZW
conformal field theories over the coset $SL(2,\real)/U(1)$~\cite{witten}. The
logarithmic behaviour in this case may be attributed to an exactly
marginal deformation of the black hole background connected with the
$W_\infty$-symmetry of the target space~\cite{km}. Again, the logarithmic
operators which arise in this case~\cite{lcftgeneral} are very similar
to those of section~\ref{LogConstruct}. More recently, logarithmic
behaviour has been discovered in the coset model $SU(2)_k/U(1)\times
U(1)_{-k}$~\cite{Sfetsos1} (and also in more general cosets
$G_k/H\times U(1)_{-k}$~\cite{Sfetsos2}), which admits a spacetime
interpretation as an exact three-dimensional plane wave solution of
supergravity in a correlated Penrose limit which involves taking
$k\to\infty$. The main ingredients of the construction of logarithmic
operators in this case, namely the usage of a small deformation
parameter $\epsilon=1/k$ and a free time-like boson field, are exactly
the same as those used in section~\ref{LogConstruct}.

\subsection*{Acknowledgments}

R.J.S. thanks J.-S.~Caux and J.~Wheater for helpful discussions. A
preliminary version of this paper was presented by N.E.M. at the Workshop
``Non-Unitary/Logarithmic CFT'' which was held at the Institute des
Hautes Etudes Scientifiques, Paris, France, June 10--14 2002. The
work of R.J.S. was supported in part by an Advanced Fellowship from
the Particle Physics and Astronomy Research Council~(U.K.).


\begin{thebibliography}{99}

\baselineskip=12pt

\bibitem{gurarie} V. Gurarie, ``Logarithmic Operators in Conformal
  Field Theory'', Nucl. Phys. {\bf B410} (1993) 535 [{\tt
  hep-th/9303160}].

\bibitem{lcftfurther} M.A.I. Flohr, ``On Modular Invariant Partition
  Functions of Conformal Field Theory with Logarithmic Operators'',
  Int. J. Mod. Phys. {\bf A11} (1996) 4147 [{\tt hep-th/9509166}];
  ``On Fusion Rules in Logarithmic Conformal Field Theories'',
  Int. J. Mod. Phys. {\bf A12} (1997) 1943 [{\tt hep-th/9605151}];
  ``Operator Product Expansion in Logarithmic Conformal Field
  Theory'', Nucl. Phys. {\bf B634} (2002) 511 [{\tt
  hep-th/0107242}];\\ M.R. Rahimi-Tabar, A. Aghamohammadi and
  M. Khorrami, ``The Logarithmic Conformal Field Theories'',
  Nucl. Phys. {\bf B497} (1997) 555 [{\tt hep-th/9610168}];\\
  S. Moghimi-Araghi, S. Rouhani and M. Saadat, ``Logarithmic Conformal
  Field Theory through Nilpotent Conformal Dimensions'',
  Nucl. Phys. {\bf B599} (2001) 531 [{\tt hep-th/0008165}].

\bibitem{ckt} J.-S.~Caux, I.I.~Kogan and A.M.~Tsvelik, ``Logarithmic
  Operators and Hidden Continuous Symmetry in Critical Disordered
  Models'', Nucl.\ Phys. {\bf B466} (1996) 444 [{\tt hep-th/9511134}].

\bibitem{disorder} V.~Gurarie, M.A.I.~Flohr and C.~Nayak, ``The
  Haldane-Rezayi Quantum Hall State and Conformal Field Theory'',
  Nucl.\ Phys. {\bf B498} (1997) 513 [{\tt cond-mat/9701212}];\\
M.J.~Bhaseen, J.-S.~Caux, I.I.~Kogan and A.M.~Tsvelik, ``Disordered
Dirac Fermions: The Marriage of Three Different Approaches'', Nucl.\
Phys. {\bf B618} (2001) 465 [{\tt cond-mat/0012240}].

\bibitem{liouv} A. Bilal and I.I. Kogan, ``On Gravitational Dressing of
$2D$ Field Theories in Chiral Gauge'', Nucl. Phys. {\bf B449} (1995) 569
[{\tt hep-th/9503209}].

\bibitem{km} I.I.~Kogan and N.E.~Mavromatos, ``Worldsheet Logarithmic
Operators and Target Space Symmetries in String Theory'', Phys.\
Lett. {\bf B375} (1996) 111 [{\tt hep-th/9512210}].

\bibitem{kmw} I.I.~Kogan, N.E.~Mavromatos and J.F.~Wheater, ``D-Brane
  Recoil and Logarithmic Operators'', Phys.\ Lett.\ {\bf B387} (1996) 483
[{\tt hep-th/9606102}].

\bibitem{ms} N.E.~Mavromatos and R.J.~Szabo, ``Matrix D-Brane
Dynamics, Logarithmic Operators and Quantization of Noncommutative
Spacetime'', Phys.\ Rev.\ {\bf D59} (1999) 104018 [{\tt hep-th/9808124}].

\bibitem{logads} I.I. Kogan, ``Singletons and Logarithmic CFT in
  $AdS$/CFT Correspondence'', Phys. Lett. {\bf B458} (1999) 66 [{\tt
  hep-th/9903162}];\\ Y.S. Myung and H.W. Lee, ``Gauge Bosons and the
  $AdS_3$/LCFT$_2$ Correspondence'', J. High Energy Phys. {\bf 9910}
  (1999) 009 [{\tt hep-th/9904056}];\\ I.I. Kogan and D. Polyakov,
  ``Worldsheet Logarithmic CFT in $AdS$ Strings, Ghost-Matter Mixing
  and M-Theory'', Int. J. Mod. Phys. {\bf A16} (2001) 2559 [{\tt
  hep-th/0012128}];\\ S. Moghimi-Araghi, S. Rouhani and M. Saadat,
  ``On the $AdS$/CFT Correspondence and Logarithmic Operators'',
  Phys. Lett. {\bf B518} (2001) 157 [{\tt hep-th/0105123}].

\bibitem{LCFTRev} M.A.I.~Flohr, ``Bits and Pieces in Logarithmic
  Conformal Field Theory'', [{\tt hep-th/0111228}];\\ M.R. Gaberdiel,
  ``An Algebraic Approach to Logarithmic Conformal Field Theory'',
  [{\tt hep-th/0111260}];\\ M.R. Rahami-Tabar, ``Disordered Systems
  and Logarithmic Conformal Field Theory'', [{\tt cond-mat/0111327}].

\bibitem{GabKausch1} M.R. Gaberdiel and H.G. Kausch, ``Indecomposable
  Fusion Products'', Nucl. Phys. {\bf B477} (1996) 293 [{\tt
  hep-th/9604026}]; ``A Local Logarithmic Conformal Field
  Theory'', {\it ibid.} {\bf B538} (1999) 631 [{\tt hep-th/9807091}];
  ``A Rational Logarithmic Conformal Field Theory'', Phys. Lett. {\bf
  B386} (1996) 131 [{\tt hep-th/9606050}].

\bibitem{Gab1} M.R. Gaberdiel, ``Fusion Rules and Logarithmic
  Representations of a WZW Model at Fractional Level'',
  Nucl. Phys. {\bf B618} (2001) 407 [{\tt hep-th/0105046}].

\bibitem{fuchs} J.~Fjelstad, J.~Fuchs, S.~Hwang, A.M.~Semikhatov and
I.Yu.~Tipunin, ``Logarithmic Conformal Field Theories via Logarithmic
Deformations'', Nucl. Phys. {\bf B633} (2002) 379 [{\tt hep-th/0201091}].

\bibitem{CKLT} J.-S. Caux, I.I. Kogan, A. Lewis and A.M. Tsvelik, ``Logarithmic
Operators and Dynamical Extension of the Symmetry Group in the Bosonic
$SU(2)_0$ and SUSY $SU(2)_2$ WZNW Models'', Nucl. Phys. {\bf B489} (1997) 469
[{\tt hep-th/9606138}].

\bibitem{KAG} M. Khorrami, A. Aghamohammadi and A.M. Ghezelbash, ``Logarithmic
$N=1$ Superconformal Field Theories'', Phys. Lett. {\bf B439} (1998) 283 [{\tt
hep-th/9803071}].

\bibitem{MavSz} N.E. Mavromatos and R.J. Szabo, ``D-Brane Dynamics and
Logarithmic Superconformal Algebras'', J. High Energy Phys. {\bf 0110} (2001)
027 [{\tt hep-th/0106259}].

\bibitem{FQS} D.H. Friedan, Z. Qiu and S.H. Shenker, ``Superconformal
  Invariance in Two Dimensions and the Tricritical Ising Model'',
  Phys. Lett. {\bf B151} (1985) 37.

\bibitem{Kausch1} H.G. Kausch, ``Curiosities at $c=-2$'', [{\tt
    hep-th/9510149}]; ``Symplectic Fermions'', Nucl. Phys. {\bf B583}
    (2000) 513 [{\tt hep-th/0003029}].

\bibitem{KogNich1} I.I. Kogan and A. Nichols, ``Stress-Energy Tensor
  in LCFT and the Logarithmic Sugarawa Construction'', J. High Energy
  Phys. {\bf 0201} (2002) 029 [{\tt hep-th/0112008}].

\bibitem{rohsiepe} F. Rohsiepe, ``On Reducible but Indecomposable
  Representations of the Virasoro Algebra'', [{\tt hep-th/9611160}].

\bibitem{Gerasimov} A. Gerasimov, A.Yu. Morozov, M.A. Olshanetsky,
  A.~Marshakov and S.L.~Shatashvili, ``Wess-Zumino-Witten Model as a
  Theory of Free Fields'', Int. J. Mod. Phys. {\bf A5} (1990) 2495.

\bibitem{wakimoto} M. Wakimoto, ``Fock Representations of the Affine
  Lie Algebra $A_1^{(1)}$ '', Commun. Math. Phys. {\bf 104} (1986) 605.

\bibitem{nichols} I.I.~Kogan and A.~Nichols, ``$SU(2)_0$ and
  $OSp(2|2)_{-2}$ WZNW Models: Two Current Algebras, One Logarithmic
  CFT'', Int. J. Mod. Phys. {\bf A17} (2002) 2615 [{\tt hep-th/0107160}].

\bibitem{lcftgeneral} I.I. Kogan and A. Lewis, ``Origin of Logarithmic
  Operators in Conformal Field Theories'', Nucl. Phys. {\bf B509}
  (1998) 687 [{\tt hep-th/9705240}].

\bibitem{giribet} A. Nichols and S. Sanjay, ``Logarithmic Operators in
  the $SL(2,\real)$ WZNW Model'', Nucl. Phys. {\bf B597} (2001) 633
  [{\tt hep-th/0007007}];\\ G.~Giribet, ``Prelogarithmic Operators and Jordan
Blocks in $SL(2)_k$ Affine Algebra'', Mod. Phys. Lett. {\bf A16}
(2001) 821 [{\tt hep-th/0105248}].

\bibitem{DFMS} L. Dixon, D.H. Friedan, E.J. Martinec and S.H. Shenker, ``The
Conformal Field Theory of Orbifolds'', Nucl. Phys. {\bf B282} (1987) 13.

\bibitem{gezel} A.M. Ghezelbash and V. Karimipour, ``Global Conformal
  Invariance in $d$ Dimensions and Logarithmic Correlation
  functions'', Phys. Lett. {\bf B402} (1997) 282 [{\tt
  hep-th/9704082}];\\ M. Khorrami, A. Aghamohammadi and
  M.R. Rahimi-Tabar, ``Logarithmic Conformal Field Theories with
  Continuous Weights'', Phys. Lett. {\bf B419} (1998) 179 [{\tt
  hep-th/9711155}];\\ A.M. Ghezelbash, M. Khorrami and
  A. Aghamohammadi, ``Logarithmic Conformal Field Theories and $AdS$/CFT
  Correspondence'', Int. J. Mod. Phys. {\bf A14} (1999) 2581 [{\tt
  hep-th/9807034}].

\bibitem{FMS1} D.H. Friedan, E.J. Martinec and S.H. Shenker, ``Conformal
Invariance, Supersymmetry and String Theory'', Nucl. Phys. {\bf B271} (1986)
93.

\bibitem{gsw} M.B. Green, J.H. Schwarz and E. Witten, {\it Superstring
    Theory} (Cambridge University Press, 1987).

\bibitem{fermspin} D.H. Friedan, E.J. Martinec and S.H. Shenker, ``Covariant
Quantization of Superstrings'', Phys. Lett. {\bf B160} (1985) 55;\\ V.G.
Knizhnik, ``Covariant Fermionic Vertex in Superstrings'', Phys. Lett. {\bf
B160} (1985) 403.

\bibitem{KLLSW} V.A. Kostelecky, O. Lechtenfeld, W. Lerche, S.~Samuel
  and S.~Watamura, ``Conformal Techniques, Bosonization and Tree-Level
  String Amplitudes'', Nucl. Phys. {\bf B288} (1987) 173.

\bibitem{Saleur1} H. Saleur, ``Polymers and Percolation in
  Two-Dimensions and Twisted $N=2$ Supersymmetry'', Nucl. Phys. {\bf
  B382} (1992) 486 [{\tt hep-th/9111007}].

\bibitem{LMRS} F. Lesage, P. Mathieu, J. Rasmussen and H. Saleur,
  ``The $\widehat{su}(2)_{-1/2}$ WZW Model and the $\beta\gamma$
  System'', [{\tt hep-th/0207201}].

\bibitem{wdoubl} F.~Yu, ``The Super-KP Origin of Super-$W_{1+\infty}$
  Algebra and its Topological Version'', Nucl.\ Phys. {\bf B375} (1992) 173.

\bibitem{twist} T.~Eguchi and S.K.~Yang, ``$N=2$ Superconformal Models
  as Topological Field Theories'', Mod.\ Phys.\ Lett. {\bf A5} (1990)
  1693;\\ S.~Nojiri, ``Superstring in Two-Dimensional Black Hole'',
Phys.\ Lett. {\bf B274} (1992) 41 [{\tt hep-th/9108026}].

\bibitem{coulomb} V.G. Kac and I.T. Todorov, ``Superconformal Current
  Algebras and their Unitary Representations'', Commun. Math.
Phys. {\bf 102} (1985) 337.

\bibitem{swzw} P.~Di Vecchia, V.G.~Knizhnik, J.L.~Petersen and P.~Rossi,
``A Supersymmetric Wess-Zumino Lagrangian in Two-Dimensions,''
Nucl.\ Phys. {\bf B253} (1985) 701;\\ J. Fuchs, ``Superconformal Ward
Identities and the WZW Model'', Nucl.\ Phys. {\bf B286} (1987) 455;
``More on the Super WZW Theory'', {\it ibid.} {\bf B318} (1989) 631;\\
E. Kiritsis and G. Siopsis, ``Operator Algebra of the $N=1$ Super
Wess-Zumino Model'', Phys.\ Lett. {\bf B184} (1987) 353 [Erratum: {\it
  ibid.} {\bf B189} (1987) 499];\\ S. Nam, ``Superconformal and Super
Kac-Moody Invariant Quantum Field Theories in Two-Dimensions'', Phys.\
Lett. {\bf B187} (1987) 340.

\bibitem{terao} H.~Terao, ``The Coulomb Gas Representation for $SU(2)$
  Wess-Zumino-Witten Model in Superspace'', Mod.\ Phys.\ Lett. {\bf
  A5} (1990) 1731.

\bibitem{nemesh} D. Nemeschansky, ``Feigin-Fuks Representation of
  $su(2)_k$ Kac-Moody Algebra'', Phys.\ Lett. {\bf B224} (1989) 121.

\bibitem{FF1} B.L. Feigin and D.B. Fuks, ``Verma Modules over the
  Virasoro Algebra'', Funct. Anal. Appl. {\bf 17} (1983) 241.

\bibitem{screen} M.A.~Bershadsky, V.G.~Knizhnik and M.G.~Teitelman,
``Superconformal Symmetry in Two-Dimensions'', Phys.\ Lett. {\bf B151}
(1985) 31.

\bibitem{Cardy1} J.L. Cardy, ``Logarithmic Correlations in Quenched
  Random Magnets and Polymers'', [{\tt cond-mat/9911024}];\\
  V. Gurarie and A.W.W. Ludwig, ``Conformal Algebras of $2D$
  Disordered Systems'', J. Phys. {\bf A35} (2002) L377 [{\tt
  cond-mat/9911392}].

\bibitem{nicholsmult} A. Nichols, ``Extended Multiplet Structure in
  Logarithmic Conformal Field Theories'', [{\tt hep-th/0205170}].

\bibitem{grav} E. Gravanis and N.E. Mavromatos, ``Higher-Order
  Logarithmic Conformal Algebras from Robertson-Walker $\sigma$-Model
  Backgrounds'', J. High Energy Phys. {\bf 0206} (2002) 019 [{\tt
  hep-th/0106146}].

\bibitem{Felder1} G. Felder, ``BRST Approach to Minimal Models'',
  Nucl. Phys. {\bf B317} (1989) 215.

\bibitem{Corrigan1} E.F. Corrigan and D.I. Olive, ``Fermion-Meson
  Vertices in Dual Theories'', Nuovo Cim. {\bf A11} (1972) 749;\\
  E.F. Corrigan and P. Goddard, ``Gauge Conditions in the Dual Fermion
  Model'', Nuovo Cim. {\bf A18} (1973) 339.

\bibitem{witten} E.~Witten, ``On String Theory and Black Holes'',
Phys.\ Rev. {\bf D44} (1991) 314.

\bibitem{Mald1} J.M. Maldacena and H. Ooguri, ``Strings in $AdS_3$ and
  the $SL(2,\real)$ WZW Model 1: The Spectrum'', J. Math. Phys. {\bf
  42} (2001) 2929 [{\tt hep-th/0001053}].

\bibitem{figueroa} J.M.~Figueroa-O'Farrill and S.~Stanciu,
``$N=1$ and $N=2$ Cosets from Gauged Supersymmetric WZW Models'',
[{\tt hep-th/9511229}];
``Supersymmetric Cosets from Gauged SWZW Models'',
Mod.\ Phys.\ Lett. {\bf A12} (1997) 1677.

\bibitem{Sfetsos1} I. Bakas and K. Sfetsos, ``PP-Waves and Logarithmic
  Conformal Field Theories'', Nucl. Phys. {\bf B639} (2002) 223 [{\tt
  hep-th/0205006}].

\bibitem{Sfetsos2} K. Sfetsos, ``String Backgrounds and LCFT'',
  Phys. Lett. {\bf B543} (2002) 73 [{\tt hep-th/0206091}].

\end{thebibliography}
\end{document}